\documentclass[11pt, twoside, reqno]{amsart}
\usepackage[thm_section]{macros}
\usepackage{float}
\newfloat{prompt}{tbhp}{lom}[section]
\floatname{Prompt}{Prompt}
\crefname{prompt}{Prompt}{Prompts}

\begin{document}
\begin{titlepage}
\begin{spacing}{1}
\title{\textbf{\Large Counting Clinical Trials:}\\
\Large{New Evidence on Pharmaceutical Sector Productivity}}
\author{
\begin{tabular}[t]{c@{\extracolsep{2em}}c@{\extracolsep{2em}}c} 
\large{Maya M. Durvasula}$^\star$ &  
\large{Sabri Eyuboglu}$^\dagger$ &
\large{David M. Ritzwoller}$^\ddagger$
\vspace{-0.3em}\vspace{-0.7em}
\end{tabular}%
\\
}
\date{%
\today\\ 
$\star$ Stanford Department of Economics and Stanford Law School, \url{maya.durvasula@stanford.edu}\\
$\dagger$ Stanford Department of Computer Science, \url{eyuboglu@stanford.edu}\\
$\ddagger$ Stanford Graduate School of Business, \url{ritzwoll@stanford.edu}\\
 We thank Marcella Alsan, Nicholas Bloom, Agnes Cameron, Jiafeng Chen, Liran Einav, Matthew Gentzkow, Han Hong, Guido Imbens, Ramesh Johari, Charles Jones, Neale Mahoney, Evan Munro, Lisa Larrimore Ouellette, Christopher Ré, Joseph Romano, Brad Ross, Bhaven Sampat, Dean Stratakos, Michael Weisman, Heidi Williams, Brandon Yang, Frank Yang, James Zou, and seminar audiences at the NBER Innovation Information Initiative and the Stanford Department of Medicine for helpful comments and conversations. We are especially grateful to Arjun Desai and Karan Goel for their work on the technical infrastructure that enabled this project. We gratefully acknowledge financial support from the National Science Foundation through grant DGE-1656518 (Durvasula, Eyuboglu, and Ritzwoller), the Knight Hennessy Scholars Program, the Stanford Law School John M. Olin Program in Law and Economics, the National Bureau of Economic Research Innovation Information Initiative Summer Fellows Program, and the OpenAI Researcher Access Program. The data produced in this paper are available at the link: \url{https://github.com/DavidRitzwoller/pubmed_clinical_trials}.}
                      
\begin{abstract}
\normalsize{We develop a method for assigning high-quality labels to unstructured text. This method is based on fine-tuning an efficient, open-source language model with data extracted from a large, proprietary language model. We apply this method to construct a census of published clinical trials. With these data, we revisit a literature that contends that pharmaceutical sector productivity is declining. Central to this conclusion are measurements of substantial increases in the quantity of clinical trials over time, unmatched by trends in measures of output. In our data, the quantity, quality, and composition of clinical trials are stable since 2010. We show that previous measurements are an artifact of biases introduced by shifts in the composition of other forms of research.}
\end{abstract}
\end{spacing}
\end{titlepage}
\maketitle
\thispagestyle{empty}
\clearpage
\setcounter{page}{1}

\begin{spacing}{1.6}
\section{Introduction}\label{sec:introduction}

For nearly thirty years, policy discussions about the pharmaceutical industry have been colored by the concern that sectoral productivity is in a state of decline \citep[see e.g.,][]{engelberg1998special,harris2011federal}. A series of influential analyses, drawing on industry reports and U.S. Food and Drug Administration records, suggest that the resources required to bring a new drug to market have been increasing substantially year-on-year. \cite{scannell2012diagnosing}, for example, estimate that the productivity of the pharmaceutical sector has halved roughly every decade since 1950; \cite{ruffolo2006has}, \cite{pammolli2011productivity}, \cite{myers2019endogenous}, and \cite{bloom2020ideas} arrive at similar conclusions.

The policy consequences of these findings have been substantial. Proposals that would restructure public subsidies administered by the National Institutes for Health (NIH) \citep[e.g.,][]{NSTC2019,hamiltonproductivity2020,economist2022escape}, relax the regulatory standards for approval of new medicines by the U.S. Food and Drug Administration (FDA) \citep[e.g.,][]{roy2021high,chertman2023slow,Graboyes2017}, and prevent the Centers for Medicare and Medicaid Services (CMS) from negotiating drug prices \citep[e.g.,][] {york2023inflation,long2023hidden,inflation_reduction_act_2022} are premised, in no small part, on the concern that existing policy is contributing to this pattern of decline.\footnote{In recent years, discussions of declining medical and pharmaceutical research productivity have been especially extensive in the popular press. \cite{CollisonNielsen2018}, \cite{Broad2023}, \cite{Thompson2021}, and \cite{Piper2023}, for example, interpret trends in pharmaceutical research productivity, alongside patterns collected from other fields, as a `warning sign'' for the future of technological progress \citep{Piper2023}. } For the same reasons, an architect of the 1984 Hatch Waxman Act, which established the modern system for pharmaceutical regulation in the United States, called for an overhaul of that regulatory system just fifteen years after its inception \citep{engelberg1998special}. These patterns also spurred the 2011 establishment of the National Center for Translational Sciences, a federal agency with a roughly one billion dollar annual budget and a mandate to speed up the commercialization of life-saving medicines---on the grounds that, left to private firms, this process had become too inefficient.\footnote{The fiscal year 2025 budget request for NCATS is \$926.1 million. For details on its founding, see \cite{harris2011federal}. For details on its budget, see \url{https://ncats.nih.gov/about/budget}.} 

Though this declining pharmaceutical sector productivity is often characterized as an empirical regularity on par with Moore's Law in computing, its empirical support is quite thin.\footnote{Moore's Law is the observation that the number of transistors on an integrated circuit will double every two years with minimal rise in cost. See \cite{moore1965cramming}. \cite{scannell2012diagnosing} coined ``Eroom's Law'' (Moore, spelled backwards) to describe pharmaceutical sector productivity.} On some measures, the quantity of resources devoted toward the commercialization of new medicines appears to be sharply increasing over time. Given that the number of innovative drugs approved by the FDA has been roughly constant, year-on-year, since 1990, this divergence between inputs and outputs generates the conclusion of declining productivity. 

This paper constructs new data on pharmaceutical sector inputs. We develop a method for assigning high-quality labels to unstructured text. This method is based on fine-tuning an efficient, open-source language model with data extracted from a large, proprietary language model. We apply this method to construct a new census of clinical trials disclosed in scientific publications. We document that the quantity, quality, and composition of published clinical trials have remained remarkably stable year-over-year. On the basis of these facts, we find that there is insufficient evidence to reject the null hypothesis that pharmaceutical sector productivity has remained stable since 2010. 

We show that reported increases in the quantity of published clinical trials, offered as recent evidence of declining productivity in \cite{bloom2020ideas}, result from imprecise classification methods. These methods misattribute shifts in the composition of biomedical research to trends in clinical trial production. When our data are combined with evidence from \cite{sertkaya2024costs}---that clinical trial costs, and broader drug development expenses, have remained roughly constant since 2000---our findings leave little in the way of evidence for the conventional wisdom of a persistent decline in pharmaceutical sector productivity.

This paper makes two primary contributions. First, we revise a fact central to debates on biomedical productivity. Second, we demonstrate how advances in generative AI can address longstanding challenges in measuring innovation. The paper is, thus, organized in two parts. 

In the first part, we introduce a procedure for constructing task-specific large language models, particularly suited to data construction problems that involve classification.\footnote{We include, as a supplemental appendix to this paper, a more detailed summary of this procedure, which may be useful for future work.} The class of methods that we consider has two steps. First, a comparatively small number of labels are extracted from a large, typically proprietary, language model.\footnote{By proprietary large language model, we are referring, at present, to models such as OpenAI's GPT 3.5 and GPT-4. Alternative models include Anthropic's Claude 3.5 Opus, Google's Gemini, among many others.} Second, these labels are used as training data for a smaller, more efficient, open-source language model.\footnote{Here, we are referring to classes of open-source language models such as RoBERTa, based on Google's BERT model, among others.} This process is an instance of model distillation, a growing area of methodological research in computer science \citep{taori2023alpaca,xu2024survey}. This procedure allows researchers to capture the benefits of complex, frontier models, while retaining the resource efficiency and transparency of lighter-weight open-source models. 

We apply this method in stages. Our objective is to identify all records in the National Library of Medicine's PubMed/MEDLINE database, from 2010 forward, that disclose the results of a clinical trial that studied the effects of a medicine in human subjects. We aim to classify records based only on the contents of the publication's abstract, with an accuracy close to that of a human labeler. We devise an interface to iteratively revise model prompts. With the highest performing prompts, we use proprietary language models, OpenAI’s GPT-3.5 and GPT-4, to classify a large number of records \citep{achiam2023gpt,bubeck2023sparks}. These noisy labels are used as training data to fine-tune a set of open-source models. An ensemble of the fine-tuned models matches the performance of the best proprietary model, with false positive and false negative rates below 5 percent, and produces a sample of approximately 150,000 publications that report the results of clinical trials from 2010 to 2022. 

The second part of this paper interprets trends in these data, in light of existing evidence. We begin by establishing that trends in clinical trials, as measured in publication data, are central to conclusions of declining biomedical and pharmaceutical productivity. We show that standard methods for data classification identify clinical trials in publication data imprecisely. These imprecise measurements indicate, spuriously, that the quantity of clinical trials has increased sharply since 2010. With improved classification, we find that the quantity, quality, and composition of clinical trials have been remarkably stable over this period. 

We collect a set of ex-ante and ex-post quality measures, derived from scientific publication data. On these dimensions, we find considerable heterogeneity that is, to our knowledge, new to the literature. Roughly half of clinical trial publications in our sample are never cited by a leading medical journal. Nearly 15 percent are never cited by any other scientific publication. However, we find that the distribution across clinical trials of measures of quality, importance, and geography is unchanging over time. 

Our data offer more suggestive evidence on a second piece of evidence cited in support of declining research productivity---that the quantity of scientific publications is increasing sharply over time. We document evidence of large compositional shifts in \textit{other} biomedical publications, including a shift in the geographic composition of research and a large increase in the number of articles that summarize existing literature. In line with the arguments of a long-standing literature on similar shifts in patent data \citep[e.g.,][]{griliches1990patent,griliches1994}, these patterns cast doubt on whether simple counts of scientific publications are stable, informative measures of research effort. These insights may be more generally applicable to discussions about how large increases in the quantity of publications---as documented in \cite{park2023papers} and \cite{chu2021slowed}---ought to be interpreted. 

\subsection{Related Literature}
We draw on, and contribute to, a long literature on the measurement of invention. This paper considers, in effect, a binary classification task, of a kind that has long been a challenge for studies of innovation. The proliferation of large-scale corpora of patents and scientific publications has, seemingly, expanded opportunities to measure instances of research and invention \citep{bryan2021innovation}. Patents and publications are attractive proxies for innovation, as they necessarily relate to instances of invention and research. Their utility is limited, however, by heterogeneity in their technical content and value \citep[see e.g.,][]{pakes1986patents,griliches1990patent,griliches1994}. Consequently, trends in counts of documents are often confounded by trends in their composition \citep{lerner2022use}. Large-scale classification efforts have focused on mapping pieces of unstructured text to economic quantities of interest.\footnote{The most prominent example of such a classification effort is the construction of the National Bureau of Economic Research (NBER) Patent Data, which, among other things, maps patent records to technology classes \citep{hall2001nber}. Earlier, \cite{scherer1984using} classified, by hand, over 15,000 patents according to their technological category.} In this paper, we demonstrate that large language models can facilitate classification tasks of this kind, achieving an accuracy and precision comparable to hand-made labels at a research-relevant scale and budget.

More specifically, within studies of medical innovation, we contribute to a body of work that uses clinical trials as a measure of research investment. Trials---discrete, regulated experiments that generate structured statistical information on new and existing health technologies \citep{chavez2016randomized}---are an especially attractive proxy for the level and composition of medical research. Since the early 2000s, clinical trials have been reported in a consistent, standardized format across scientific publications and, with less consistency, in administrative databases \citep{devito2020compliance,devito2021evaluation}. Trials have, thus, been widely used to measure relative and absolute differences in the composition of research investment \citep[see e.g.,][]{finkelstein2004static,yin2008market,budish2015firms}. Recent studies of information production and disclosure have examined the substantive content of trials (see e.g., \cite{alsan2024representation}, \cite{oostrom2024funding}, and \cite{kao2024representation}). 

Given limitations in publicly-available administrative trial registries \citep{zarin2017update,devito2020compliance,devito2021evaluation}, recent empirical studies in economics have used trial data extracted from scientific publications \citep[e.g.,][]{lichtenberg2018impact,bloom2020ideas,kao2024information,guzman2024}. The challenge, for each, has been again one of classification: of the millions of records in large-scale publication databases, which disclose the results of a clinical trial? Manually identifying clinical trial records at a scale sufficient to study aggregate trends is infeasible.\footnote{A small number of papers curate select samples of clinical trial data, for specific research questions. For example, \cite{oostrom2024funding} collects records of psychiatric clinical trials as identified in prominent meta-analyses for a study of the relationship between sources of clinical trial funding and reported drug efficacy.} We show that standard methods for classification---based on publication data and machine learning tools---perform poorly. In particular, we show that keyword-based searches, including methods used in a productivity analysis in \cite{bloom2020ideas}, are very imprecise and generate spurious measurements of increases in the quantity of clinical trials.\footnote{Specifically, \cite{bloom2020ideas} categorize many records as clinical trials that discuss trials (e.g., meta-analyses and literature reviews). We show, in Section 4, that changes in the scientific publishing landscape increase the size of this categorization error over time.} Specialized classifiers developed for use in the health services literature, as in \cite{thomas2021machine}, similarly lack the precision necessary to recover meaningful trends. We introduce both a new method and new data that address this problem. 

The idea that patterns like those considered in this paper---rising research quantities and stagnant outcomes---signal the ``exhaustion of inventive opportunities'' \citep{griliches1994} is old and closely tied to the development of methods for the measurement of innovation. Early appearances of this concern in the literature include Robert K. Merton's 1935 discussion of the ``possibility of a slackening in the rate of technologic advance'' \citep{merton1935fluctuations}, Alfred B. Stafford's 1952 paper titled ``Is the Rate of Invention Declining?" \citep{stafford1952rate} and Jacob Schmookler's various investigations \citep{schmookler1952changing,schmookler1954level,schmookler1966invention} into patterns suggestive of diminishing returns to research. Our treatment of these patterns hews closely to that of earlier literatures, which view the evidence base on productivity decline as inconclusive, given the shortcomings of available data \citep[e.g.,][]{pakes1980patents,griliches1990patent,griliches1994,cockburn2006pharmaceutical}.\footnote{\cite{griliches1990patent} summarizes his version of the conclusion that any finding of diminishing returns to innovative efforts are limited by the availability of high-quality data more prosaically: ``One can always worry that the world is coming to an end. Someday it undoubtedly will, but it does not look as if the end is already upon us, at least not yet.''} This paper differs on one important dimension. We show that advances in data construction technology, if deployed carefully, can allow researchers to resolve some long-standing data construction challenges.

A nascent literature investigates the use of large language models in economics \citep{dell2024deep,korinek2023generative,ludwig2025large}. We join a small set of papers---notably, \cite{bartik2023costs} and \cite{dell2024deep}---that use generative AI as a tool to construct high-quality economic data. Our approach is distinct, in that we fine-tune an open-source language model using data extracted from a larger, proprietary model. In doing so, we show that, with model distillation, it is feasible for researchers to achieve the quality of hand-constructed data at a practically interesting scale, with a reproducible and transparent methodology, at a feasible cost. Prominent examples of model distillation in the computer science literature focus on the construction of chat bots by fine tuning open source models with data extracted from proprietary models \citep{taori2023alpaca,chiang2023vicuna,xu2023baize}. Other papers have fine-tuned models for simple completion tasks like arithmetic \citep{liu2023goat} or querying an API \citep{patil2023gorilla}. We are distinguished from this literature by considering a problem with, arguably, a higher-degree of complexity: the construction of data of sufficient quality to be of substantive interest to social science.

Our use of large language models is also distinct from other applications considered in papers such as \cite{korinek2023generative}, \cite{manning2024automated}, \cite{bybee2023ghost}, \cite{horton2023large}, and \cite{mullainathan2024predictive}, which focus on the use of generative AI for other aspects of the scientific process, including hypothesis identification, survey participation, idea generation, and scientific writing. It is worth emphasizing that, in this paper, we develop a procedure that is applicable to a binary classification problem, albeit one that is highly complex. The viability of these methods, and the extent of required modifications from the process outlined here, for application in more complicated classification problems of interest to economists, including those outlined in \cite{dell2024deep}, is a valuable direction for further research.

More generally, we contribute to the large empirical literature that uses text as data \citep{gentzkow2019text}. We anticipate that our procedure for the construction of new, precise data on scientific research may be applicable to other contexts where unstructured text is available and manual construction of data is infeasible. We highlight a set of steps---in validation, prompt design, and fine-tuning of open source models---that yield  high-quality data at comparatively low cost.

\section{Counting Clinical Trials}\label{sec:identifying-trials}

We construct a new census of clinical trials disclosed in scientific publications. We begin, in \cref{sec: preliminaries} and \cref{sec: existing}, by providing context on the objectives of our data construction exercise and by detailing the deficiencies of existing, or alternative, approaches to measuring this form of medical innovation. In \cref{sec: objective,sec: llm main} we propose a method, based on distilling data from a large language model, for identifying relevant records at scale and with high fidelity. The performance of the method is assessed in \cref{sec: performance}.

\subsection{Context\label{sec: preliminaries}} 

Brief institutional context on the existing evidence related to pharmaceutical research productivity is helpful both in motivating our data construction and in interpreting our findings. 

Declining pharmaceutical research productivity has been a topic of policy concern since the late 1990s. In the past thirty years, discussions have largely centered on two sets of empirical patterns, which \cite{cockburn2006pharmaceutical} summarizes as follows: ``More and more money is being invested in R\&D, but the rate at which new drugs are introduced is failing to keep pace.'' In these exercises, the empirical object of interest is a measure of investments in preclinical and, especially, clinical studies intended to translate discoveries made in basic science into commercialized medicines suitable for use in human patients. 

The ``puzzle'' as characterized by \cite{cockburn2006pharmaceutical} and others centers on the apparent decline in the efficiency of pharmaceutical firms tasked with translational research. Hypotheses for the decline are wide-ranging. \cite{ruffolo2006has} emphasizes shifts in managerial incentives. \cite{cockburn2006pharmaceutical} highlights  ``vertical disintegration'' of the pharmaceutical industry. \cite{scannell2012diagnosing} points to rising regulatory thresholds. \cite{myers2019endogenous} and \cite{bloom2020ideas} suggest that the ``low-hanging fruit''---high-efficacy, easy-to-test drug candidates---have already been brought to market. Though the candidate mechanisms in these discussions differ, declining productivity in the pharmaceutical sector is treated as a fact \citep{bloom2020ideas,goldin2024productivity}.\footnote{\cite{scannell2012diagnosing} coin the term ``Eroom's Law''---Moore's Law in computing, spelled backwards---to described this trend. Policy analyses and academic surveys (e.g., \cite{goldin2024productivity}) have characterized both Eroom's Law and Moore's Law as well-supported stylized facts.} 

The measures of industry productivity central to these debates are somewhat crude. For the pharmaceutical industry, a standard measure of output is the number of ``new molecular entities''---innovative drugs---approved each year by the U.S. Food and Drug Administration.\footnote{Technically, a new molecular entity is a drug that contains no previously approved active moiety. In practice, this means that every active element of the drug (e.g., acetaminophen and caffeine in the commonly used over-the-counter migraine medicine Excedrin) must be novel. This is the same output measure used in discussions of changes in federal research policy. See, for example, \cite{harris2011federal}.} As \cite{bloom2020ideas} note, that this count has been stagnant or declining over time is viewed as an established fact.\footnote{In data on new molecular entities approved by the FDA between 2005 and 2021, we observe a small increase in the number of approvals, largely driven by an uptick in approvals in 2020 and 2021. In earlier periods of data, we confirm that there is no discernible trend in new molecular entity approvals.} 

Measures of inputs are, typically, some estimate of R\&D expenditures. The most widely cited analysis of pharmaceutical research productivity, \cite{scannell2012diagnosing}, uses a measure of annual, aggregated industry-wide R\&D expenditures, published by the trade organization Pharmaceutical Research and Manufacturers of America (PhRMA).\footnote{These PhRMA estimates are used more widely. See e.g.,  \cite{myers2019endogenous} for use in an economics paper on pharmaceutical productivity. See e.g., \cite{harris2011federal} for a discussion in the popular press.} There are few other sources of R\&D expenditure data for the pharmaceutical industry. \cite{dimasi2003price} and \cite{dimasi2016innovation} provide, perhaps, the most commonly cited estimates of drug development costs, derived from internal records of a select set of firms. In an editorial, \cite{frank2003new} describes these estimates as "a matter of heated debate since they were first made public."\footnote{\cite{Light2005}, for example, criticizes these estimates derived from industry data as "nonrandom and small" and characterizes the findings as based on "unverifiable industry data."} 

Given these substantial limitations of industry-wide aggregate estimates, a natural alternative is to decompose aggregate R\&D into its component parts. The simplest such decomposition would develop estimates of the cost of pre-clinical and clinical studies and scale this quantity by the number of such clinical trials. In practice, nearly all work focuses on clinical trials and assumes that pre-clinical trials contribute a fixed proportion of total development costs \citep{dimasi2003price,dimasi2016innovation,sertkaya2024costs}.\footnote{Ideally, estimates of clinical trial costs would be inflation-adjusted. To date, there are no price indices that are appropriate for such an adjustment. See \cite{berndt2013price} for an effort to construct such a price index. See \cite{sertkaya2016key} for a discussion of efforts to construct cost estimates, including limitations associated with this type of adjustment.} Two recent estimates of these objects come from \cite{bloom2020ideas} and \cite{sertkaya2024costs}. In an influential analysis, \cite{bloom2020ideas} document a sharp increase in the number of clinical trials over time, using data constructed using scientific publications. In \cite{bloom2020ideas} and in a recent survey article, \cite{goldin2024productivity}, this fact is central to the argument that pharmaceutical productivity is declining. Using granular data on clinical trial site contracts and internal records of the U.S. Food and Drug Administration, \cite{sertkaya2024costs} find both that the cost of clinical trials and the cost of drug development were roughly constant between 2000 and 2018.\footnote{The data used in the \cite{sertkaya2024costs} analysis have previously been used by a small set of other researchers---notably, \cite{azoulay2004capturing} and \cite{berndt2013price}. As of 2019, the firm that collects data on clinical trial contracts has been acquired. As of 2022, the data are no longer available for research use. Correspondence on file with the authors.} 

Arguments of declining productivity in the pharmaceutical industry, then, depend substantially on the finding that the quantity of clinical trials is rising. 

\subsection{Existing Approaches\label{sec: existing}} Although many datasets report information on clinical trials---and have been increasingly used by economists to measure innovation \citep[see e.g.,][]{budish2015firms,cunningham2021killer}---no existing dataset, used off-the-shelf, yields accurate, transparent aggregate quantity measures.

\subsubsection{Trial Registries and Proprietary Databases}

Nearly all clinical trials are conducted under the auspices of a regulator that requires both pre-registration and results disclosure. In the United States, the Food and Drug Administration Amendments Act (FDAAA), as passed in 2007 and revised in a "Final Rule" that took effect in 2017, imposes these requirements on sponsors of regulated trials. Technically, affected sponsors are required to register their studies on ``ClinicalTrials.gov.'' For nearly all studies, sponsors must also report the results of a trial within a designated period after completion. Moreover, increasingly, medical journals require registration as a pre-condition of publication \citep{ICMJE}. 

In practice, there is widespread noncompliance with these regulations and requirements---by private firms, academic institutions, and the federal government itself.\footnote{There have been many extremely careful, detailed analyses of ClinicalTrials.gov non-compliance. See \cite{anderson2015compliance} for one comprehensive analysis. The FDAAA Trials Tracker---\url{https://fdaaa.trialstracker.net/}---provides up-to-date, record-by-record details on non-compliance.} Deborah Zarin, then-director of ClinicalTrials.gov, notes that ``funders, sponsors, and institutional review boards continue to allow unregistered trials or trials with late registration to be conducted, and some journals continue to allow the results of such trials to potentially be published'' \citep{zarin2017update}. Noncompliance with federal reporting requirements is similarly widespread. In a series of papers assessing compliance over time, \cite{devito2020compliance} and \cite{devito2021evaluation} find that roughly 60 percent of trials registered on ClinicalTrials.gov fail to report required data elements at the time of their reporting deadline.\footnote{Note that noncompliance in this industry persists despite the existence of high statutory penalties. Sponsors can be fined up to \$10,000 per day for failing to register studies and disclose results. See the FDA Amendments Act of 2007 (FDAAA) and its Final Rule issued in 2021, at 42 U.S.C. § 282(j)(5)(C)(i). To date, no fines have been collected, but the FDA now sends notices of noncompliance to delinquent sponsors \citep{stephenson2021first}. Notices are available here: {https://www.fda.gov/science-research/fdas-role-clinicaltrialsgov-information/clinicaltrials gov-notices-noncompliance-and-civil-money-penalty-actions}}

From the perspective of our study, there are three issues with the use of registry data. First, requirements for study registration have changed over time. \cref{tab:ctgov policies} lists five changes between 2000 and 2020 that altered the set of studies that should appear in a complete version of the database. \cite{oostrom2024funding} documents, in a case study of psychiatric drugs, large changes in reporting propensity over this time period.\footnote{More details on the \cite{oostrom2024funding} comparison of registry data and publication data appear in Section 4.} Second, sponsors engage in considerable ``back-filling'' of past studies. \cite{zarin2017update} find that a large proportion of trials are registered after the trial start date.\footnote{\label{fn: back-fill}Often back-filling relates to changes in organizational disclosure policy. In a version of ClinicalTrials.gov current through 31 May 2023, Boehringer Ingelheim---the largest privately held pharmaceutical company in the world---appears to have posted 804 studies in 2014, only 72 of which were started in 2014. In 2013, Boehringer registered 101 studies. In the fifteen preceding years, since the creation of the database, Boehringer posted 872  records in total. We selected this case study based on a discussion in \cite{zarin2017update}.}  Third, not all records in ClinicalTrials.gov and similar registries are, in fact, clinical trials \citep{tse2018avoid}. As a result, it is difficult to determine whether changes in the number of records reflect actual growth in scientific research or a change in the propensity to register studies (trials and non-trial studies).

In \cref{sec: ctgov overview}, we illustrate the difficulty of recovering accurate measures of trends in the quantity of clinical trials from this database. In particular, we show that small changes in data construction yield meaningfully different estimates of trial levels and that many records are not in fact clinical trials. Moreover, due to widespread data gaps resulting from noncompliance in reporting of results and study attributes, there is little complete information on characteristics of registered studies.\footnote{See \cite{alsan2024representation} for a discussion of gaps in the reporting of the racial composition of studies.}

Other databases, such as the PDQ Cancer Information database maintained by the U.S. National Cancer Institute and used in papers including \cite{budish2015firms}, are disease-specific and thus insufficiently general for our purposes. A small number of firms (for example, GlaxoSmithKline) have disclosed all clinical trials over a longer period of time.\footnote{See GlaxoSmithKline's registry here: https://www.gsk-studyregister.com/. \cite{gibson2004glaxosmithkline} provides an overview of the lawsuit that led to the development of this registry.} Of course, records from a single firm are not informative about aggregate trends, especially because many investigational drugs change hands frequently during the course of development \citep{cunningham2021killer}.

Researchers, instead, often draw on proprietary databases. Although these databases are extremely useful for certain research questions---for example, regarding changes in the ownership of investigational drugs \citep[][using Pharmaprojects data]{cunningham2021killer} or on the link between insurance policy and drug development \citep[][using Cortellis data]{agha2022insurance}---they are unsuitable for analyses of either research trends (as in this paper) or summaries of the evidence base relevant for clinical decision-making (as in the health services literature). Our aim is to construct a census that allows for comparison of quantities across points in time. First, we are unable to determine, with specificity, how records are identified and cataloged based on available documentation. Second, these databases take public registries as inputs and, thus, risk adopting the same shifts in reporting over time \citep[see e.g., the list of registry inputs for one frequently-used database,][]{cortellislabs}. 

\subsubsection{Publication Data}
Publication data provide a natural alternative.\footnote{In Section 4, we discuss the two natural concerns with measures constructed from publication data---publication bias and shifts in reporting norms over time. Our data, a recent analysis by \cite{oostrom2024funding}, and the relevant institutional context suggest, together, that neither is likely to threaten our interpretation of \textit{trends} in publication data.} The database PubMed / MEDLINE indexes roughly $34$ million publications, a near universe of published biomedical research.\footnote{See \cref{sec: pubmed overview} for further details on our treatment of PubMed data.} The challenge is to classify the specific forms of research captured by this corpus. Of these $34$ million records, which are clinical trials? 

The existing literature takes several approaches. In health sciences---where data on the universe of clinical trials are relevant for the construction of meta-analyses and the development of prescribing guidelines---researchers typically employ overly-inclusive methods characterized by high true positive rates. To use one prominent example, Cochrane, a British organization that synthesizes the findings of medical research, developed the ``Cochrane RCT Classifier,'' a machine learning-based classifier for retrieving randomized controlled trials. The Cochrane approach successfully identifies trials with a 99 percent true positive rate. For every eight true positives, however, it returns 92 false positives. \cite{thomas2021machine} provides details on the Cochrane approach.

Approaches used by researchers in social science, instead, largely draw on metadata contained in PubMed.\footnote{\cite{oostrom2024funding} and \cite{kao2023transparency} are two exceptions. \cite{oostrom2024funding} uses data from a meta-analysis of published clinical trials. \cite{kao2023transparency} collect trial records from the proceedings of scientific conferences.}  \cite{lichtenberg2018impact} and \cite{bloom2020ideas} collect all records assigned a ``publication type''---in the PubMed indexing process---of ``clinical trial.'' \cite{feldman2019quantifying} collect all records with a publication type similar to clinical trial (e.g., also including ``randomized controlled trial''). On inspection, these tags are imprecise. Both approaches retrieve large numbers of records that are not, in fact, clinical trials, while excluding potentially relevant records. 

\begin{figure}[t]
\begin{centering}
\caption{Universe of Potential Clinical Trials}
\label{fig: universe}
\medskip{}
\begin{tabular}{c}
\includegraphics[scale=0.4]{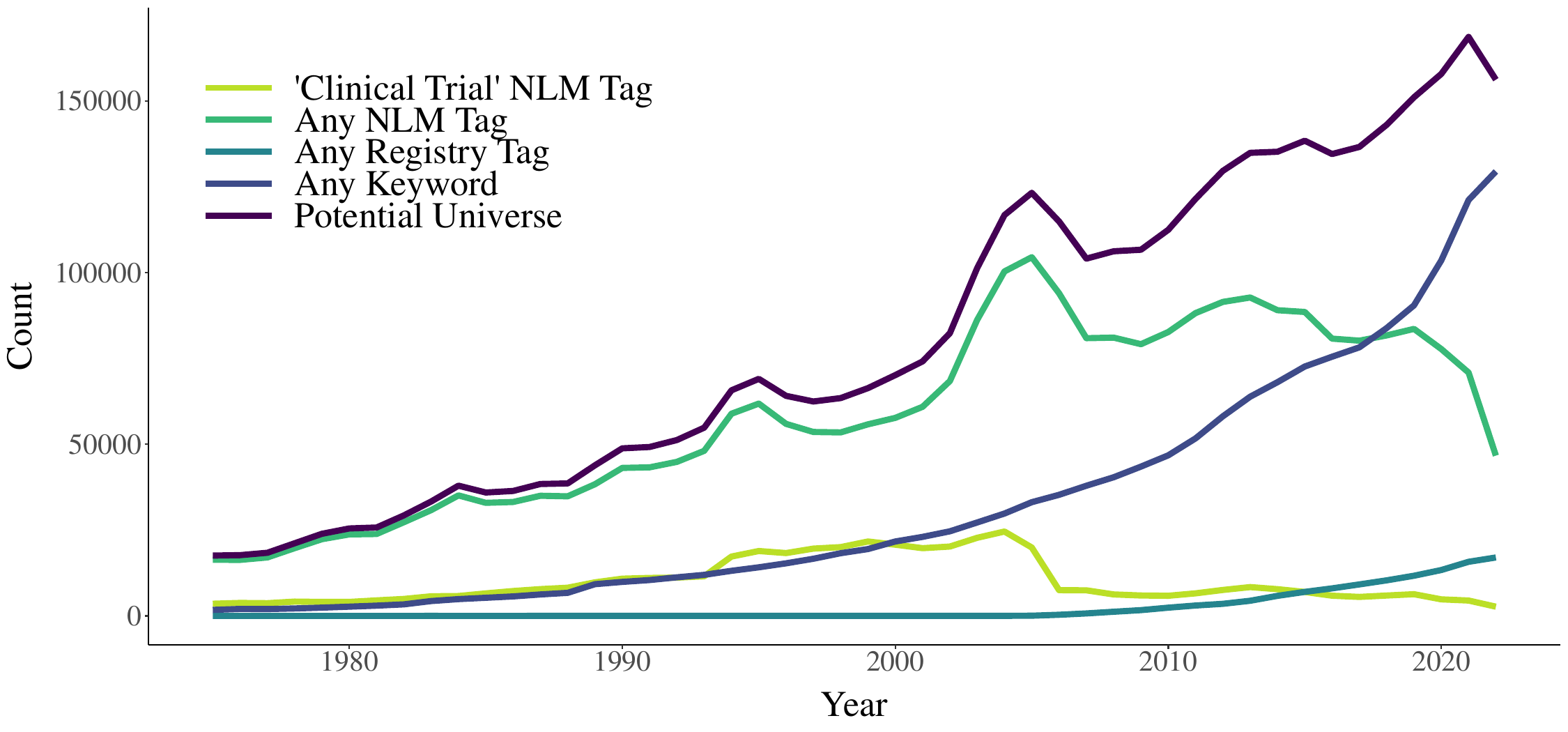}\tabularnewline
\end{tabular}
\par\end{centering}
\medskip{}
\justifying
{\footnotesize{}Notes: \cref{fig: universe} displays counts of the number of clinical trials indexed in PubMed / MEDLINE over time, constructed using alternative search strategies. The National Library of Medicine (NLM) categorizes each publication into a ``pubtype.'' The light green line (`Clinical Trial' NLM Tag) displays the number of publications in the ``Clinical Trial'' pubtype. The teal line (Any NLM Tag) displays the number of publication whose pubtype is an element of a set of 18 categories likely to include clinical trials. The blue line (Any Registry Tag) gives the number of publications that report, in their abstract, an identifier associated with one of the four largest international trial registries. The dark blue line (Any Keyword) indicates the number of publications whose abstract contains a keyword indicative of a clinical trial. The purple line (Potential Universe) displays the number of clinical trials in the union of the sets of publications identified with the other lines. See \cref{sec: definition app} for further details.}{\footnotesize\par}
\end{figure}

\cref{fig: universe} contrasts counts of published trials constructed in various ways. The light green line (`Clinical Trial' NLM Tag) plots the number of publications, in each calendar year, indexed with the publication type ``clinical trial.'' This light green line is the measure reported in \cite{bloom2020ideas}. The teal line (Any NLM Tag) plots counts of records indexed with any of 18 types that are likely to include clinical trials or related medical research.\footnote{On inspection, it appears as though the NLM shifted from primarily using the tag ``Clinical Trial'' to ``Randomized Controlled Trial'' and added more  specific tags (e.g., ``Clinical Trial, Phase 1''). There are 19,937 records tagged with ``Clinical Trial'' in 2005 and 7,518 with this same tag in 2006.} Alternative approaches to identifying trials are similarly imprecise. Medical journals increasingly require trials to be registered in open registries, e.g., ClinicalTrials.gov, as a condition of publication. The blue line (Any Registry Tag) counts records that report, in their abstract, an identifier associated with one of the four largest international trial registries.  Language in a record's abstract can indicate that it might be a clinical trial if, for example, it references a ``treatment group'' and ``control group.'' The dark blue line (Any Keyword) plots the count of records with such keywords over time. 

We define the ``potential universe'' of clinical trials as those publications with any of the following: a publication type variable similar to clinical trial, a registry identifier reported in its abstract, or a keyword reported in its abstract. The purple line (Potential Universe) plots this trend over time. 

\cref{fig: universe} highlights that alternative approaches to data construction, in this setting, yield meaningfully different conclusions about levels, trends, and composition. See \cref{sec: definition app} for further details on the construction of these series. Given these difficulties, we describe a procedure, below, to recover a sample of clinical trials from this corpus of text in a manner that is consistent and transparent.

\subsection{Objective\label{sec: objective}} What exactly are we trying to recover when we search for clinical trials? We argue that the object of interest is a census of clinical trials that study the effects of a medicine in human subjects. 

\begin{defn}\label{def:main}
The sample of interest is composed of all publications that report the results of a prospective, interventional clinical trial that evaluates the effects of investigational or approved drugs in a setting with exclusively human subjects.
\end{defn}

\noindent \cref{def:main} embeds several restrictions, intended to yield precise measures of clinical trials relevant to the approval of new pharmaceuticals for use in human patients. Studies involving animals, literature summaries, and re-analyses of existing data, for example, are excluded. 
We take the ``potential universe'' plotted in \cref{fig: universe} as our baseline sample of potential clinical trials. 

We further restrict attention to the 1,821,429 records published in or after 2010, as there is a trade-off between data coverage and quality in this setting. In the 1990s and 2000s, medical experts created a standardized system for clinical trial reporting in scientific publications. By the time of the most recent Consolidated Standards of Reporting Trials (CONSORT) in 2010, nearly all medical journals required authors to comply with a standardized structure for all clinical trial publications \citep{schulz2010consort}. As a result, most trial publications have a distinctive format and include a pre-defined set of elements. Although our procedure can be applied to any type of scientific publication text, the standardization required by the CONSORT Statement changes the relationship between clinical trial research and publications in the early 2000s. To answer the research question at the center of our empirical analysis---is the quantity of clinical trials rising over time?---we elect to proceed with a narrower sample to avoid this issue of interpretability. 

To quantify our ability to identify the clinical trials in this set of publications, we hand-label approximately 3,000 randomly selected publications based on the content of their abstracts. We develop a custom labeling interface that reduces the time required to label 100 records, for the authors, by a factor of seven. \cref{sec: hand label} describes this tool and its application to this context in detail. Of the hand-labeled publications, 11.2\% meet the criteria for inclusion in our sample. The hand-labeled data are split into three subsets---validation, training, and testing---based on their eventual use.

\begin{figure}[t]
\begin{centering}
\caption{Performance of Standard Machine Learning Methods}
\label{fig: ml perform}
\medskip{}
\begin{tabular}{c}
\includegraphics[scale=0.4]{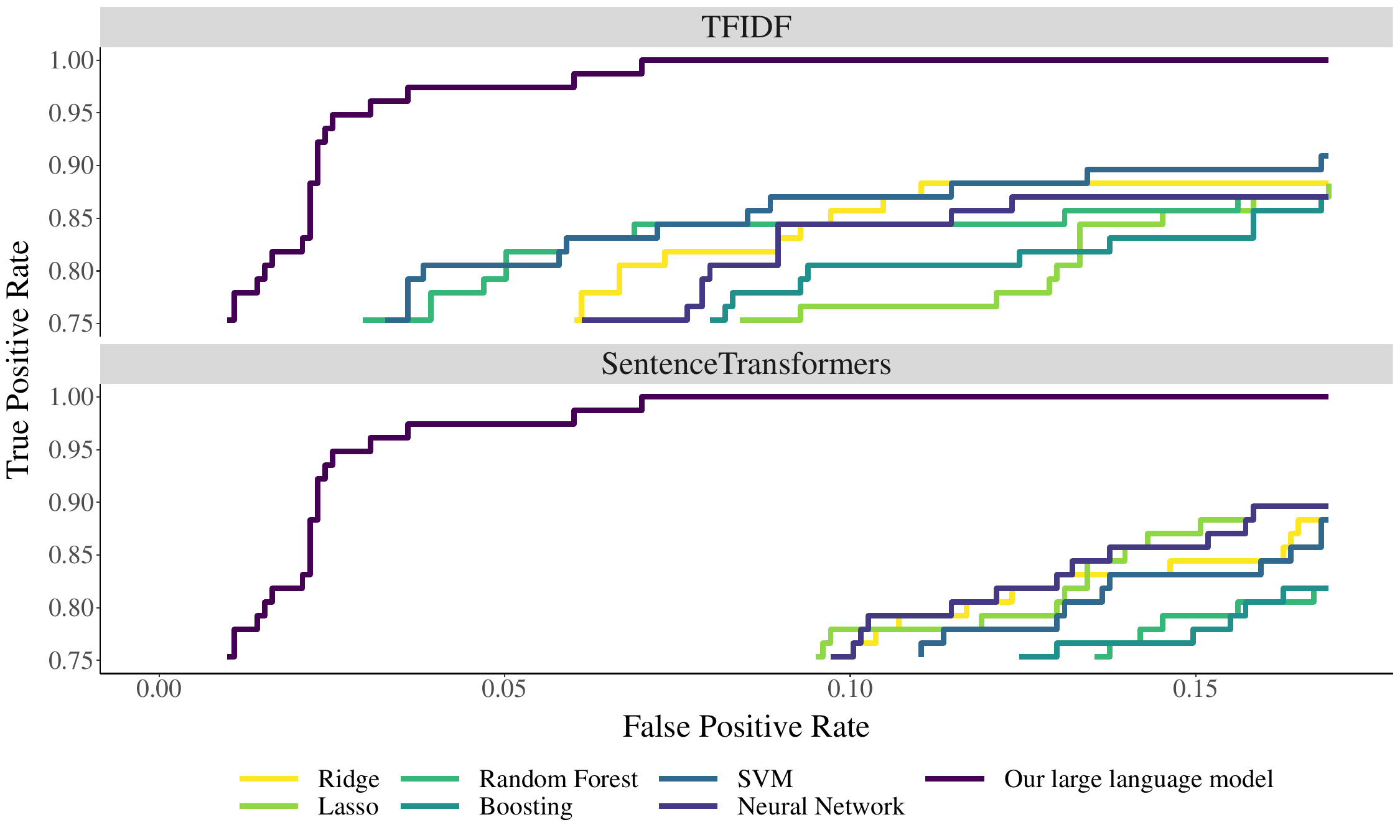}\tabularnewline
\end{tabular}
\par\end{centering}
\medskip{}
\justifying
{\footnotesize{}Notes: \cref{fig: ml perform} displays the receiver operating characteristic of several machine learning models, trained on two types of embeddings, for classifying whether a publication satisfies the restrictions enumerated in \cref{def:main}. The hyper-parameters of each model are determined with 10-fold cross-validation. The models are trained in the training and validation data set. Error rates are estimated in the testing data set. In addition, we display the performance of our, ensemble, fine-tuned large language model in purple. This is the same curve labelled ``Ensemble'' in \cref{fig: roc model}.}{\footnotesize\par}
\end{figure}

As a baseline, we assess the performance of several standard machine learning algorithms. The results are displayed in \cref{fig: ml perform}. Each algorithm is estimated on the hand-labels assigned to the publications in the training and validation sample splits. Performance is measured in the testing sample split. For feature vectors, we use either TF-IDF embeddings computed in the corpus of abstracts in the hand-labeled sample or the embeddings of each abstract obtained from the \textsc{SentenceTransformer} language model \citep{reimers2019sentencebert}. We find that the performance of the standard machine learning algorithms is again unsuitable for our application. At a 90\% true positive rate, the best performing model identifies 50 true positives for every 50 false positives.

\subsection{Model Distillation\label{sec: llm main}} 

We construct a large language model optimized for our task. This is accomplished in three steps. First, we iteratively construct a set of prompts that exhibit good performance when posed to proprietary models---OpenAI's GPT-3.5 and GPT-4. \textit{A priori}, we expect GPT-4 to output labels of slightly lower quality than hand-labeled output and GPT-3.5 to produce slightly noisier labels than GPT-4. The extraction of GPT-3.5 labels is substantially faster and cheaper. Second, we extract noisy labels for a moderate number of publications in our sample by querying the proprietary models. Third, the noisy labels are used to train an off-the-shelf large language model. The resulting model is then used to identify clinical trials in our baseline sample. This process for building specialized large language models is referred to as model distillation \citep{xu2024survey}.

\subsubsection{Prompt Design} 

To produce these labels at scale, each of the two models must be appropriately prompted. That is, each pre-trained model must be provided with a block of text as an input and asked to return the appropriate completion of this input. \textit{A priori}, it is not clear what prompt structure will work well. The relative infancy of this area of research renders it difficult to identify a set of ``best practices.''\footnote{See, for one example of this evolving area of work, a guide to ``prompt engineering'' from OpenAI: https://platform.openai.com/docs/guides/prompt-engineering.} All of the prompts that we consider are displayed as figures in \cref{sec: prompt repository}.

We identify three general prompt formats, which differ both in the amount of detail provided about our classification task and in the structure of the requested model completion. The simplest prompt provides a version of our sample definition, \cref{def:main}, and the text of an abstract, and asks the model to return `TRUE' if the abstract satisfies these criteria. Otherwise, it returns `FALSE'. The text of this prompt is displayed in \cref{prompt: 1.0}. We refer to this prompt as Prompt 1.0. A second, more complicated prompt provides the same definition, with a set of examples of publication characteristics that do and do not satisfy the definition. Here, the prompt asks the model to return either `TRUE' or the name of a specific excluded category. The text of this prompt is displayed in \cref{prompt: 2.0}. We refer to this prompt as Prompt 2.0. A third, more complex prompt provides the same definitions and examples, but asks the model to return either `TRUE' or an explanation of why the record does or does not satisfy our sample definition. The text of this prompt is displayed in \cref{prompt: 3.0}. We refer to this prompt as Prompt 3.0. We devise initial language for each prompt iteratively, using a small number of records from our hand-labelled validation data. 

We test each of these three prompts in our 1000-record hand-labelled validation dataset, using both GPT-3.5 and GPT-4. We conduct a detailed error analysis, reported in \cref{tab:Error-Analysis}. For each instance in which a model returns a label (TRUE/FALSE) that differs from that in the hand-labelled dataset, we inspect the record. We categorize errors into types and sub-types. Details are given in \cref{sec: prompt design}.

This exercise is instructive on three margins. First, it highlights the highest performing prompts. Second, it indicates particular types of errors in categorization---which suggest opportunities for more precise language in a prompt. Third, it draws attention to differences in the performance of the two models. Observe that, in the ``Other'' error type, we include a sub-type called ``overly literal interpretation of inclusion criteria.'' We primarily record errors of this type for Prompt 3, which asked the model to return a completion that described why, or why not, a record was classified as being in our sample. Here, GPT-3.5 makes 30 such errors---classifying records as FALSE by recapitulating the sample definition provided. GPT-4 makes two errors. The insight that emerges from this exercise, then, is that different prompt structures may be preferable depending on the model used, and that, by and large, GPT-4 is better suited for tasks that require (the resemblance of) more sophisticated reasoning.

We revise each class of prompt based on these findings. For Prompt 1.0, the simplest true/false prompt, we consider three variants. We refer to these revised prompts as Prompts 1.1, 1.2, and 1.3. The text of these prompts is displayed in \cref{prompt: 1.1,prompt: 1.2a,prompt: 1.3a}. For Prompts 2 and 3, we consider one variant each. We refer to these revised prompts as Prompts 2.1 and 3.1. The text of these prompts is displayed in \cref{prompt: 2.1a,prompt: 3.1a}. These changes reflect the differences in performance catalogued in \cref{tab:Error-Analysis}. 

\cref{tab:prompt-performance-in-validation-data} reports estimates of the true positive rate and false positive rate for each prompt, in both models, computed in the validation sample. Observe---for example, with Prompts 1.1, 1.2, and 1.3 queried to GPT-4---that even small changes in the text of a prompt yield substantial differences in performance. We attempt a second round of prompt iteration (Prompt 1.3). We observe that performance deteriorates with even small modification. Thus, we select the two highest performing prompts---one for each proprietary model. We use Prompt 2.0 to extract weak labels using GPT-3.5, and Prompt 1.2 to extract weak labels using GPT-4.

\begin{table}
\renewcommand{\arraystretch}{1.1}
\begin{centering}
\caption{Prompt Performance in Validation Data\label{tab:prompt-performance-in-validation-data}}
\begin{tabular}{c}
\textit{Panel A: GPT-3.5}\tabularnewline
\begin{tabular}{cccc}
Prompt Type & Prompt Sub-Type & False Positive Rate & True Positive Rate\tabularnewline
\midrule
\midrule 
\multirow{4}{*}{1} & 0 & 0.247 & 0.788\tabularnewline
 & 1 & 0.172 & 0.898\tabularnewline
 & 2 & 0.037 & 0.584\tabularnewline
 & 3 & 0.037 & 0.489\tabularnewline
\midrule 
\multirow{2}{*}{2} & 0 & \textbf{0.162} & \textbf{0.876}\tabularnewline
 & 1 & 0.176 & 0.905\tabularnewline
\midrule 
\multirow{2}{*}{3} & 0 & 0.248 & 0.722\tabularnewline
 & 1 & 0.171 & 0.883\tabularnewline
\bottomrule
\end{tabular}
\tabularnewline
\tabularnewline
\textit{Panel B: GPT-4}\tabularnewline
\begin{tabular}{cccc}
Prompt Type & Prompt Sub-Type & False Positive Rate & True Positive Rate\tabularnewline
\midrule
\midrule 
\multirow{4}{*}{1} & 0 & 0.202 & 0.971\tabularnewline
 & 1 & 0.065 & 0.949\tabularnewline
 & 2 & \textbf{0.049} & \textbf{0.934}\tabularnewline
 & 3 & 0.056 & 0.912\tabularnewline
\midrule 
\multirow{2}{*}{2} & 0 & 0.081 & 0.964\tabularnewline
 & 1 & 0.072 & 0.964\tabularnewline
\midrule 
\multirow{2}{*}{3} & 0 & 0.167 & 0.971\tabularnewline
 & 1 & 0.068 & 0.956\tabularnewline
\bottomrule
\end{tabular}
\end{tabular}
\tabularnewline

\par\end{centering}
\medskip{}
\justifying
{\footnotesize{}Notes: \cref{tab:prompt-performance-in-validation-data} records the performance of each of eight prompt variants in a sample of 1,000 validation dataset records. Panel A reports performance associated with the proprietary model GPT-3.5. Panel B reports analogous statistics for model GPT-4. Prompt 1 asks the model to return TRUE or FALSE. Prompt 2 asks the model to return TRUE or the name of a specific, excluded category. Prompt 3 asks the model to return TRUE or an explanation of why the record should be excluded. Prompt sub-types correspond to various iterations. Sub-type 0 is the initial version of the prompt. Sub-types 1-3, where applicable, are subsequent iterations. \cref{sec: prompt repository} records the text of each prompt. We use Prompt 2, Sub-Type 0, to extract weak labels using GPT-3.5, and Prompt 1, Sub-Type 2 to extract weak labels using GPT-4. The performance measurements for these prompts are displayed in boldface.}{\footnotesize\par}
\end{table}

The performance of GPT-4 represents a substantial improvement over existing methods. However, practical and substantive considerations mitigate the applicability of proprietary models for classification of the complete baseline sample. Practically, it is---and likely will remain---prohibitively expensive, both computationally and financially, to deploy GPT-4 at this scale.\footnote{We incur a cost of roughly \$4,500 to extract noisy labels for 64,000 abstracts. Extrapolating this figure to the full sample of 1.8 million abstracts gives a price of approximately \$130,000.} Substantively, proprietary models are black boxes. Their details and substance are not public and are known to change at regular intervals. 

\subsubsection{Fine-Tuning} We compute noisy labels for a moderate number of randomly selected publications in our baseline sample using the best performing prompts for both GPT-3.5 and GPT-4. These noisy labels are used as data to train off-the-shelf \textsc{BERT} models from two architecture classes: (1) \textsc{BigBird}~\citep{zaheer2021big} and (2) \textsc{BioMedBERT}~\citep{Gu_2021}.\footnote{There are many open-source large language models (\textit{e.g.}, LLaMa, Mistral, Pythia). Many are several orders of magnitude (\textit{e.g.}, 7-70 billion parameters) larger than \textsc{BERT} ($\sim$ 350 million parameters). We selected \textsc{BigBird} and \textsc{BioMedBERT} via trial-and-error that included testing the performance of these larger models, including LLaMA (both 7 and 70 billion parameter versions, trained with QLoRA). Both models exhibited comparable performance, but \textsc{BigBird} was substantially faster and simpler to train.}
Both models use medium-scale Transformer architectures, i.e., between 100 and 300 million parameters, pre-trained with the masked-language modeling (MLM) objective~\citep{vaswani2023attention,devlin2019bert}. The models differ in their architectural details and pre-training corpora.

\textsc{BigBird} uses a sparse attention mechanism to reduce the computational cost of processing long text sequences.\footnote{The computational complexity of regular attention is quadratic in sequence length, while \textsc{BigBird} reduces this to linear complexity.} This enables efficient handling of long documents up to 4,096 tokens. In contrast, standard \textsc{BERT} models can only process 512 tokens. \textsc{BigBird} is useful because abstracts of clinical trials regularly exceed 512 tokens. Prior to pre-training, the model was warm-started from the \textsc{RoBERTa} checkpoint \citep{liu2019roberta} and then pre-trained on the standard \textsc{BERT} corpus.\footnote{ The standard  \textsc{BERT} corpus consists of the Books \citep{zhu2015aligning}, CC-News \citep{guu2020realm}, Stories \citep{trinh2019simple}, and Wikipedia \citep{wikidump} datasets.} We evaluate two model sizes: \textsc{BigBird} Base (125 million parameters) and \textsc{BigBird} Large (355 million parameters).

\textsc{BioMedBert} uses a domain-specific pre-training corpus sourced from articles on PubMed~\citep{Gu_2021}. It uses the same model architecture as RoBERTa, but a different, domain-specific token vocabulary optimized for PubMed articles. Like \textsc{BigBird}, we evaluate two model sizes: \textsc{BioMedBert} Base (125 million parameters) and \textsc{BioMedBert} Large (355 million parameters).

We fine-tune the pre-trained architectures to classify abstracts according to the inclusion and exclusion criteria enumerated in \cref{def:main}. We replace the language modeling classification head with a binary classification head consisting of a single linear layer and softmax activation. We fine-tune the full model (\textit{i.e.} not just the classification head) with cross-entropy loss.\footnote{We use the Adam optimizer with learning rate $1\times 10^{-4}$, $\beta_1 = 0.9$, and $\beta_2 = 0.999$ (determined via hyperparameter sweep). We use a maximum sequence length of 4096 tokens. If an abstract does not fit into the maximum sequence length provided by a model, the abstract is truncated.} All models were trained on a single $V100$ GPU with 16GB of HBM.

\subsection{Performance\label{sec: performance}} 

\begin{figure}[t]
\begin{centering}
\caption{Receiver Operating Characteristic by Data Size and Origin}
\label{fig: roc size}
\medskip{}
\begin{tabular}{c}
\includegraphics[scale=0.4]{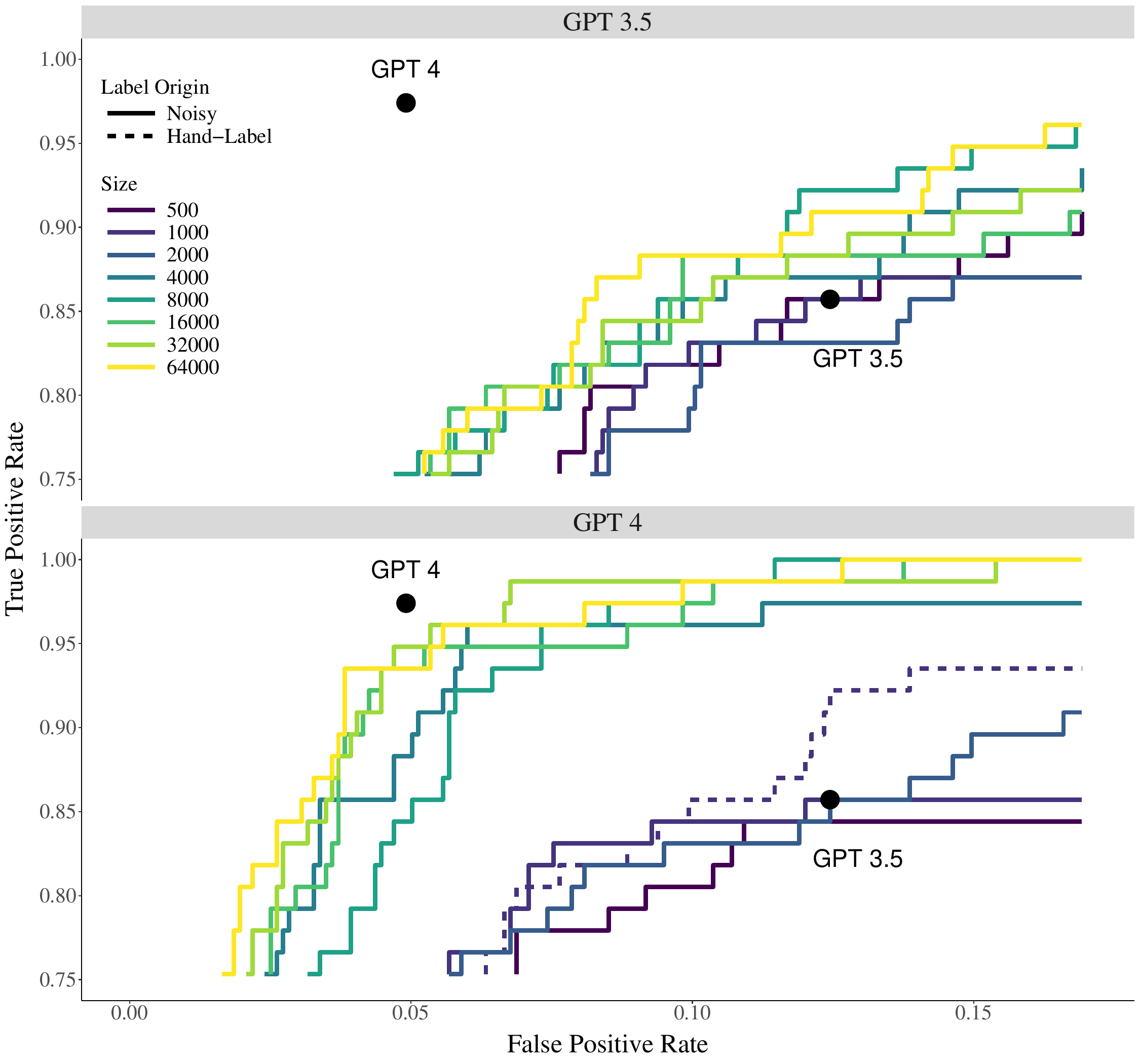}\tabularnewline
\end{tabular}
\par\end{centering}
\medskip{}
\justifying
{\footnotesize{}Notes: \cref{fig: roc size} displays estimates of the the receiver operating characteristic of several fine-tuned language models, based on the \textsc{BigBird} architecture. The models differ in terms of the source of their training data. Solid lines display the performance of models trained on noisy labels extracted from either GPT-3.5 or GPT-4. The dotted line displays the performance of a model trained on the hand-labeled data in the training dataset. Estimates of the true positive rate and false positive rate of GPT-3.5 and GPT-4 are indicated with black dots and are measured using the testing data.}{\footnotesize\par}
\end{figure}

Given the text of an abstract, our fine-tuned language models output a probability that the publication satisfies the restrictions enumerated in \cref{def:main}. Publications whose probabilities fall above a chosen threshold are classified as belonging in our sample. \cref{fig: roc size} displays estimates of true and false positive rates, computed with the test data, as we vary this threshold for labels assigned by fine-tuned version of the base \textsc{BigBird} model. The top and bottom panels are trained on noisy labels extracted from GPT-3.5 and GPT-4, respectively. We vary the quantity of noisy labels used to train the language model. Additionally, in the bottom panel, we display the performance of a fine-tuned model trained with the 1000 hand-labelled observations in the training dataset.

Fine-tuned models trained with noisy labels extracted from GPT-3.5 outperform GPT-3.5. By contrast, fine-tuned models trained with noisy labels extracted from GPT-4 match the performance of GPT-4, and significantly outperform models trained with labels extracted from GPT-3.5 or with hand-labels. There appears to be a threshold where performance dramatically improves at around 8,000 training labels.

\begin{figure}[t]
\begin{centering}
\caption{Receiver Operating Characteristic by Model Architecture}
\label{fig: roc model}
\medskip{}
\begin{tabular}{c}
\includegraphics[scale=0.4]{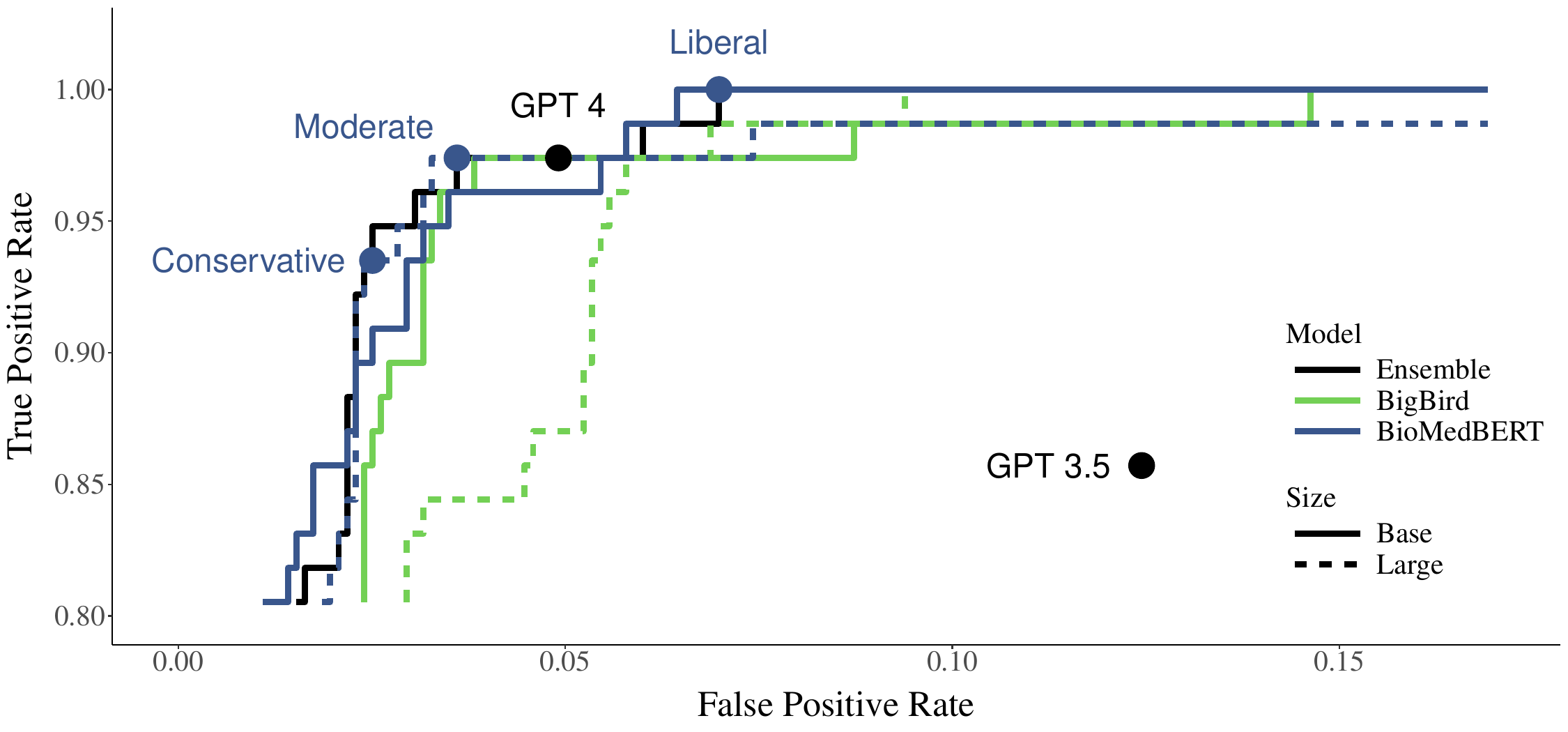}\tabularnewline
\end{tabular}
\par\end{centering}
\medskip{}
\justifying
{\footnotesize{}Notes: \cref{fig: roc model} displays estimates of the receiver operating characteristic of several models used to classify the publications indexed by PubMed / MEDLINE according to whether they satisfy the restrictions enumerated in \cref{def:main}. All metrics are computed in the testing split of the hand-labelled publications. Estimates of the true positive rate and false positive rate of OpenAI's proprietary models GPT-3 and GPT-4 are indicated  with black dots. The curves give the performance of four fine-tuned, open-source large language models, in addition to an ensemble model. The blue dots indicate the true positive rate and false positive rate of the models used to construct the ``Conservative,'' ``Moderate,'' and ``Liberal'' samples of clinical trials.}{\footnotesize\par}
\end{figure}

\cref{fig: roc model} displays analogous estimates for both sizes of the \textsc{BigBird} and \textsc{BioMedBERT} models. Both models are trained with 64,000 training labels extracted with GPT-4. Again, the fine-tuned models are able to match the performance of GPT-4 in the test data. The figure additionally displays the performance of an ensemble model estimated in the training data. This model is obtained by estimating a logistic regression of the hand-labels on the probabilities output by all four models displayed \cref{fig: roc model}. The ensemble model is used to produce our final sample. 

We choose three thresholds according to the stringency with which they enforce the sample restrictions. The estimated true and false positive rates associated with these points are displayed in \cref{fig: roc model} and labeled as ``Conservative,'' ``Moderate,'' and ``Liberal.'' Our preferred sample is associated with the conservative threshold.\footnote{The liberal model may be particularly useful for conducting literature reviews, where a near-perfect true positive rate is needed.} For every 82 true positives, the conservative model identifies 18 false positives.\footnote{In the test data, the conservative model assigns incorrect labels to 27 of 993 papers. We conduct an error analysis. See \cref{sec: final error analysis} for further details. In 13 cases, there is a clear error. In 14 cases, however, errors are associated with records that were difficult to categorize for a human labeller. For example, PubMed record 32737793 is flagged as satisfying \cref{def:main} with all three thresholds, but is a literature review. By contrast, PubMed record 27880726 was categorized as satisfying \cref{def:main} twice by a human labeller. It is excluded from both the conservative and moderate model-generated samples. On inspection, the abstract does not explicitly state that the study enrolled human subjects, but hints that it may have been conducted in an animal model. Review of the associated full text confirms that this study did, in fact, enroll only rats.} All results that follow are robust to the use of the moderate and liberal thresholds. Our final, conservative, sample consists of 152,027 publications classified as satisfying the sample restrictions. 

Each table displayed in \cref{sec: pubmed} compares the contents of our sample to the counts of potential clinical trials plotted in \cref{fig: universe}. Although certain elements of PubMed metadata---including the union of all NLM tags---capture many of the records in our final sample, we confirm that they miss many records that we flag as trials and include many records that do not satisfy \cref{def:main}. No combination of existing search strategies, then, classifies records with the same accuracy or precision as our final sample. 

\section{Disentangling Trends in Publication Data}\label{sec:trends}

\cref{fig: universe} indicates that, across measures, the total quantity of potentially-relevant records in PubMed has increased substantially over time, including since 2010.\footnote{Here, we focus on patterns in the production of publications that are similar to those that disclose the results of a clinical trial. We find similar patterns in PubMed as a whole, and in alternative cuts of records. Across fields, the total number of scientific papers, the total number of unique authors listed on scientific papers, the total number of citations given and received by scientific papers, the total number of scientific journals, and the total number of pages in scientific journals exhibit similar patterns.} These patterns are consistent with documented increases in the quantity of scientific papers in \cite{bloom2020ideas} and \cite{park2023papers}, as well as economy-wide increases in research effort reported in \cite{goldin2024productivity}. In this section, we use our census of clinical trials to decompose this trend. We contrast trends in the production of clinical trials with those documented for other, similar forms of medical research.

There are many potential ways to construct a sample to contrast with data on clinical trial production. Our preferred approach defines this category as the collection of publications that \textit{cite} clinical trials. This sample includes publications that ``look like'' clinical trials---a fact that we confirm on inspection---and which necessarily engage with the findings of these studies. Thus, in an imprecise classification task, it is precisely this set of records that are likely to be erroneously included. For example, a keyword search of publication abstracts for ``clinical trials'' may capture studies that summarize the findings of clinical trials, without reporting new information. Based on our inspection of these records, these types of errors are especially likely for publications in which the writing quality is low, for which it may be challenging---even for a human labeler---to determine whether the record reports novel results or summarizes existing findings. 

Alternative methods of constructing a measure of ``other'' research introduce additional, complicating conceptual problems. Approaches that rely on keyword searches, NLM tags, etc., as Section  \ref{sec:identifying-trials} and its appendices make clear, capture different sets of records over time, as a consequence of changes in database structure. Approaches that examine other publications in the same journals as clinical trials will mechanically suggest shifts in total quantities, if the number of pages in journals change over time or if certain journals become online-only.\footnote{\cite{ioannidis2023rapid} document that the number of medical journals is increasing. In 2024, the \textit{Annals of Emergency Medicine}, the highest impact journal in the medical sub-field of emergency medicine, shifted to an online-only format. \cite{brainard2020dozens} discusses dozens of instances in which medical journals transitioned to online-only formats, then ceased publication entirely.}

The primary challenge in using a citation-based measure is one of truncation: constructing measures of papers that cite clinical trials requires determining the window in which such citation must have occurred. To account for concerns about citation delays, we construct various measures, using 2--6 year citation counts. Two-year citation counts, thus, count publications that cite clinical trials published in the preceding two years. As our clinical trial sample begins in the year 2010, these measures are available only beginning in 2012. It is possible, and common, for clinical trials to cite other trials. We drop from our ``citing'' sample any records that we also flag as reporting the results of a trial.

\subsection{Decomposing Trends}

\begin{figure}[t]
\begin{centering}
\caption{Growth in Clinical Research, Stability of Clinical Trials}
\label{fig: trend}
\medskip{}
\begin{tabular}{c}
\includegraphics[scale=0.4]{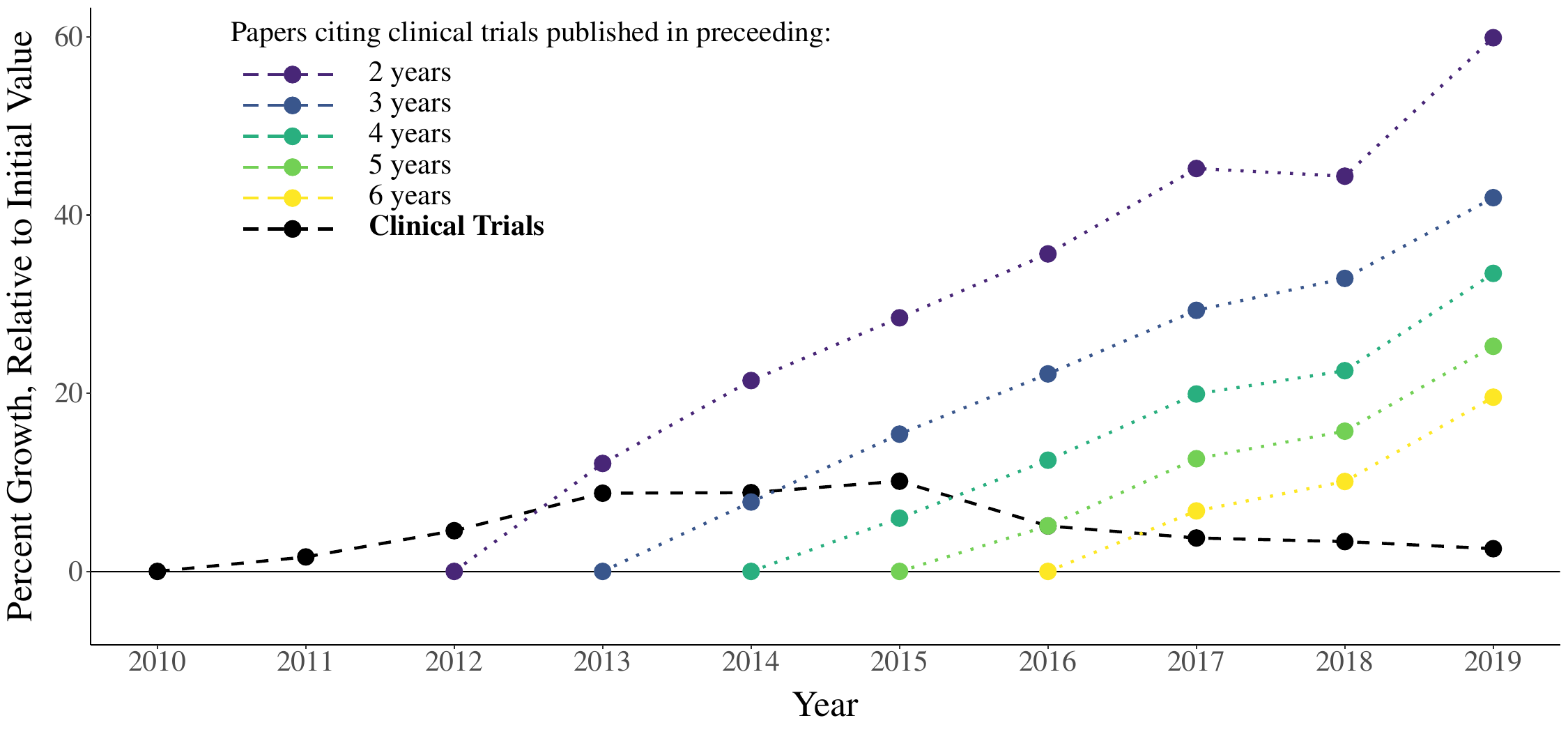}\tabularnewline
\end{tabular}
\par\end{centering}
\medskip{}
\justifying
{\footnotesize{}Notes: \cref{fig: trend} displays measurements of the number of clinical trials, and papers that cite clinical trials, published in each calendar year. Each series is reported in terms of the percent change relative to its initial value. The sample of published clinical trials is constructed with the conservative model. To address truncation, we report the number of publications that cite clinical trials published in the preceding $t$ years for each $t$ between 2 and 6.}{\footnotesize\par}
\end{figure}

\cref{fig: trend} documents that the quantity of clinical trials---the black line---has remained essentially constant over the past ten years. In contrast, the quantity of other non-trial research---the colored lines---has increased substantially. Observe that we display the percent \textit{growth} in the quantity of clinical trials in our sample published in each calendar year, alongside the \textit{growth} in the set of papers that cite clinical trials, constructed in various ways to account for the truncation discussed above. Each series is normalized to begin at zero. This presentation masks substantial differences in the levels of these series. There are 10,903 clinical trials in our census in 2010, and 30,841 publications that cite clinical trials (using three-year citations) in 2013. 

\subsection{Clinical Trials as a Productivity Indicator\label{sec:trials_as_productivity}}

The trends in \cref{fig: trend}, on their own, allow us to draw two conclusions. First, existing estimates that suggest sharply rising quantities of clinical trials \citep[e.g.,][]{bloom2020ideas} are a consequence of technical challenges in the measurement of research. Second, existing estimates of rising quantities of medical and scientific research \citep[e.g.,][]{bloom2020ideas,chu2021slowed,park2023papers} are \textit{not} driven by shifts in the quantity of clinical trials. 

\begin{figure}[t]
\begin{centering}
\caption{Public Funding and Never-Cited Research}
\label{fig: public}
\medskip{}
\begin{tabular}{c}
\includegraphics[scale=0.4]{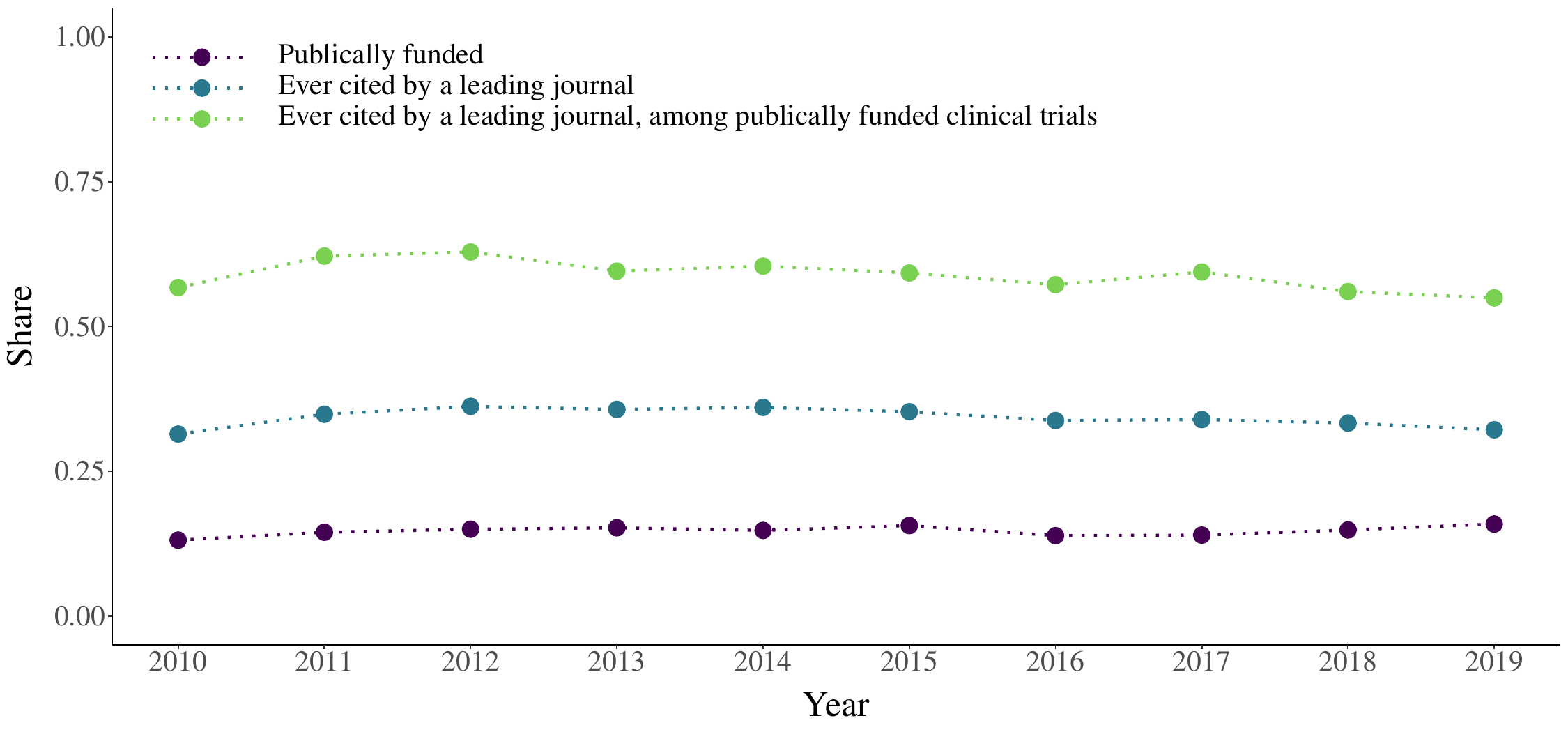}\tabularnewline
\end{tabular}
\par\end{centering}
\medskip{}
\justifying
{\footnotesize{}Notes: \cref{fig: public} displays three time series: the proportion of clinical trials that are publicly funded, the proportion of clinical trials that are ever cited by a leading journal, and the proportion of publicly funded clinical trials that are ever cited by a leading journal. The sample of published clinical trials is constructed with the conservative model.}{\footnotesize\par}
\end{figure}

\cref{fig: public} documents that---on two additional dimensions---clinical trial production has been stable over the past decade. In particular, \cref{fig: public} displays the share of publications in our census with the following characteristics: any funding from a government agency (purple), any (three-year) citations from a leading journal in medicine (teal), or any citations from a leading journal \textit{conditional} on having public funding (green). 

We designate a publication as having public funding if it appears in the National Institutes of Health RePORTER data, as linked to a funded research grant, or if PubMed reports a source of U.S.-based research funding.\footnote{Technically, PubMed indicates whether a publication received research funding from government agencies and private philanthropies outside of the United States. We restrict consideration to those sources of funding based in the United States because there appears to be a shift in the reporting of certain sources of non-U.S. funding beginning in 2017. The total number of publications linked to a source of public funding increases sharply---by roughly 50 percent from 2016---in 2017. On inspection, other trends are smooth during this time, including trends in public funding for the United States.  We are unable to determine if certain sets of authors began to disclose sources of funding in 2017, if certain journals changed their reporting requirements, or if PubMed changed its data construction process. We thus focus on the United States, where we find no evidence of large changes in data reporting practices.} To designate a set of the ``top'' journals in medicine, we follow \cite{angrist2020inside}. We collect all citations originating from  the ``trunk journals'' in medicine, the Journal of the American Medical Association and the New England Journal of Medicine, to records in PubMed between 2010 and 2022. We designate a journal as being ``leading'' if it received at least 100 citations from a trunk journal over this time period. This yields a list of 84 journals.\footnote{See \cref{sec: leading} for details. \cref{sec: additional} reproduces this figure using five-year citations. The trend is unchanged.}

Two facts are worth highlighting. First, all three measures are stable, essentially unchanging, over our time period. Second, nearly 70 percent of clinical trials are \textit{never} cited by a leading journal, suggesting substantial, but time-invariant, heterogeneity across trials.

Stability in public funding suggests, on its own, that the allocation of public funding to clinical trials has not considerably shifted over this time period. One can imagine alternative explanations for this stability. For example, perhaps the share of all projects receiving public funding has remained stable, but the cost of each trial has increased substantially. Alternatively, perhaps public agencies are funding a higher or lower share of studies, but that these changes are perfectly offset by changes in either study success or publication bias. 

We view the first potential explanation as unlikely for two reasons. Somewhat suggestively, NIH RePORTER data allow us to tabulate total expenditures associated with grants acknowledged in these publications. These grant-based expenditure measures, too, are stable over this period. Note, however, that linking research grant dollars to specific scientific papers is an imprecise exercise. See \cite{li2017expertise} for an extended discussion of this measurement challenge. More directly, recent work by \cite{sertkaya2024costs} suggests that the costs of drug development---measured using the contracts associated with particular study sites in clinical trials---have been essentially constant from 2000 to 2018. The second potential explanation is unlikely from a more conceptual perspective. We find essentially complete stability in this measure over time: any changes in publication bias or trial failure rates must, then, perfectly offset changes in public funding patterns, to yield the flat series here. 

\begin{figure}[t]
\begin{centering}
\caption{Citation Distribution Across Time, Clinical Trials}
\label{fig: heterogeneity}
\medskip{}
\begin{tabular}{c}
\includegraphics[scale=0.4]{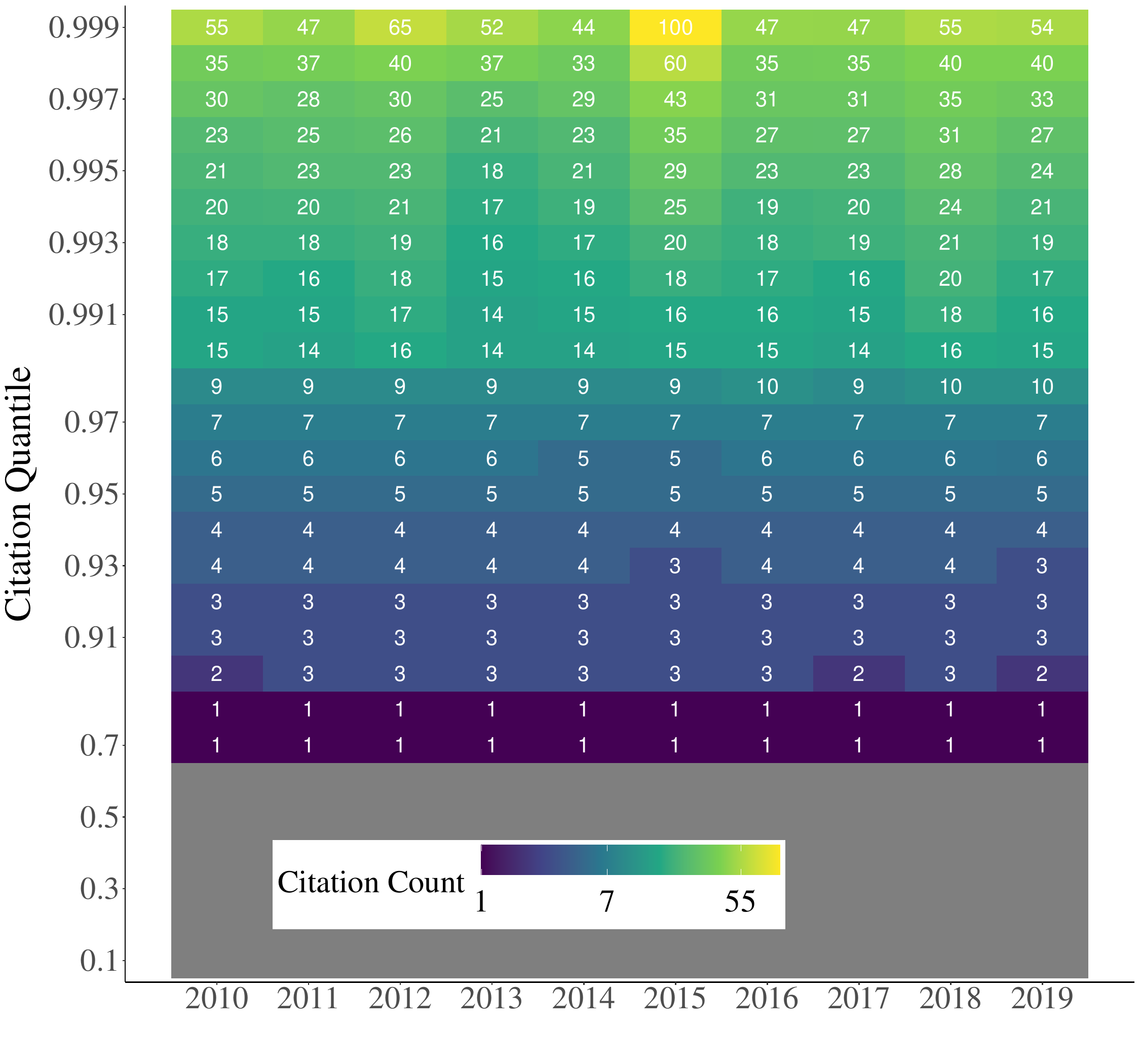}\tabularnewline
\end{tabular}
\par\end{centering}
\medskip{}
\justifying
{\footnotesize{}Notes: \cref{fig: heterogeneity} displays a heat map measuring the quantiles of the distribution of citations received by clinical trials in each calendar year from leading journals. The $y$-axis has been stretched to elongate the right-tail of the citation distribution. Colors are displayed in a log scale. The sample of published clinical trials is constructed with the conservative model.}{\footnotesize\par}
\end{figure}

\cref{fig: heterogeneity} provides a more granular view of this heterogeneity. We plot quantiles of the distribution of the number of (three-year) citations to publications in our census from publications in leading journals, for each year between 2010 and 2019.\footnote{\cref{sec: additional} presents these results using alternative cuts of our data and five-year citations. The patterns remain unchanged.} Grey areas at the bottom of each plot represent publications that receive zero citations. This distribution is stable over our time period. Observe that we elongate the y-axis to highlight the right-tail of the citation distribution. The quantiles corresponding to the right tail of the citation distribution are constant across years. The left-tail is similarly stable: the share of records receiving zero citations is essentially constant, year-on-year.

We interpret this set of facts as evidence that trends in clinical trial production have been stable over this time period. At the very least, we find that there is insufficient evidence to reject the null that pharmaceutical research productivity has been stable over this time period. There are four potential concerns with this interpretation. 

First, if the costs of clinical trials have increased substantially during this time, the total resources devoted to the commercialization of drugs will have increased, although \cref{fig: trend} suggests that the quantity of trials produced has been stable. In our view, this is unlikely to be the case. The most up-to-date, comprehensive evidence on clinical trial and drug development costs is given in \cite{sertkaya2024costs}, which uses internal data from the U.S. Food and Drug Administration and contracts associated with clinical trial sites to compute estimates for the period from 2000 to 2018. In these data, they find that over this period, ``the cost of drug development remained relatively stable or may have even decreased.''\footnote{This finding revises a frequently cited claim made in popular press interviews \citep[e.g.,][]{herper2022biotech}, in academic articles that rely on proprietary data sources \citep[e.g.,][]{dimasi2016innovation}, and in surveys on research productivity \citep{goldin2024productivity}---which suggest that clinical trial and drug development costs are sharply increasing over time. This revision is in the spirit of an argument made by \cite{cockburn2006pharmaceutical}, in response to an earlier debate about drug development costs.}  Thus, we do not consider the prospect of rising study prices a concern for the interpretation of this pattern. 

A distinct and somewhat related concern is that the cost of basic scientific research ``upstream'' of clinical trials could have increased substantially, either because more studies are necessary to produce a candidate drug or because the cost of inputs has increased. To be clear, our interest in this paper is on the productivity of the pharmaceutical industry, not biomedical research as a whole. Debates about the productivity of the pharmaceutical industry hinge on the idea that efforts to commercialize medicines are slowing---because firms are too inefficiently managed to run new studies \citep{ruffolo2006has}, because concerns about commercial viability encourage firms to shelve valuable drug candidates \citep{scannell2012diagnosing}, or because regulation has become too onerous \citep{scannell2012diagnosing}. Changes in the productivity of basic science are interesting and important, to be sure, but not the central object. 

Nonetheless, in our read, there is essentially no empirical evidence indicating either decreases in the productivity of basic scientific studies or increases in their cost. A small number of papers examine frictions in markets for basic research. See, for example, \cite{hill2021race} and \cite{myers2020elasticity}. To our knowledge, this literature has not provided evidence of substantial shifts in the quantity of inputs or factor prices over time, which in turn would increase the cost or decrease the productivity of basic science. In fact, changes in the production of basic scientific inputs around the year 2000---enabled by breakthroughs such as the Human Genome Project and the advent of high-throughput screening for drug discovery \citep[see][]{scannell2012diagnosing}---suggest, if anything, that costs of basic science inputs have fallen. In fact, the pharmaceutical productivity literature often characterizes the productivity ``puzzle'' in terms of this decrease in costs \citep[see e.g.,][]{cockburn2006pharmaceutical}. 

There is some evidence that the total quantity of scientific papers has increased over this time period, which has been interpreted in turn as evidence of increasing (basic) scientific research effort---notably in \cite{bloom2020ideas} and \cite{park2023papers}. In our view, moving from these trends in publication data to a conclusion of decreasing productivity is premature. In the next section, we document that measured increases in the quantity of scientific publications capture, in substantial part, a sharp increase in the quantity of low-quality papers over time.\footnote{\cite{park2023papers} document the same: in their data, the quantity of low-impact research has increased substantially, while the quantity of high-impact research has remained essentially stable over time. Their measure of impact is a measure of ``disruptiveness''---how much a given piece of research deviates from existing work.} If measured increases in the quantity of papers reflect, say, an increase in the likelihood that the marginal paper is published, perhaps because the cost of publication has decreased, publication trends alone are largely uninformative on the question of scientific productivity. 

Our data also provide a more suggestive piece of evidence on trends in upstream research. We find remarkable stability in trial quantity, quality, and composition. \cite{sertkaya2024costs} finds analogous stability in cost. If there were large changes in upstream productivity or cost, we  might expect to see some indication in these downstream markets. In the absence of such indicators, we conclude that large shifts in the productivity of upstream markets are unlikely to be relevant considerations for the interpretation of our findings.\footnote{Constructing a comprehensive measure of biomedical research productivity, which grapples with these various input markets, remains an open, important challenge. Our work does not speak directly to this challenge, but does highlight limitations in existing measures of productivity and challenges with the use of publication data.}

Second, one might be concerned about how to interpret the x-axis in our plots. In particular, the relationship between clinical trial completion dates and clinical trial start dates varies systematically within and across disease classes \citep{budish2015firms}. Here, we follow existing papers that use completion dates affixed to patents and publications as measures of the timing of research. In the context of clinical trials, the use of completion dates is reasonable as a practical matter. As \cref{fn: back-fill} suggests, records of the start dates of clinical trials are back-filled in administrative databases and, thus, available only with a considerable lag. The date of publication is more easily interpretable. 

Third, our findings may be threatened by publication bias. Specifically, the existence of publication bias means that counts of scientific publications only partially measure the quantity of clinical trials being produced in each time period \citep{andrews2019identification}. To weigh concerns about publication bias, it is useful to recall that our objective is to produce an estimate of trends in clinical trial production, sufficient to test existing hypotheses about declining productivity in this industry. Our aim is \textit{not} to produce a comprehensive census of all clinical trials. Given this objective, publication bias is a serious threat to the interpretation of our results if it is meaningfully \textit{changing} over this time period. 

Recent empirical evidence in \cite{oostrom2024funding} suggests that publication bias may have been stable over this time period. Oostrom collects a set of roughly 600 clinical trials associated with psychiatric medications, as listed in a comprehensive meta-analysis spanning five decades. For each record, the author collects any associated scientific publications and any ClinicalTrials.gov entries. Of interest to our analysis are two facts in this case study. First, Oostrom documents evidence of publication bias across years of her data, but finds that \textit{trends} in this bias become stable by the early 2000s.\footnote{See \cite{oostrom2024funding}, Figure5(b).} That is, although there is publication bias, it is not meaningfully changing in her sample of trials across our period of interest. Second, she finds that trends in ClinicalTrials.gov reporting shift sharply over our period of interest. In 2010, roughly 40 percent of the trials in her sample were registered on ClinicalTrials.gov. By 2019, roughly 80 percent were registered. Taken together, her findings lend credence to our assumption that publication bias and publication reporting norms are relatively stable during this period, whereas database reporting is not. 

In our data, we find that the quantity, quality, and composition of published clinical trials have been stable over time. If a trend were in fact present, and our measure were confounded by publication bias, then it would need to be the case that the trend was perfectly offset by the bias. Cancellation in this way---across all three margins we consider---is unlikely. 

Fourth, and finally, we must assume that the number of publications associated with each clinical trial is constant across our period of interest. We view this assumption as reasonable for three reasons. First, by 2010, nearly all scientific journals in our sample had adopted rules requiring authors of clinical trial publications to adhere to a standardized format when reporting trial results \citep{schulz2010consort}. Under these rules, there is little scope for authors to, for example, split results across several publications. Second, when constructing our sample of clinical trials, we excluded publications that re-analyze data generated by an existing study. Third, more suggestively, we find no evidence of a change in this relationship between the size of NIH grants and the quantity of resulting publications reporting the results of a clinical trial during our period of interest.\footnote{We inspect this relationship by collecting records from the NIH RePORTER database, for years 2010 to 2019. Specifically, we collect records of the total award associated with funded projects and the quantity of any publications flagged by our procedure as clinical trials. For the median NIH grant, we find that an award of roughly \$100,000 per clinical trial, and a mean of roughly \$200,000 per trial. This is stable across years of data.}

Taken together, we view the trends in our data, when set against existing empirical evidence, as indicating that clinical trial publication has been stable over this time period. 

\subsection{Why is the Quantity of Research Rising?}

The quantity of papers citing clinical trials has increased, per \cref{fig: trend}, by nearly a factor of two across our period of interest. In this section, we examine various measures of publication composition and quality to shed light on the source of this increase. Our aim is to determine the source of the categorization errors that yield measures such as those in \cref{fig: universe}. In doing so, we provide suggestive evidence on whether measured increases in the quantity of scientific research correspond to ``real'' increases in the quantity of underlying research or whether they capture changes in the ease, frequency, and composition of publication.

\subsubsection{Publication Content\label{sec: content}}

We inspect the technical content of ``non-clinical-trial'' papers in three ways. We hand-label random samples of 100 records, drawn from the beginning (publication year 2013) and end (publication year 2019) of our sample of papers that cite clinical trials. In this pooled set of 200 records, we identify one record that should have been in our sample of clinical trials.\footnote{The erroneously labeled publication is PMID 23797691, which reports the results of a clinical trial studying a drug to treat HIV. Note that this exercise suggests a false negative rate of 0.005, consistent with estimates reported previously.} The remainder of the records can be split into seven sets: editorial records (e.g., comments on other papers, errata, editorials, 2.5\%), publications reporting preclinical studies (e.g., in-laboratory and in-animal tests, 21.5\%), clinical trial protocols (1\%), randomized trials studying something other than a drug (2.5\%), case studies and case reports (4.5\%), observational studies (30.5\%), and review articles (e.g., meta-analyses and literature reviews, 37\%).\footnote{In three cases, review articles appeared with the heading ``Comment'' or ``Editorial.'' We categorize these records as falling into the ``Review'' category.} We find a 30 percent increase in the total number of review articles between 2013 and 2019, while the quantity of observational studies remains essentially constant (31 in 2013, and 30 in 2019). The increase in the quantity of meta-analyses is consistent with increases documented elsewhere \citep{ioannidis2013geometric}. Of course, in our small sample of records, it is difficult to draw authoritative conclusions about the role of growth in this type of literature. 

We collect two measures of changes in the quantity of meta-analyses, based on the hypothesis that this may be a quantitatively important driver of the changes in our sample. First, we implement a simple search of the abstracts of publications in this ``citing'' sample. To select keywords for this search, we inspect the set of hand-labeled records, used as the ``training data" for our fine-tuning procedure in \cref{sec: llm main}, as many were flagged as out-of-sample because they were literature reviews or meta-analyses.\footnote{\label{fn: meta keywords}We searched for the following keywords: meta-analysis, metaanalysis, metaanalyses, systematic review, systematic reviews, systematically review, systematic search, review of published data, literature review, literature search, search of databases, review all literature, reviewed all literature, narrative review, systemic review. We also included any record that mentioned at least two of the following databases, which---based on our hand-labeling experience---are often referenced as part of the methods involved in a literature review or meta-analysis: MEDLINE, EMBASE, CINAHL, PubMed, Cochrane Central Register of Controlled Trials, BioMedCentral.} As expected, given the stringency of our search terms, this yields a small number of records: in our three-year citing trials sample, there are 1,606 records that satisfy these criteria in 2013 and 3,372 in 2019, corresponding to a 109 percent increase.

Our second method generates---as the discussion in \cref{sec:identifying-trials} suggests---what might be viewed as an upper bound on the quantity of meta-analyses and literature reviews in our data. We collect all records indexed in PubMed with the NLM tags ``meta-analysis,'' ``systematic review'' and ``review.'' There are 9,407 records in our citing sample with one of these tags in 2013 and 14,991 in 2019, corresponding to a 59 percent increase.\footnote{\cref{fig: meta} displays a time series of the number of `citing' papers that satisfy both our keyword and NLM-tag based methods for identifying meta-analyses.}

From these three procedures, we draw a suggestive conclusion: there has been a quantitatively important increase in the number of meta-analyses over our period of interest. This increase is on the order of 30 to 100 percent, across measurement strategies. The specific quantity of meta-analyses is not of central interest for this paper, except insofar as it allows us to conclude that the quantity of medical research that summarizes existing papers---rather than producing new evidence---appears to be growing over time. To be clear, meta-analyses and similar forms of literature review provide important information to physicians and can be vital in the diffusion of existing information, especially as the stock of knowledge grows \citep{jones2009burden}. This change in composition, however, is relevant context when interpreting measured increases in the quantity of publications. 

\subsubsection{Publication Geography\label{sec:intext_geography}}

Changes in the production of meta-analyses have been linked \citep[e.g.,][]{ioannidis2013geometric} to shifts in the geographic distribution of medical research. Publication data allow us to inspect geographic trends in our sample. 

\begin{figure}[t]
\begin{centering}
\caption{Composition of Publications Across Countries}
\label{fig: countries}
\medskip{}
\begin{tabular}{cc}
\textit{Panel A: Clinical Trials} & \textit{Panel B: Papers Citing Clinical Trials}\tabularnewline
\includegraphics[scale=0.4]{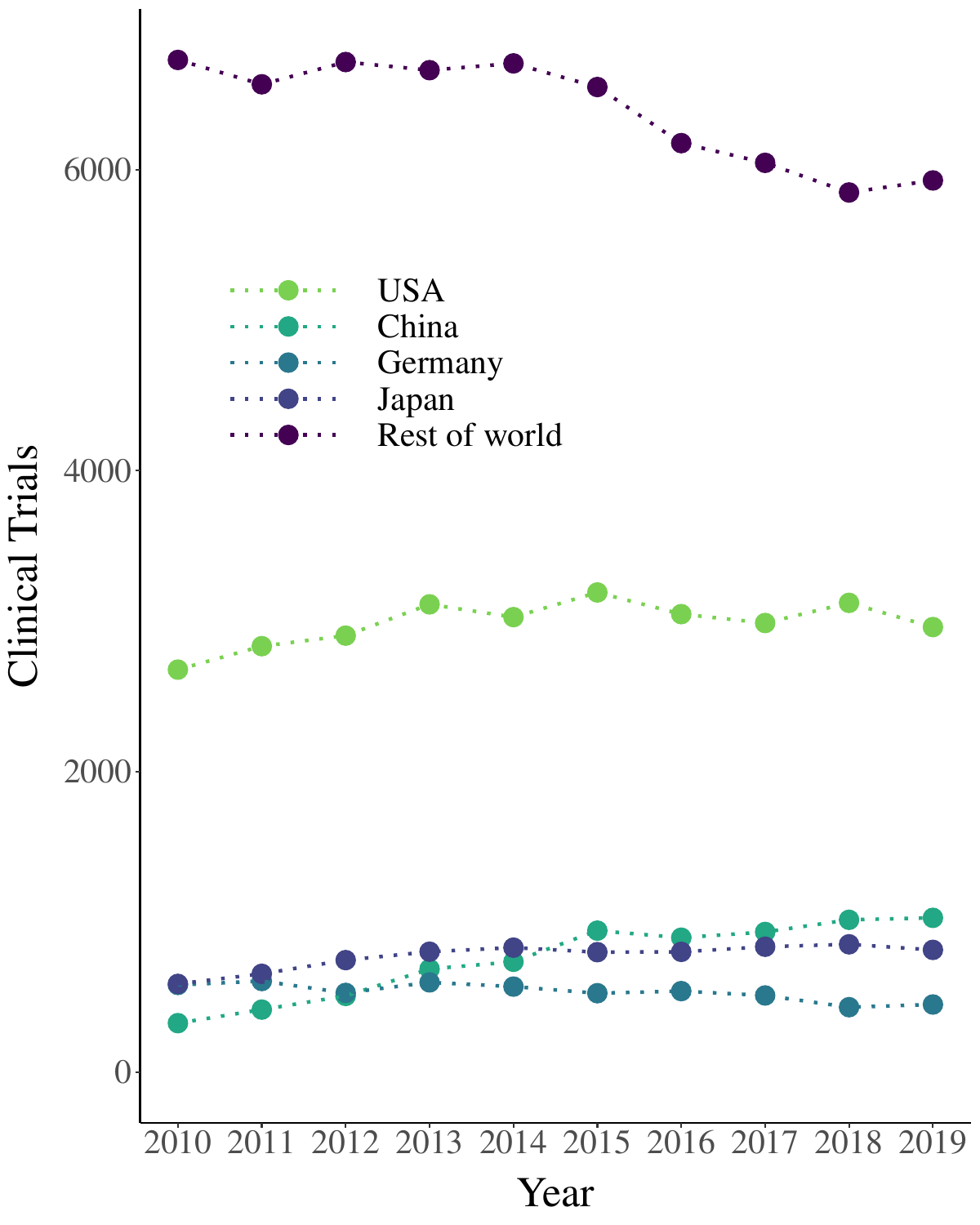} & \includegraphics[scale=0.4]{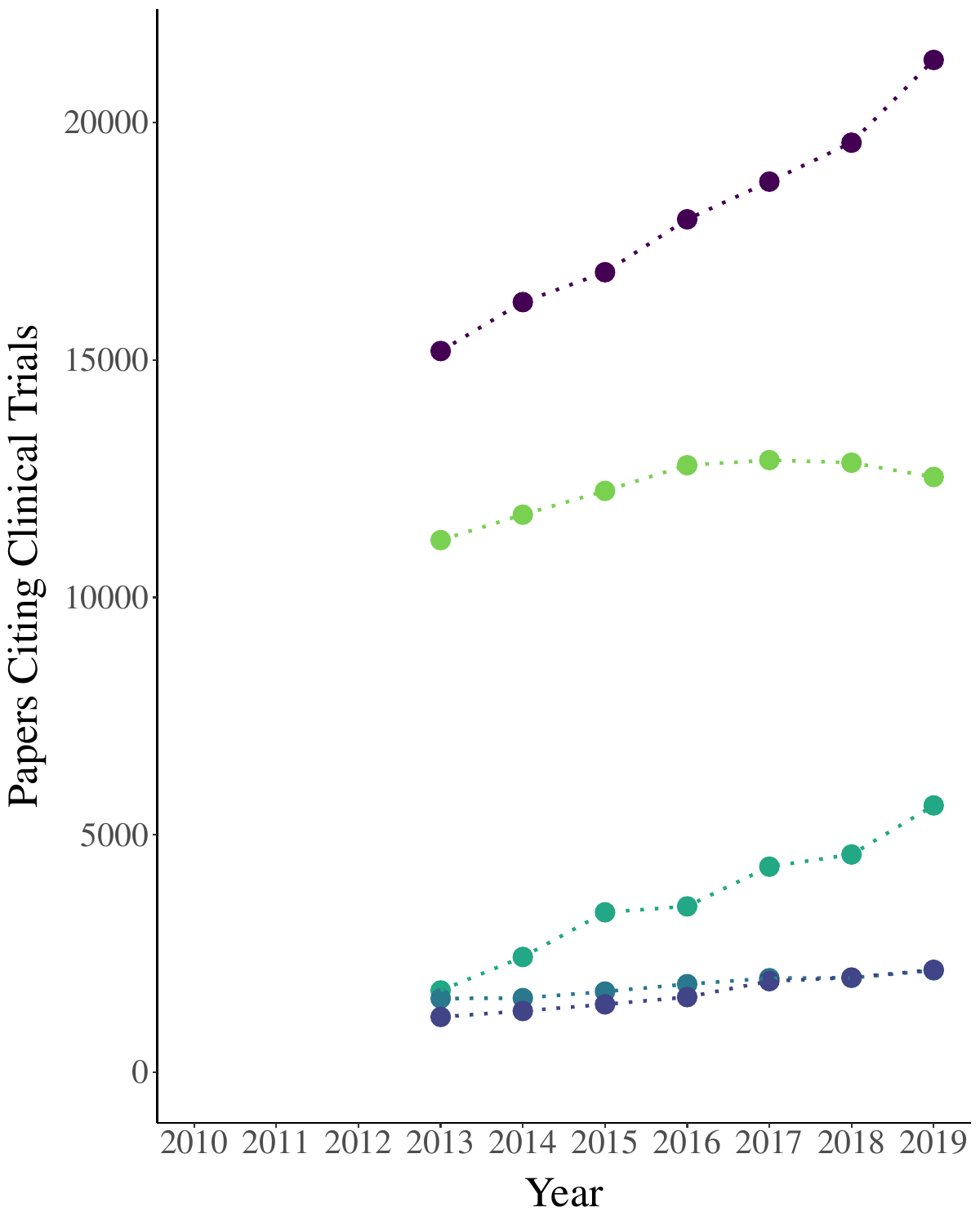}\tabularnewline
\end{tabular}
\par\end{centering}
\medskip{}
\justifying
{\footnotesize{}Notes: \cref{fig: countries} displays measurements of the number of clinical trials, and papers that cite clinical trials, published in each calendar year, by country. The sample of published clinical trials is constructed with the conservative model. To address truncation, we report the number of publications that cite clinical trials published in the preceding three years.}{\footnotesize\par}
\end{figure}

\cref{fig: countries} displays trends in the annual quantity of published clinical trials, and papers citing clinical trials, disaggregated by the location of the first-listed author.\footnote{In most domains of scientific and medical publication, last-listed authors are senior investigators. First-listed authors are typically junior investigators. \cite{agha2018local} observe that, for large-scale clinical trials, first-authors are more likely to be the principal investigator. In our context, the first- and last-author have the same listed country for 85 percent of clinical trial records and 91 percent of non-clinical trial records. We collect details on author location from Clarivate Analytics' Web of Science. See \cref{sec: wos overview} for further details.} In our sample, the top four producers of published clinical trials, and papers citing clinical trials, are, in order, the United States, China, Germany, and Japan. 

Clinical trial quantities---across geographies---are stable over our period of interest.\footnote{Note that Panel A of \cref{fig: countries} displays measurements of the location of clinical trial publication authors, not the location of clinical trial sites. We cannot rule out that, for example, trials authored by researchers with United States mailing addresses were conducted elsewhere. Thus, we do not interpret these as facts on clinical trial ``offshoring.'' See \cite{petryna2007clinical} and \cite{durvasula2023inclusive} for longer discussions of the geography of clinical trial \textit{sites}.} Roughly 30 percent of published clinical trials originate in the United States. This share is unchanging across periods. By contrast, Panel B indicates a large shift in the medical research ecosystem. Since 2013, there has been a small increase in the quantity of other medical research published by authors in the United States. This small change is dwarfed by increases---on the order of 30-230 percent---in the quantity of other medical research from China and from countries outside of the top four producers of medical research.

Institutional details are helpful in making sense of the differential patterns across these cuts of data. Regulatory requirements and local capabilities are key factors influencing where clinical trials are conducted and published. Sites must be capable of producing data sufficient to persuade regulators in high-value markets. Certain regulators require that sponsors of new drug applications submit trial evidence collected from a domestic population.\footnote{Until December 2023, Japan required domestic phase I clinical trials of drugs developed overseas before Japanese individuals could participate in international phase III trials for pharmaceutical regulatory approval. This policy reflected, in part, differences in disease burden in Japan, relative to countries where drugs are routinely tested. This may explain why Japan is a top producer of clinical trials over our sample period. See \cite{namba2024clinical} for one discussion.} In recent years, the U.S. Food and Drug Administration (FDA)---responding to concerns from patient groups---has indicated that it may not be willing to approve drugs on the basis of evidence collected from exclusively foreign sites \citep[for a longer discussion, see][]{alsan2024representation}. In one especially high-profile example, a cancer drug was rejected by the FDA after being tested exclusively in China \citep{kolata2022fda}. As long as the U.S. market continues to have outsized value for pharmaceutical firms---a consequence of the especially high prices paid by American consumers---it may be unsurprising to find that trial evidence disproportionately originates in the United States, even as global production of other forms of medical research continues to grow. Observational studies, meta-analyses, and case reports are, by contrast, cheap and not subject to the same financial and regulatory pressures. Digital tools, including search engines and the large language models deployed in this paper, will likely further depress these costs, potentially further widening this gap.

Four countries are the most prolific producers of clinical trials and other forms of medical research. Of interest here, however, are those countries that experienced the largest growth in their production of medical research over this period. On this measure, China, Poland, Japan, and Spain top the ranking in our data. In \cref{sec: geography}, we provide more details on the set of countries with the largest growth in this form of medical research over our period of interest. Several details are worth noting. First, Chinese production of records in our citing sample increases by roughly 225 percent, producing 5,620 papers in our citing set in 2019. The next country in this ranking, Poland, experiences 96 percent growth and produces 311 papers in our citing set in 2019. The top 23 countries in terms of growth, listed in the appendix, produce 93 percent of all 2019 papers in this set. Per the World Bank income classification scheme, nearly every country in this fast-growing group is high income. The only countries categorized differently are China, India, Brazil, Iran, and Turkey, which the World Bank classifies as either upper (China, Brazil, Iran, Turkey) or lower (India) middle-income.\footnote{https://datatopics.worldbank.org/world-development-indicators/the-world-by-income-and-region.html} That is, this trend captures, in part, an increase in the quantity of diffusion-focused literature in high-income countries that have long had robust health infrastructure. Below, we establish that on several quality measures, these publications appear to be declining in value over time---consistent with the idea that some of this increase in production may be a response to changing scientific incentives for publication. 

\subsubsection{Publication Quality}

\begin{figure}[t]
\begin{centering}
\caption{Citation Distribution Across Time, Papers that Cite Clinical Trials}
\label{fig: other heterogeneity}
\medskip{}
\begin{tabular}{c}
\includegraphics[scale=0.4]{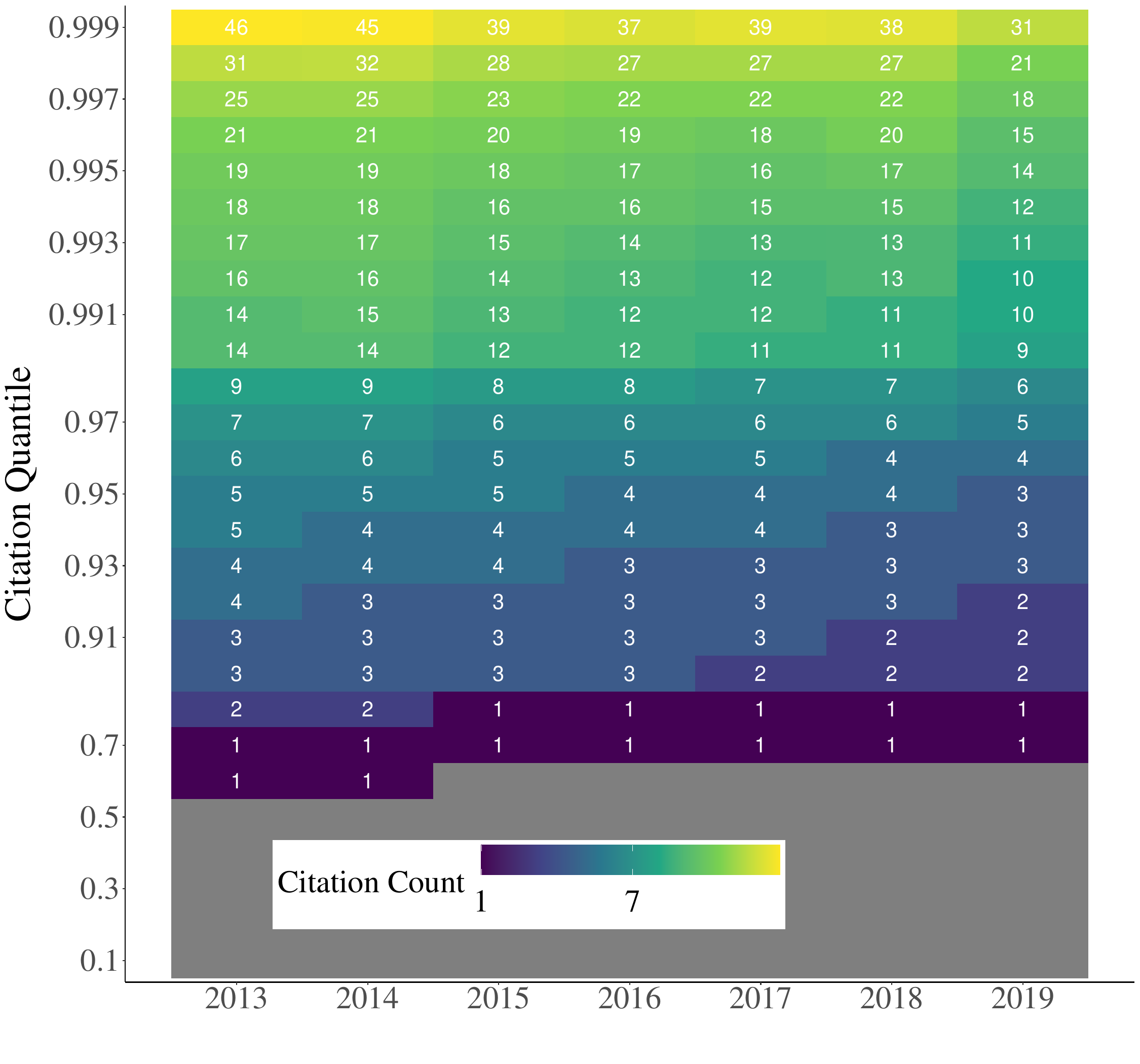}\tabularnewline
\end{tabular}
\par\end{centering}
\medskip{}
\justifying
{\footnotesize{}Notes: \cref{fig: other heterogeneity} displays a heat map measuring the quantiles of the distribution of citations received by papers that cite clinical trials, in each calendar year, from leading journals. The $y$-axis has been stretched to elongate the right-tail of the citation distribution. Colors are displayed in a log scale. The sample of published clinical trials is constructed with the conservative model. The sample of papers that cite clinical trials is constructed using a three-year window.}{\footnotesize\par}
\end{figure}

Next, we collect a set of facts about the quality of this growing set of ``citing'' papers. We begin in \cref{fig: other heterogeneity} by replicating \cref{fig: heterogeneity} for the sample of papers that cite clinical trials. There are several elements of this plot that are informative. First, the share of ``citing'' papers with a small number of citations appears to grow over time. That is---the left tail of the citation distribution for this set is both long and growing. Note also that the right tail of the citation distribution appears to shrink over time. That is, the proportion of highly cited papers appears to be shrinking. These patterns provide one indication that the central tendency of the quality of non-clinical-trial medical research may be decreasing over time.

Citations are imperfect measures of the ex post quality of a scientific publication. Our data allow us to examine several other measures that provide suggestive insight into the ex ante quality of these publications. Unfortunately, shifting geographic trends in this sample and the low quality of measures of public funding for countries outside of the United States render it difficult to use a measure of public funding as a proxy for quality. We do find a 10 percentage point decrease in the share of publications in this set that record having received (U.S.-based) public funding, but this is likely best interpreted as a mechanical consequence of the geographic heterogeneity displayed in \cref{fig: countries}. 

Instead, we focus on author characteristics, constructed using publication data, that also shed some light on research quality. Scientific author norms typically mean that first-listed authors are junior scientists and last-listed authors are senior, lead scientists. We collect facts about these groups separately. 

We begin by examining the productivity of lead scientists who authored papers in our citing sample. We collect all publications indexed in PubMed authored by lead scientists (last-authors) who write at least one paper in our citing sample. The median scientist in this set writes five publications per year. We observe opposing trends in annual publication quantity and citation patterns during this time. We find that the median quantity of papers written by a scientist per year increases from 3 to 5 over this period. The median number of citations to those papers from leading journals drops, over the same period, from 3 to 1. At the 90th percentile, the quantity of per scientist publications per year is stable, at 9 per scientist between 2013 and 2019. Citations from leading journals are similarly stable, at 10 per publication.\footnote{We observe sharp declines in the likelihood that a scientist has ever received funding from the NIH or from another U.S. federal agency. As noted above, however, we hesitate to place much weight on this fact, as it likely reflects changes in the composition of the sample.} For the median senior scientist who appears in our citing sample in 2013, approximately 80 percent of their previously published papers (i.e., published before 2013) had received at least one citation. For scientists who first appear in our citing sample in 2019, the share of previous publications with a non-zero number of citations falls to roughly 70 percent.  

For junior scientists, we focus on a measure of attrition: after authoring a paper in our citing sample, do we ever observe another publication written by that junior scientist?\footnote{\cite{hill2021race} use a similar measure of attrition, when considering the impacts of losing a priority race for scientists' careers.} We collect all publication records in the Web of Science associated with these junior (first-listed) authors. In 2013, the likelihood that a junior scientist published another paper in 2014 was roughly 15 percent. By 2019, this falls to roughly 8 percent. We observe similar trends when two- and three-year publication patterns are observed. We interpret this attrition as evidence that junior scientists who author papers in our citing sample may be less likely to remain research productive in the future.

Taken together, we interpret these facts about scientist research productivity as suggestive evidence of declining average research quality in our citing sample over time. When interpreted in light of the citation distribution trends in  \cref{fig: other heterogeneity}, we conclude that measured increases in the quantity of this ``other'' medical research may be accompanied by a decline in its average quality.

\section{Discussion}\label{sec:conclusions}

Studies of innovation and productivity have long grappled with the prospect of diminishing returns to research \citep[e.g.,][]{merton1935fluctuations,stafford1952rate,schmookler1952changing,schmookler1954level,schmookler1966invention,bloom2020ideas}. Answers to the question at the center of this literature---is the cost of sustained technological progress increasing over time?---have been unsatisfying, as fundamental challenges in the measurement of technological innovation have limited our ability to do more than determine that available metrics do not yield conclusive answers \citep{griliches1990patent,griliches1994}. In one sense, this paper fits in this tradition, in that we find that, in the context of pharmaceutical research, recent concerns about declining productivity are an artifact of confounded measurements of research quantities. 

On the other hand, enabled by advances in generative AI, we develop a procedure for labeling unstructured text that allows us to produce data with an accuracy, precision, and scale sufficient to make progress on the measurement challenges inherent to empirical studies of innovation. We apply this procedure to identify records of clinical trials from a corpus of scientific publications. The resulting census corrects errors in previous measures of clinical trial research. We show that trends in the quantity, quality, and composition of clinical trials have been stable since 2010. 

We contrast these trends with analogous patterns in the production of other forms of medical research. We find that, since 2010, the quantity of other forms of medical research has roughly doubled, a pattern consistent with estimates reported in other papers. On inspection, the increase in quantity is driven by a large increase in the number of papers authored by scientists in China and in the number of papers that synthesize existing literature (e.g., literature reviews and meta-analyses). Growth in these forms of research coincides with a decline in various measures of average publication quality. These facts provide evidence on the source of classification errors in existing measures of clinical trial production and, more generally, provide context on compositional trends relevant to studies that use publications as measures of research. 

These patterns are, in many ways, unsurprising. In recent years, editorials in prominent medical journals \citep{harvey2020we} have observed that changes in the incentive structures for physicians, scientists, and even medical students, across countries, have led to ``an extreme of quantity at the expense of quality'' in medical publishing \citep{siegel2018publish}.\footnote{See, as one example of how these incentives have shifted, program certification requirements from the Accreditation Council for Graduate Medical Education in \cite{ACGME2017}. The ACGME provides formal accreditation for training programs and, as part of this process, lays out expectations of minimum quantities of publication for medical students and clinical faculty. Note that certain researchers have put forth alternative explanations for the rapid increase in the total number of publications. A trend toward evidence-based decision-making in medicine has increased demand for articles that synthesize large bodies of research \citep{lohr2004rating}.} Recent work documents a substantial increase in the quantity of meta-analyses and systematic reviews---summaries of existing clinical trials---driven by scientists outside of the United States \citep{ioannidis2016mass}.  In parallel, there has been a rapid increase in the number, and scale, of ``mega-journals''---scientific journals published frequently, with a large and growing number of pages \citep{ioannidis2023rapid}. 

Despite these changing incentives, substantial constraints---financial, ethical, and practical---keep firms, physicians, and scientists from increasing the quantity of clinical trials. Clinical trials are infrastructure-intensive. Trials are run at specific sites, which must identify patients who satisfy all relevant eligibility criteria and monitor them for periods ranging from six months to twenty years \citep{piantadosi2024clinical}. Sites may struggle to find patients who meet these criteria, especially in small disease markets \citep{kolata2017cancer}, and the costs of accruing large samples for certain studies can be prohibitive \citep{alsan2024representation}. 

Our findings focus on one aspect of the larger biomedical research ecosystem: investments in commercialization efforts in the pharmaceutical industry, which bring products to market in the form of new medicines. Development of new measures of upstream research productivity---centered on basic discoveries in biomedical research and translational efforts to map these discoveries to products safe for testing in human subjects---are a natural area for future research. 

Our findings, also, apply to roughly one decade. We impose this restriction because a set of policy changes in the early 2000s substantially shifted incentives for disclosure and publication of clinical trial results \citep{ICMJE,schulz2010consort}. To be clear, there is nothing about our method that limits its applicability to data reported in the particular form required by these policy changes. Our concern here is about the interpretability of the resulting series. As noted previously, proxies for scientific research provide very little insight if the relationship between one unit of the proxy and units of underlying, unobserved research shift over time. Policy changes in this sector are such that we risk introducing this source of bias with a longer series. Thus, we cannot speak to trends that precede this period. From the perspective of contemporary public policy, however, trends from 2010 to the present are likely to be most relevant. 

Our revision of estimates of clinical trial production is both, from a policy perspective, optimistic and pessimistic. On the one hand, we find evidence consistent with a market in equilibrium: clinical trial production, on a variety of margins, appears stable. This directly calls into question the stylized fact of diminishing returns to drug development, which is viewed in some settings (e.g., \cite{scannell2012diagnosing} and \cite{goldin2024productivity}) as an empirical regularity on par with Moore's Law for semiconductor development. 

On the other hand, it is worth remarking, briefly, on the implications of completely stable clinical trial trends. The early twenty-first century is often described by scientists as a ``golden age'' for medicine. Breakthroughs, such as the completion of the Human Genome Project and the diffusion of DNA sequencing technologies one billion times faster than their predecessors, rapidly expanded the set of available scientific opportunities \citep[e.g.,][]{scannell2012diagnosing}. Technological advances were matched by institutional support, including the doubling of the budget of the U.S. National Institutes of Health between 1998 and 2003. Against these factors that seemingly forecast a period of rapid innovation, the facts documented in this paper---stability in the publication of clinical-trials, but rising quantities of other forms of translation research---highlight the importance of understanding frictions that affect the translation of basic scientific insights into medical technologies relevant to patients. 

Specifically, these patterns suggest that there is likely value in efforts to increase the productivity and decrease the costs of clinical trials, and in efforts to reconcile incentives in academic publishing with social value. A long-standing discussion in medicine and statistics, spurred by \cite{altman1994scandal}, centers on the argument that ``we need less research, better research, and research done for the right reasons.'' Our data indicate that nearly half of all clinical trials published in our sample are never cited by leading medical journals, and roughly 15 percent are never cited at all. 
\end{spacing}
\clearpage 
\newpage
\begin{spacing}{1.2}
\bibliographystyle{apalike}
\bibliography{refs.bib}
\end{spacing}
\newpage

\begin{appendix}
\renewcommand\thefigure{\thesection.\arabic{figure}}    
\setcounter{page}{1}
\setcounter{figure}{0} 
\begin{center}
\begin{spacing}{1}
\large{\it Supplemental Appendix to:}
\vskip0.2cm
\textbf{\Large Counting Clinical Trials:\vspace{0.5em} \\
\Large{New Evidence on Pharmaceutical Sector Productivity}\protect\daggerfootnote{\textit{Date}: \today\\$\star$ Stanford Department of Economics and Stanford Law School, \url{maya.durvasula@stanford.edu}\\
$\dagger$ Stanford Department of Computer Science, \url{eyuboglu@stanford.edu}\\
$\ddagger$ Stanford Graduate School of Business, \url{ritzwoll@stanford.edu}}}\\\vspace{1em}
\begin{tabular}[t]{c@{\extracolsep{2em}}c@{\extracolsep{2em}}c} 
\large{Maya M. Durvasula$^\star$}&  
\large{Sabri Eyuboglu$^\dagger$}&
\large{David M. Ritzwoller$^\ddagger$}
\vspace{-0.3em}\vspace{-0.7em}
\end{tabular}%
\end{spacing}
\end{center}
\begin{spacing}{1.13}
\DoToC
\end{spacing}
\newpage

\begin{spacing}{1.4}

\section{Publication Data}\label{sec: pubmed}

\subsection{The PubMed / MEDLINE Database}\label{sec: pubmed overview}

Our version of the PubMed / MEDLINE database contains roughly $34$ million records and is current through December 2022. These records were constructed by parsing bulk MEDLINE XML files.\footnote{We are grateful to Heidi Williams for sharing this processed data.}  Technically, PubMed and MEDLINE are different products. MEDLINE, a subset of PubMed, is the U.S. National Library of Medicine's (NLM) bibliographic database, which contains references to journal articles in life sciences, with a primary focus on biomedicine. A committee at the NLM determines the set of indexed journals, meaning that only journals that meet certain quality and content standards are indexed. In practice, this means that so-called predatory journals are excluded, as are pre-prints and non-peer reviewed articles. PubMed includes a broader set of records, including pre-prints and publications deposited through alternative processes. Official documentation for MEDLINE is available at the link: \url{https://www.nlm.nih.gov/databases/download/pubmed_medline_documentation.html}. Throughout the text, and from this point forward in the Appendix, we refer to this database as ``PubMed.''

There are 16 records, across all years of data, that are not indexed with tags describing their contents (NML tags). From 2010 forward, roughly 92 percent of records have associated abstracts. We randomly inspect roughly 150 records with no abstracts. None correspond to publications that satisfy our definition of a clinical trial. We assume, throughout this paper, that records with missing citation data have zero associated citations. We confirm that this is generally accurate in two ways. First, we search for Google Scholar records corresponding to roughly 100 randomly selected records with zero citations. In each case, Google Scholar indicates that the paper has no more than two citations. Second, we link records in PubMed to Clarivate Analytics' Web of Science database. We compare citation counts constructed using information in the two databases. We find that the two measures have a correlation of 0.9. 

\subsection{The Universe of Potential Clinical Trials}\label{sec: definition app} 

We construct our sample using records drawn from the universe of publications indexed in PubMed. The initial sample consists of 34,957,127 unique publications indexed in PubMed, as of 15 April 2023. We drop 2,335,653 records that are missing information on publication year, to yield a base sample of 32,621,474 records. From 2010 forward, at least 92 percent of scientific publications published in each year have associated abstracts. We also drop 14,179 publications with publication year 2023, as we have incomplete data for 2023. 

\subsubsection{NLM Tags\label{sec: nlm}} The National Library of Medicine (NLM) assigns each publication a `pubtype.' In the entirely of PubMed, there are 16 records missing pubtype tags. To the best of our knowledge, there have been no efforts to validate the PubMed indexing process used to generate these flags. We flag records with each of the following pubtype (or associated unique identifiers, reported in the field `pubtypeUI') tags: 
\begin{quote}
Adaptive trial; Clinical conference; Clinical study; Clinical trial;
Clinical trial protocol; Clinical trial, Phase 1; Clinical trial,
Phase 2; Clinical trial, Phase 3; Clinical trial, Phase 4; Comparative
study; Controlled clinical trial; Equivalence trial; Evaluation study;
Observational study; Pragmatic clinical trial; Randomized controlled
trial; Twin study; Validation study.
\end{quote}
These categories are chosen to include all pubtypes with the potential to contain a publication satisfying the restriction enumerated in \cref{def:main}. In particular, we follow \cite{feldman2019quantifying}, who use multiple pubtype fields to retrieve a sample of records. Here, two authors reviewed the list of potential pubtypes to identify categories likely to include records of interest. \cref{tab:XML-tags} reports the frequency of each NLM tag across all records in PubMed and subset to those published after 1 January 2010.

\begin{table}[t]
\renewcommand{\arraystretch}{1.1}
\caption{Composition of PubMed by NLM Tag \label{tab:XML-tags}}
\medskip{}

\begin{centering}
\begin{tabular}{ccccccccc}
 & \multicolumn{2}{c}{\textbf{A. All records}} &  & \multicolumn{2}{c}{\textbf{B. 2010-2022}} &  & \multicolumn{2}{c}{\textbf{C. Conserv. Sample}}\tabularnewline
 & Frequency & \% &  & Frequency & \%&  & Frequency & \%\tabularnewline
\hline 
\hline 
adaptive trial & 36  & 0.00  &  & 36  & 0.00 & & 11&0.01\tabularnewline
clinical conference & 7,045  & 0.02 &  & 1,724  & 0.01 & &12 &0.01\tabularnewline
clinical study & 5,053  & 0.02  &  & 5,050  & 0.04 & & 641&0.42 \tabularnewline
clinical trial & 498,722  & 1.53 &  & 79,029 & 0.55 & & 13,969&9.19 \tabularnewline
clinical trial protocol & 9,542  & 0.03  &  & 9,542  & 0.07 & & 124&0.08\tabularnewline
clinical trial, phase 1 & 23,704  & 0.07  &  & 14,139  & 0.10 & & 10,951&7.20\tabularnewline
clinical trial, phase 2 & 37,623  & 0.12  &  & 22,749  & 0.16 & & 17,166&11.29\tabularnewline
clinical trial, phase 3 & 20,597  & 0.06 &  & 15,647  & 0.11 & & 11,112&7.31\tabularnewline
clinical trial, phase 4 & 2,251  & 0.01  &  & 1,790  & 0.01 & & 1,250&0.82\tabularnewline
comparative study & 1,752,380 & 5.37  &  & 424,681  & 2.97 & & 22,521&14.82\tabularnewline
controlled clinical trial & 88,132  & 0.27  &  & 13,839  & 0.10 & & 2,181&1.43\tabularnewline
equivalence trial & 1,047  & 0.00 &  & 1,047  & 0.01 & & 505&0.33\tabularnewline
evaluation study & 243,290  & 0.75 &  & 123,897  & 0.87 & & 998&0.66\tabularnewline
observational study & 127,461  & 0.39 &  & 127,370 & 0.89 & & 5,996&3.94\tabularnewline
pragmatic clinical trial & 2,112  & 0.01  &  & 2,112 & 0.01 & & 253&0.17\tabularnewline
randomized controlled trial & 549,711  & 1.69  &  & 284,774  & 1.99 & & 79,907&53.57\tabularnewline
twin study & 9,030  & 0.03 &  & 4,964  & 0.03 & & 14&0.01\tabularnewline
validation study & 101,817  & 0.31 &  & 62,261 & 0.44 & & 495&0.33\tabularnewline
\hline 
\hline 
 & \multicolumn{2}{c}{$N$=32,621,474} &  & \multicolumn{2}{c}{$N$=14,316,494} &  & \multicolumn{2}{c}{$N$=151,997}\tabularnewline
\hline 
\end{tabular}
\par\end{centering}
\medskip{}
\justifying
{\footnotesize{}Notes: \cref{tab:XML-tags} reports the frequency and percentage of records indexed in PubMed that have been categorized by the NLM as falling into each of 18 categories. These categories are selected for their potential to contain a publication satisfying the restriction enumerated in \cref{def:main}.}{\footnotesize\par}
\end{table}

\subsubsection{Clinical Trial Registry Identifiers\label{sec: registry identify}} In 2004, the International Committee of Medical Journal Editors recommended that research journals decline to publish outcomes associated with trials not pre-registered in some repository.  \cite{ICMJE} summarizes these policies, based on a 2007 revision, in more detail. The U.S. Food and Drug Administration Amendments Act of 2007, Section 801, mandates registration of all clinical trials regulated by the FDA in ClinicalTrials.gov. Many countries have adopted similar guidance.

Registry identifiers are distinctive strings of letters and numbers. We flag records that include clinical trial registry identifiers in their abstract text. In particular, we search for records containing acronyms associated with the following registries:
\begin{quote}
ClinicalTrials.gov (NCT); European Union Drug Regulating Authorities
Clinical Trials Database (EUDRACT); International Traditional Medicine
Clinical Trial Registry (ISRCTN); Australian New Zealand Clinical
Trials Registry (ACTRN).
\end{quote}
We note the potential to overcount records using these searches. We
focus on instances where each trial identifier prefix is followed
by numbers, letters, or punctuation (e.g., NCT12345, ISRCTN: 12345).
However, we may collect records that include these characters in other
settings (e.g., the world disti\textbf{nct}). \cref{tab:Trial-identifiers} reports the frequency of each registry
identifier flagged across all records in PubMed and subset to those published after 1 January 2010. \cref{fig: universe} plots the number of trials with any registry identifier over time. Reassuringly, we observe registry identifiers in PubMed data beginning around 2010, as registration mandates were implemented, and find a steady, smooth increase over time.

\begin{table}[t]
\caption{Composition of PubMed by Trial Registry Identifiers in Abstract Text\label{tab:Trial-identifiers}}
\renewcommand{\arraystretch}{1.1}
\begin{centering}
\begin{tabular}{ccccccccc}
 & \multicolumn{2}{c}{\textbf{A. All records}} &  & \multicolumn{2}{c}{\textbf{B. 2010-present}}&  & \multicolumn{2}{c}{\textbf{C. Conserv. Sample}}\tabularnewline
 & Frequency & \% &  & Frequency & \%&  & Frequency & \%\tabularnewline
\hline 
\hline 
NCT & 97,923 & 0.30 &  & 94,279 & 0.66 & &30,656 &20.17\tabularnewline
EUDRACT & 2,835  & 0.0 &  & 2,816 & 0.00& & 1,564&1.03\tabularnewline
ISRCTN & 12,261 & 0.03 &  & 11,325 & 0.07& & 1,275&0.84\tabularnewline
ACTRN & 5,818 & 0.01 &  & 5,656 & 0.03& & 553&0.36\tabularnewline
\hline 
 & \multicolumn{2}{c}{$N$=32,621,474} &  & \multicolumn{2}{c}{$N$=14,316,494}&  & \multicolumn{2}{c}{$N$=151,997}\tabularnewline
\hline 
\end{tabular}
\par\end{centering}
\medskip{}
\justifying
{\footnotesize{}Notes: \cref{tab:Trial-identifiers} reports the frequency and percentage of records indexed in PubMed that include a string associated with a clinical trial registry identifier in their abstract text. That is, the first row counts the number and percentage of publications that contain the string ``nct.''}{\footnotesize\par}
\end{table}

\subsubsection{Keywords\label{sec: keyword}} We flag records that include any keyword likely to indicate that the record reports the results of a clinical trial in their abstract text. We selected these keywords after several reviewing roughly 200 abstracts flagged through the pubtype process, described above. These keywords are:
\begin{quote}
Randomized; Controlled trial; Control trial; Clinical trial; Treatment
group; Control group; Intervention; Clinical study.
\end{quote}
\cref{tab:Keywords} reports the frequency of each keyword flagged across all records in across all records in PubMed and subset to those published after 1 January 2010.

\begin{table}[t]
\renewcommand{\arraystretch}{1.1}
\caption{Composition of PubMed by Keywords in Abstract Text\label{tab:Keywords}}
\begin{centering}
\begin{tabular}{ccccccccc}
 & \multicolumn{2}{c}{\textbf{A. All records}} &  & \multicolumn{2}{c}{\textbf{B. 2010-present}}&  & \multicolumn{2}{c}{\textbf{C. Conserv. Sample}}\tabularnewline
 & Frequency & \% &  & Frequency & \% &  & Frequency & \%\tabularnewline
\hline \hline
randomized & 550,165  & 1.69  &  & 366,378 & 2.56& & 65,637&43.18\tabularnewline
controlled trial & 107,935  & 0.33  &  & 80,789 & 0.56 & & 11,535&7.59\tabularnewline
control trial & 6,049  & 0.02 &  & 4,846  & 0.03 & & 371&0.24\tabularnewline
clinical trial & 129,510 & 0.40  &  & 92,035 & 0.64& & 17,644&11.61\tabularnewline
treatment group  & 44,627  & 0.14 &  & 27,826  & 0.19& & 4,296&2.63\tabularnewline
control group & 447,901  & 1.37 &  & 286,250  & 2.00 & & 14,245&9.37\tabularnewline
intervention & 629,237  & 1.93  &  & 463,373  & 3.24 & & 10,460&6.88\tabularnewline
clinical study & 32,652  & 0.10 &  & 18,589  & 0.13 & & 2,411&1.59\tabularnewline
\hline \hline
 & \multicolumn{2}{c}{$N$=32,621,474} &  & \multicolumn{2}{c}{$N$=14,316,494}&  & \multicolumn{2}{c}{$N$=151,997}\tabularnewline
\hline 
\end{tabular}
\par\end{centering}
\medskip{}
\justifying
{\footnotesize{}Notes: \cref{tab:Keywords} reports the frequency and percentage of records indexed in PubMed that include each of a collection of keywords in their abstract text. That is, the first row counts the number and percentage of publications that contain the string ``randomized.''}{\footnotesize\par}
\end{table}

\subsubsection{Intersection} There are 3,925,958 records with at least one of the
following attributes: a clinical-trial indicative NLM tag; a clinical trial registry identifier; a clinical-trial indicative keyword in the abstract text. 
The three categories overlap. \cref{tab:Overlap-attributes} records the size of the overlap of each category. In the main text, we restrict attention to this sample for the years 2010-2022. This sample includes 1,821,429 publications.

\subsubsection{Novel Census} We add columns to \cref{tab:Trial-identifiers,tab:Keywords,tab:XML-tags,tab:Overlap-attributes} that document the frequency of each flag in the data constructed for this paper (we use the conservative sample). Several facts are worth noting. First, \cref{tab:Trial-identifiers} suggests that---between 2010 and 2022---less than 25 percent of records identified as reporting the results of a clinical trial reported a registry identifier in their abstract. Although, in principle, registry identifiers may be reported in publication full-texts, rather than abstracts, the low frequency of identifiers in our sample suggests non-compliance with requirements, consistent with patterns in ClinicalTrials.gov documented in \cite{devito2020compliance}. Second, \cref{tab:XML-tags} sheds light on the differences in trial composition that we observe in \cref{fig: universe}: NLM tags that, ostensibly, capture clinical trials have little overlap with one another and do not fully capture the records in our sample. Of the roughly 150,000 records in our census, approximately 100,000 could be identified using NLM tags for ``clinical trial" (and its variants) and ``randomized controlled trial." Observe, however, that use of such flags captures a much larger number of records than one would intend to include. Finally, \cref{tab:Overlap-attributes} suggests that although various text-based flags that may indicate a record is a clinical trial, in the sense of \cref{def:main}, are correlated, there is less overlap than one might expect. 

\begin{table}[t]
\renewcommand{\arraystretch}{1.1}
\caption{Overlap of Publication Attributes\label{tab:Overlap-attributes}}
\begin{centering}
\begin{tabular}{cccccccc}
 Records with any:&  that have any: &  \textbf{A. All records} &  \textbf{B. 2010-present} &  \textbf{C. Conserv. Sample}\tabularnewline
\hline \hline 
Registry ID &  &  116,564  & 111,837 & 33,040\tabularnewline
 & NLM tag  &  81,894 & 78,366 & 26,885\tabularnewline
 & Keyword  &  73,164 & 70,414 & 19,103\tabularnewline
NLM tag  &  &  2,858,924  & 1,804,081 & 124,260\tabularnewline
 & Registry ID &  81,894  & 3,528 & 26,885\tabularnewline
 & Keyword  &  515,755  & 220,681 & 67,153\tabularnewline
Keyword  &  &  1,564,659  & 1,044,461 & 91,349\tabularnewline
 & Registry ID  &  73,164  & 70,414 & 19,103\tabularnewline
 & NLM tag  &  515,755  &  295,074 & 67,153\tabularnewline
\hline \hline 
\end{tabular}
\par\end{centering}
\medskip{}
\justifying
{\footnotesize{}Notes: \cref{tab:Overlap-attributes} reports the size of the intersection between three sets of records indexed in PubMed. These categories are defined by the properties: Possession of a clinical-trial indicative NLM tag, inclusion of a clinical trial registry identifier in abstract text, and inclusion of a clinical-trial indicative keyword in abstract text. These properties are defined in \cref{sec: nlm,sec: registry identify,sec: keyword}, respectively.}{\footnotesize\par}
\end{table}

\subsection{Determining the Set of Leading Medical Journals\label{sec: leading}} \cite{angrist2020inside} propose a strategy to identify the leading journals in a field, which we adapt to this context. In medicine, the ``trunk journals,'' per \cite{angrist2020inside}, are the Journal of the American Medical Association and the New England Journal of Medicine. We collect all instances in which papers published in these journals cite other records in PubMed, between 2010 and 2022. In \cite{angrist2020inside}, the authors select the fifty journals receiving the most cites from these trunk journals. As medicine contains a large number of subfields, we instead collect a list of journals that receive at least 100 citations from a trunk journal over this time period, which yields a list of 84 journals. 

In our data, papers published in trunk journals cite records in 3,136 unique journals.\footnote{We use unique entries in the field \textit{medlineta}---which correspond to distinct MEDLINE abbreviations for journal names---to identify distinct journals. See \url{https://www.nlm.nih.gov/bsd/serfile_addedinfo.html} for details on name abbreviations in this database. We confirm that the count of journals is identical if we, instead, use journals identified by the National Library of Medicine's unique journal identification number.} The median journal is cited twice by these two trunk journals over our time period. The most frequently cited journals are, perhaps unsurprisingly, the trunk journals themselves: JAMA receives 3,143 citations, and the New England Journal of Medicine receives 4,302 citations. The next ten journals, by frequency of citations, are: Lancet (1,457), Circulation (789), Journal of Clinical Oncology (734), Nature (660), Blood (651), PLoS One (645), Journal of the American College of Cardiology (593), Clinical Infectious Diseases (589), BMJ (586), and Annals of Internal Medicine (586). 

\subsection{The Web of Science}\label{sec: wos overview}

We collect supplementary information about papers in our sample from Clarivate Analytics' Web of Science.\footnote{We use a copy of the Web of Science licensed to Stanford University.} While PubMed includes some details on authors, including limited institutional affiliation and address information, the Web of Science provides standardized author addresses for roughly 75 percent of observations in our sample. For each paper in our sample, we extract countries from available mailing addresses for first- and last-listed authors. First- and last-authors have the same listed country for 85 percent of clinical trial records and 91 percent of non-clinical trial records. Thus, we impute to each record the country associated with the first-listed author. In \cref{fig: countries}, we plot geographic trends over time. We use the Web of Science to collect details about author career and publication histories, which allow us to investigate patterns in the production of medical research. The Web of Science assigns to authors a unique, persistent identifier (\textit{DAIS ID}), which allows us to collect information on authors' output before and after the publication focal records in our sample. To our knowledge, there have been no large-scale efforts to validate the quality of these researcher identifiers. On inspection, it appears as though the identifiers perform best, perhaps unsurprisingly, for researchers with distinctive names. When researchers have names that are more frequent in the population, our spot-checking (against online publication records) suggests that there are more likely to be errors. 

\section{ClinicalTrials.gov}\label{sec: ctgov overview}

ClinicalTrials.gov is a registry of clinical trials, run by the United States National Library of Medicine. As of 1 August 2024, ClinicalTrials.gov lists more than 500,000 studies in all 50 states and in 222 countries.\footnote{See https://clinicaltrials.gov/about-site/trends-charts.} ClinicalTrials.gov was available for study registration beginning 29 February 2000. At this time, there were 1,255 registered studies.\footnote{In 1988, in response to public pressure surrounding perceived lags in the development of therapeutics and preventives to treat HIV/AIDS, U.S. Congress passed the Health Omnibus Programs Extension Act (Public Law 100-607), which required the creation of a database of AIDS Clinical Trials Information Services (ACTIS). Publicly-funded clinical trials that were associated with these previous data registration efforts were included in ClinicalTrials.gov at its inception.} By the beginning of 2006, there were 24,822 studies and by the beginning of 2012, 118,020. In January 2024, there were 477,227. The geographic composition of these studies has shifted considerably over time, as have reporting patterns from various firms and universities. 

In this Appendix, we use data from ClinicalTrials.gov to illustrate its limitations as a source for the type of high-quality records necessary for studies of productivity. Given the parallels between this paper and a long literature that studies the usefulness of patent data for studies of innovation, it is worth highlighting the important, perhaps obvious, distinction between ClinicalTrials.gov as a government database and records collected by the U.S. Patent and Trademark Office (USPTO). Records of U.S. patents are collected by the USPTO itself. Disclosure is widely considered a pivotal component of the patent system: in exchange for a short-term quasi-monopoly, information on patented inventions is placed in the public domain. Although ClinicalTrials.gov is described in similarly forceful terms in the laws that authorize its creation, disclosure has been---and continues to remain---effectively voluntary, in the sense that legal tools have never been used by a federal agency to compel the registration of a study or disclosure of its findings. 

Institutional details suggest that the ``voluntariness'' of this reporting has shifted over time, in ways that likely affected sponsors' propensity to register.\footnote{In guidance to researchers on the use of ClinicalTrials.gov, \cite{tse2018avoid} collect a list of the ten challenges associated with using these data, chief among which is the fact that changes in medical journal reporting requirements and federal law make it extremely difficult to determine if increases in the size of the database are related to actual increases in the amount of scientific research.} A coalition of medical journal editors began encouraging registration in 2005 \citep{ICMJE}, registration requirements were imposed for certain FDA-regulated trials in 2008, and the FDA began threatening to impose fines on noncompliant sponsors in 2021. In \cref{tab:ctgov policies}, we collect a list of five changes in the past 25 years that changed sponsors' requirements for trial registration. 

\begin{table}[t]
\renewcommand{\arraystretch}{1.1}
\caption{Policy Changes Affecting the Composition of ClinicalTrial.gov\label{tab:ctgov policies}}
\begin{centering}
\begin{tabular}{>{\raggedright}p{7.5cm}>{\raggedright}p{7.5cm}}
\toprule 
Policy (Effective Date) & Registration Requirement\tabularnewline
\midrule
\midrule 
Food and Drugs Administration Modernization Act (21 November 1997;
ClinicalTrials.gov registry available starting on 29 February 2000) & Clinical trials of investigational new drugs for serious or life threatening
conditions and diseases\tabularnewline
\midrule 
International Committee of Medical Journal Editors (1 July 2005 for
newly initiated trials and 13 September 2005 for ongoing trials) & All clinical trials as a condition for consideration of publication
of results\tabularnewline
\midrule 
Food and Drugs Administration Amendments Act (FDAAA) (27 September
2007) & Non-phase 1 clinical trials of FDA regulated drug and biological products,
and non-feasibility trials of FDA regulated device products\tabularnewline
\midrule 
Final rule implementing FDAAA in Title 42, Part 11 of the Code of
Federal Regulations (42 CFR Part 11) (18 January 2017) & Non-phase 1 clinical trials of FDA regulated drug and biological products,
and non-feasibility trials of FDA regulated device products\tabularnewline
\bottomrule
\end{tabular}
\par\end{centering}
\medskip{}
\justifying
{\footnotesize{}Notes: \cref{tab:ctgov policies} collects a list of five policy changes in the past 25 years that changed sponsors' requirements for trial registration. This table is adapted from \cite{tse2018avoid}.}{\footnotesize\par}
\end{table}

\begin{figure}[t]
\begin{centering}
\caption{Number of Entries in Clinical Trials.gov, by Sample Restrictions}
\label{fig: ctgov}
\medskip{}
\begin{tabular}{c}
\includegraphics[scale=0.4]{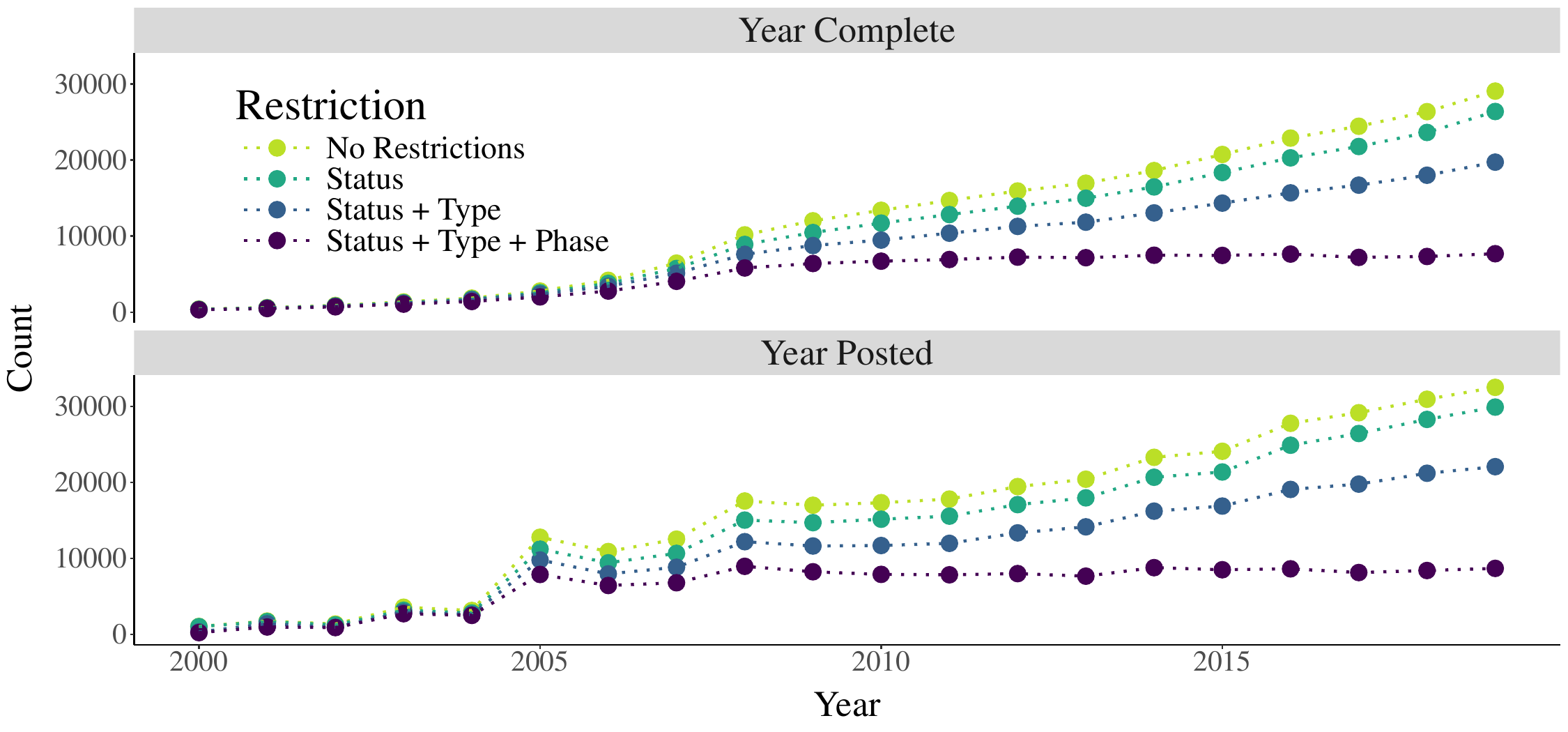}\tabularnewline
\end{tabular}
\par\end{centering}
\medskip{}
\justifying
{\footnotesize{}Notes: \cref{fig: ctgov} displays the counts of the number of records in ClinicalTrials.gov that satisfy various sample restrictions. The $x$-axis of the top panel is the year in which the clinical trial was completed. The $x$-axis of the bottom panel is the year in which the clinical trial was registered on ClinicalTrials.gov.
}{\footnotesize\par}
\end{figure}

We inspect a version of the Aggregate Analysis of ClinicalTrials.gov (AACT) data, a publicly available relational database that contains information on all studies registered on ClinicalTrials.gov, downloaded on 2023 May 31. Of interest, for us, is the question of whether ClinicalTrials.gov can provide an accurate count of the total quantity of clinical trials and, if so, at what date such an accurate count might be available. \cref{fig: ctgov} displays the results of our investigation.

We plot counts of records, selected in various ways, against two x-axes. In the top panel of \cref{fig: ctgov}, we plot counts associated with the year in which clinical trials are marked complete.\footnote{These completion dates are somewhat imprecise. If a trial is registered, but no results are reported, the completion date is an estimate, based on the projected completion date in the study's protocol. If the trial is registered and has associated results, the completion date is the actual date on which data collection ended.} In the bottom panel, we plot counts based on the year in which the study was posted on ClinicalTrials.gov. Measures of trial completion are informative, insofar as they capture the date on which information production is finished. However, trials vary greatly in length---the key observation in \cite{budish2015firms}---and as a result represent investments ``sunk'' at very different earlier points in time. Measures of trial posting are analogous, in some ways, to publication, in that they capture decisions to disclose information. Even without considering the specific series in \cref{fig: ctgov}, observe that the bottom panel includes a variety of sharp changes in the count of records. These correspond to several of the policy changes in \cref{tab:ctgov policies}. We see a sharp change in all four series in 2005, when medical journal policies were changed, a sharp change in 2008, when the FDAAA took effect, and a sharp (albeit smaller) change in 2016, in the lead-up to the effective dates of the final two policy changes in \cref{tab:ctgov policies}. We observe a smaller, but discernible, change in the slope of each series in 2008 when measures of trial completion are used. We draw attention to these differences, in part, because an important challenge for the use of these records in research is determining the appropriate measure of time. 

Turning to the content of these series, we begin by examining the total number of records indexed in ClinicalTrials.gov over time (light green). We define a record as a unique ``national clinical trial identifier,'' although \cite{tse2018avoid} note that one study may be assigned multiple NCTs. There is no method, to our knowledge, that further distinguishes between records. In both panels, this series begins at zero in the year 2000 and increases sharply, to roughly 30,000 records by the end of our data, 2022. On inspection, it is clear that many of the 30,000 records included in the 2022 bin are not clinical trials. Among those that are clinical trials, many do not study medicines, but rather randomize exposure to interventions involving psychological treatments and exercise. 

We impose three sets of restrictions, which aim to bring the data closer to a consistent sample of clinical trials studying the effects of a drug. We select these restrictions based on examination of various fields in ClinicalTrials.gov and with the aim of providing an illustration of our concerns with the use of this dataset ``off the shelf.'' There are alternative ways of constructing a sample of records that may be necessary to yield relevant data for different research questions. We begin by dropping a set of records for which the study ``status'' variable in the database suggests that the study was either never started or was terminated. Specifically, we drop 40,622 records for which the status is listed as ``Withdrawn,'' ``Suspended,'' or ``Terminated.'' This status restriction is plotted in teal and has little impact on the trends in either panel. 

Next, we observe that not all records in ClinicalTrials.gov are interventional studies. We drop any study with a type other than ``interventional.'' This removes an additional 98,138 observations from our data. This status and type restriction is plotted in blue. We observe that the sharply increasing trend in the later years of our data flattens out slightly. That is---the growth in the quantity of records is driven, in part, by growth in the number of registered non-interventional studies. 

We implement a third restriction. We observe that clinical trial ``phases'' are a hallmark of drug trials: medical devices, for example, are approved via a ``stage'' process. Any record for which the trial phase is either missing or ``Not Applicable'' is then less likely to be a drug. We drop these records. This removes 151,541 observations from our data. We plot the resulting series in purple. When all three restrictions are imposed, we find steady increases in the quantity of trials leading up to 2008 and a stable series from 2008 forward. Reassuringly, this trend mirrors that documented in \cref{fig: trend}. These plots suggest a level difference of roughly 2,000 observations, where  \cref{fig: trend} is higher. This is consistent with the idea that ClinicalTrials.gov is \textit{not} a global registry and, as \cite{zarin2017update} notes, many sponsors continue not to register studies in ClinicalTrials.gov 

We emphasize that it is not obvious---in the absence of corroborating evidence, similar to that in \cref{fig: trend}---that the version of this series with all three restrictions in \cref{fig: ctgov} provides an accurate measure of trends in research, especially given the sharp changes in the early part of the series and the somewhat haphazard way that we identified a set of sample restrictions. A series of recent papers---including \cite{devito2020compliance} and \cite{devito2021evaluation}---document limitations with the quality of particular data elements, which further suggests the difficulty of using this database as a standalone source of estimates. 

\section{Prompt Design, Fine-Tuning, and Performance Assessment}\label{sec: llm}

\subsection{Hand-Labeling\label{sec: hand label}} 

A standard concern when using large language models is the quality of the output. We hand-label a set of 3,000 abstracts, selected randomly from our candidate set of records, which allow us to quantify the performance of our classification procedure. We devise an interface for abstract labeling, which allows us to review records quickly.  A labeller reads in a set of publication records and is shown one PubMed identifier (PMID) and one abstract at a time. The labeller can select that the abstract satisfies our sample criteria, or else can indicate a reason that the abstract should be excluded. The buttons for exclusion mirror the inclusion and exclusion criteria in \cref{def:main}, using short-hand convenient for labellers. The hand-labeled data are split into three subsets---validation, training, and testing---based on their eventual use. Splits are assigned randomly. We assign 1000 records to a test set, 1000 to a validation set, and 1082 to a training set. Of these records, we assign labels only to records that have abstracts. Our final training data includes 1082 records, our validation dataset includes 1000 records, and our test set includes 993 records. The loss of seven records in our test set reflects a coding error. We add additional records, randomly selected, to the training and validation datasets when abstracts are missing. We do not add extra records to the testing dataset. 

\subsection{Prompt Design and Error Analysis\label{sec: prompt design}} 

We extract weak labels for 64,000 randomly selected abstracts with two proprietary large language models: OpenAI's GPT-3.5 and GPT-4 \citep{nori2023capabilities,bubeck2023sparks}. As discussed in the main text, we identify three general prompt formats, which differ both in the amount of detail provided about our classification task and in the structure of the requested model completion. We refer to these prompts as Prompt 1.0, Prompt 2.0, and Prompt 3.0, respectively. The text of these prompts are displayed in \cref{prompt: 1.0}, \cref{prompt: 2.0}, and   \cref{prompt: 3.0}, respectively. 

We test each of these three prompts in our 1000-record hand-labelled validation dataset, using both GPT-3.5 and GPT-4. We conduct a detailed error analysis, reported in \cref{tab:Error-Analysis}. To make this concrete, suppose that GPT-3.5, given Prompt 1.0 and an abstract that reports the results of a meta-analysis, erroneously classifies the record as clinical trial, per our definition. In \cref{tab:Error-Analysis}, we flag this error under Prompt 1 for GPT-3.5, as an error of type ``meta-analysis.'' Within meta-analysis, we assign the error to a sub-type. If the abstract explicitly includes terms such as ``meta-analysis,'' ``literature search,'' or ``literature review,'' we categorize this as an explicit error. If the abstract references that the publication is summarizing existing studies, or searching a database for records and collecting their findings, we categorize this as an implicit error. We devise error types based on our inclusion/exclusion criteria, in \cref{def:main}, and select sub-types based on common categories of errors. 

We revise each class of prompt based on these findings. For Prompt 1.0, the simplest true/false prompt, we consider three variants. We refer to these revised prompts as Prompts 1.1, 1.2, and 1.3. The text of these prompts is displayed in \cref{prompt: 1.1,prompt: 1.2a,prompt: 1.3a}. For Prompts 2 and 3, we consider one variant each. We refer to these revised prompts as Prompts 2.1 and 3.1. The text of these prompts is displayed in \cref{prompt: 2.1a,prompt: 3.1a}. These changes reflect the differences in performance catalogued in \cref{tab:Error-Analysis}.

\begin{landscape}
\renewcommand{\arraystretch}{0.8}
\begin{centering}

\begin{longtable}{>{\centering}p{14cm}>{\centering}p{1cm}>{\centering}p{1cm}>{\centering}p{1cm}>{\centering}p{1cm}>{\centering}p{1cm}>{\centering}p{1cm}}
\caption{Baseline Prompt Error Analysis\label{tab:Error-Analysis}}\\
\multicolumn{1}{c}{Model} & \multicolumn{3}{c}{GPT-3} & \multicolumn{3}{c}{GPT-4}\tabularnewline
\cmidrule{2-7} \cmidrule{3-7} \cmidrule{4-7} \cmidrule{5-7} \cmidrule{6-7} \cmidrule{7-7} 
\multicolumn{1}{c}{} & \multicolumn{3}{c}{Prompt} & \multicolumn{3}{c}{Prompt}\tabularnewline
 Error Type & 1.0 & 2.0 & 3.0 & 1.0 & 2.0 & 3.0\tabularnewline
\midrule
\midrule 
\textbf{No Drug} &  &  &  &  &  & \tabularnewline
\midrule 
  vitamin/supplement & 12 & 12 & 9 & 15 & 4 & 15\tabularnewline
  surgical/medical procedure/diagnostic & 40 & 27 & 29 & 35 & 21 & 29\tabularnewline
  abstract not specific about name of medicine & 3 & 4 & 4 & 6 & 5 & 6\tabularnewline
  food/beverages & 8 & 3 & 6 & 13 & 1 & 12\tabularnewline
  surgical/medical device & 8 & 4 & 5 & 7 & 2 & 6\tabularnewline
  surgical/medical imaging & 1 & 1 & 1 & 2 & 0 & 2\tabularnewline
  behavioral/physical therapy/exercise & 40 & 22 & 36 & 25 & 8 & 13\tabularnewline
  misses the mention of a drug & 25 & 10 & 5 & 2 & 1 & 0\tabularnewline
  surgical/medical material (e.g., resin/dental filling) & 1 & 0 & 1 & 3 & 1 & 1\tabularnewline
  non-medical pollution/chemical/drug & 1 & 1 & 1 & 1 & 2 & 1\tabularnewline 
  other & 1 & 0 & 1 & 2 & 1 & 2\tabularnewline
\midrule 
\textbf{Meta-Analysis} &  &  &  &  &  & \tabularnewline
\midrule 
  explicit mention of meta-analysis/literature search/literature review & 13 & 5 & 13 & 8 & 0 & 6\tabularnewline
  mention of database searched, explicit mention of meta-analysis & 3 & 3 & 3 & 3 & 0 & 2\tabularnewline
  summary existing studies, without reference to
data search / meta-analysis & 6 & 7 & 4 & 3 & 4 & 3\tabularnewline
\midrule 
\textbf{Retrospective} &  &  &  &  &  & \tabularnewline
\midrule 
  re-analysis of previously collected data without explicitly mentioning
``retrospective'' & 8 & 2 & 6 & 6 & 5 & 4\tabularnewline
  explicitly described as ``retrospective''
or ``retrospective analysis'' & 6 & 1 & 6 & 0 & 0 & 0\tabularnewline
  miscategorizes a study conducted in the past as ``retrospective'' & 0 & 1 & 0 & 0 & 0 & 0\tabularnewline
  other &  &  &  &  &  & \tabularnewline
\midrule 
\textbf{ Observational} &  &  &  &  &  & \tabularnewline
\midrule 
  no active intervention described; does not explicitly say ``observational'' & 10 & 9 & 11 & 6 & 0 & 4\tabularnewline
  no active intervention described; explicitly says ``observational'' & 14 & 13 & 13 & 13 & 2 & 11\tabularnewline
  misses reference to active intervention & 1 & 1 & 1 & 0 & 0 & 0\tabularnewline
\midrule 
 \textbf{Protocol} &  &  &  &  &  & \tabularnewline
\midrule 
  no results reported about current study; explicitly says ``protocol'' & 10 & 9 & 11 & 6 & 0 & 4\tabularnewline
  no results reported about current study; does not explicitly say ``protocol'' & 14 & 13 & 13 & 13 & 2 & 11\tabularnewline 
  study is based on a simulation, not real world data & 1 & 1 & 1 & 0 & 0 & 0\tabularnewline
\midrule 
\textbf{No Human Subjects} &  &  &  &  &  & \tabularnewline
\midrule 
  only references to laboratory experiments, cells drawn from humans & 0 & 1 & 0 & 0 & 0 & 0\tabularnewline
\midrule 
\textbf{Animal} &  &  &  &  &  & \tabularnewline
\midrule 
  study conducted on animals & 5 & 10 & 6 & 2 & 3 & 2\tabularnewline
\midrule 
\textbf{ Other} &  &  &  &  &  & \tabularnewline
\midrule 
  overly literal interpretation of inclusion criteria & 0 & 2 & 30 & 0 & 0 & 2\tabularnewline
\midrule 
  studying outcomes/objects unrelated to a randomized trial & 1 & 1 & 1 & 1 & 1 & 1\tabularnewline
\bottomrule
\end{longtable}
\par\end{centering}
\medskip{}
\justifying
{\footnotesize{}Notes: \cref{tab:Error-Analysis} categorizes errors documented in the first round of prompt iteration, using Prompts of sub-type 0. See \cref{tab:prompt-performance-in-validation-data} for details on model performance. For each model error---an instance in which a model returned a classification that deviated from hand-labelled data---we examined the associated record. We categorized errors by Type and Sub-Type. Types are drawn from the exclusion restrictions in \cref{def:main}. Sub-types describe consistent characteristics of publications that resulted in such an error. We report counts of errors, by model and prompt, of each Type and Sub-type.}{\footnotesize\par}\end{landscape}

\subsection{Error Analysis\label{sec: final error analysis}}

The Conservative model incorrectly labels 27 papers. We inspect each of these errors. We review each abstract and aim to understand what might have generated an error. In roughly half of cases--13 of 27---there is a clear error. These clear errors include publications that explicitly report the results of observational studies (one error), that re-analyze existing data (eight errors), and that report literature reviews (one error). This set also includes clinical trials that satisfy all criteria in \cref{def:main}, except the treatment being studied is not a drug (three errors). In 14 cases, however, errors are associated with records that are difficult for human labellers to categorize, either because the content of the publication does not fit neatly into the inclusion and exclusion criteria implied by \cref{def:main} or because the publication is written in a way that makes it difficult to determine details of the study. PubMed record 27880726 is illustrative. This publication studies the effect of chrysophanic acid (CA) on benign prostatic hyperplasia. The abstract is unclear about whether this treatment is studied in human subjects or in an animal model--though the publication title and full-text make clear that this was conducted in a sample of rats. Based on the abstract alone, however, nothing suggests that this is an animal study. It is excluded from both the conservative and moderate model-generated samples, but a human-labeller flagged it as satisfying our criteria. 

\section{Additional Figures and Further Analyses\label{sec: additional}}

\subsection{Growth in Citing Papers by Geography\label{sec: geography}}

In \cref{sec:intext_geography}, we document differential trends in the production of publications by type. Specifically, we find essentially stable geographic patterns in the production of clinical trials, whereas we find that large increases in the quantity of papers in our citing sample are driven by research produced outside of the United States. Here, we provide additional details on this geographic fact. In \cref{sec: wos overview}, we provide details on the data that are used to construct these plots. There are two especially striking aspects of \cref{fig: countries}: the quantity of citing papers published by researchers in China is high and sharply rising, and the quantity of (pooled) citing papers from what we refer to as the ``rest of the world'' set is also high and rising. \cref{tab: geography} breaks down the publication of citing papers by country. We elect to focus on countries that produced at least 200 citing papers in 2019---a cut that captures 93 percent of the citing papers in our data. That is, nearly all of the papers in the citing sample are produced by this relatively small number of countries. We highlight several patterns in \cref{tab: geography}. First---consistent with its presentation as a stand-alone series in \cref{fig: countries}---the quantity of citing papers produced by Chinese researchers, the change in the size of this set, and its growth over our sample period are distinctive. While the United States continues the produce the largest number of papers in the citing sample, the growth of this set ranks last among the countries listed in \cref{tab: geography}. Whereas there is 11 percent change in the quantity of citing papers produced by United States-based researchers, there is more a more than 200 percent change for Chinese papers. 

\begin{table}[t]
\renewcommand{\arraystretch}{1.1}
\caption{Changes in the Number of Papers Citing Clinical Trials by Country\label{tab: geography}}
\begin{centering}
\begin{tabular}{lllll}
\toprule 
Country & Published, 2013 & Published, 2019 & Change, 2013 to 2019 & \% Change, 2013 to 2019\tabularnewline
\midrule
\midrule 
China & 1724 & 5620 & 3896 & 226.0\tabularnewline
Poland & 158 & 311 & 153 & 96.8\tabularnewline
Japan & 1167 & 2159 & 992 & 85.0\tabularnewline
Spain & 487 & 892 & 405 & 83.2\tabularnewline
Italy & 1241 & 2205 & 964 & 77.7\tabularnewline
Austria & 191 & 319 & 128 & 67.0\tabularnewline
Korea & 635 & 1045 & 410 & 64.6\tabularnewline
Netherlands & 617 & 1011 & 394 & 63.9\tabularnewline
Belgium & 211 & 335 & 124 & 58.8\tabularnewline
France & 695 & 1075 & 380 & 54.7\tabularnewline
Denmark & 223 & 343 & 120 & 53.8\tabularnewline
Austria & 597 & 907 & 310 & 51.9\tabularnewline
Czech Republic & 320 & 485 & 165 & 51.6\tabularnewline
Brazil & 286 & 431 & 145 & 50.7\tabularnewline
Greece & 158 & 228 & 70 & 44.3\tabularnewline
Sweden & 276 & 391 & 115 & 41.7\tabularnewline
Germany & 1555 & 2148 & 593 & 38.1\tabularnewline
Canada & 791 & 1044 & 253 & 32.0\tabularnewline
India & 302 & 393 & 91 & 30.1\tabularnewline
Iran & 167 & 215 & 48 & 28.7\tabularnewline
Great Britain & 1680 & 2104 & 424 & 25.2\tabularnewline
Turkey & 231 & 259 & 28 & 12.1\tabularnewline
U.S.A. & 11207 & 12536 & 1329 & 11.9\tabularnewline
\bottomrule
\end{tabular}
\par\end{centering}
\medskip{}
\justifying
{\footnotesize{}Notes: \cref{tab: geography} records the number of papers that cite clinical trials published in the preceding three years in 2013 and 2019, by country. We include all countries that publish more than 200 papers that cite clinical trials in 2019. The fourth column displays the change in the number of `citing' papers between 2013 and 2019. The fifth column displays the percent change in the number of `citing' papers between 2013 and 2019.}{\footnotesize\par}
\end{table}

\subsection{Quantifying Meta-Analyses \label{sec: meta-analyses}}

In \cref{sec: content}, we present three methods that allow us to identify ``review'' publications---publications that summarize the findings of existing studies---in our set of citing papers. Each approach is imperfect (for reasons that mirror the motivation for the technical contribution of this paper). Here, we provide additional details on these three approaches. 

\subsubsection{Hand-Labeling Meta-Analyses}

From the set of publications that cite clinical trials (three-year sample), we drew 100 observations randomly from each publication year. We inspected records in the first publication year (2013) and last (2019). Each record was reviewed twice. During the first round of review, we collected notes about the content of the publication. We standardized these notes and observed that these notes could be coded into seven distinct categories. We reviewed the records a second time and assigned these standardized codes. \cref{tab: hand label meta} reports the composition of these 100-record samples. 

The three largest categories in both samples are pre-clinical studies, observational studies, and meta-analyses. We estimate a decrease in the proportion of pre-clinical studies and an increase in the proportion of meta-analyses. By contrast, the proportion of observation studies appears stable.

\begin{table}[t]
\renewcommand{\arraystretch}{1.1}
\caption{Composition of Publications that Cite Clinical Trials\label{tab: hand label meta}}
\begin{centering}
\begin{tabular}{lll}
\toprule 
 & Published in 2013 & Published in 2019\tabularnewline
\midrule
\midrule 
Clinical Trial & 1 & 0\tabularnewline
Editorial & 3 & 2\tabularnewline
Pre-Clinical & 26 & 17\tabularnewline
Non-Drug RCT & 1 & 4\tabularnewline
Protocol & 1 & 1\tabularnewline
Case Study & 5 & 4\tabularnewline
Observational & 31 & 30\tabularnewline
Meta-Analysis & 32 & 41\tabularnewline
\bottomrule
\end{tabular}
\par\end{centering}
\medskip{}
\justifying
{\footnotesize{}Notes: \cref{tab: hand label meta} records the results of an exercise aimed at characterizing the composition of the set of publications that cite clinical trials (three-year sample). We randomly draw 100 publications from this sample that were published in 2013 and 2019, respectively. We categorize each publication into one of seven categories. We report the number of publications, in each year, that fall into each category.}{\footnotesize\par}
\end{table}

\subsubsection{Text Searches for Meta-Analyses\label{sec: meta keyword}}

Recall that, in the process of constructing the training data that were used to fine-tune our large language models, we identified many records in PubMed that were meta-analyses, literature reviews, etc., based on the contents of their text. We reviewed all hand-labeled records from this training process and identified a set of keywords and phrases that appeared most frequently. We constructed a list of phrases and keywords that consistently indicated, to a human labeler, that a record is a meta-analysis or review article. This list is reproduced in \cref{fn: meta keywords}. 

Using this list of search terms, we search the abstracts of all records in our citing sample. We flag any record with any of the keywords listed in \cref{fn: meta keywords} as a meta-analysis and flag any record that mentions at least two of the databases listed in \cref{fn: meta keywords} as well. 

\begin{figure}[t]
\begin{centering}
\caption{Trends in Review Articles and Meta-Analyses}
\label{fig: meta}
\medskip{}
\begin{tabular}{cc}
\textit{Panel A: Counts} &  \textit{Panel B: Proportions}
\tabularnewline
\includegraphics[scale=0.4]{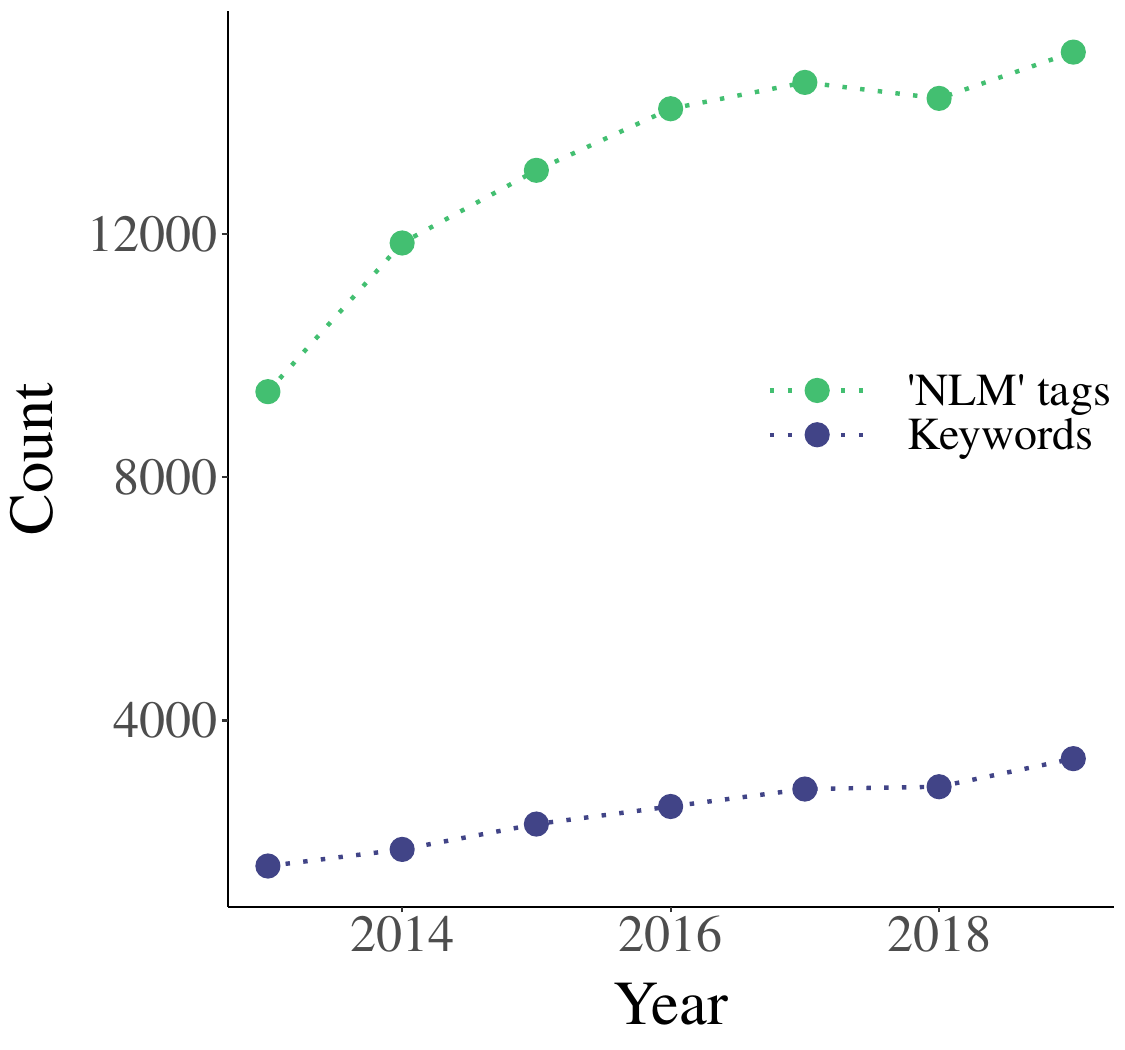}
& \includegraphics[scale=0.4]{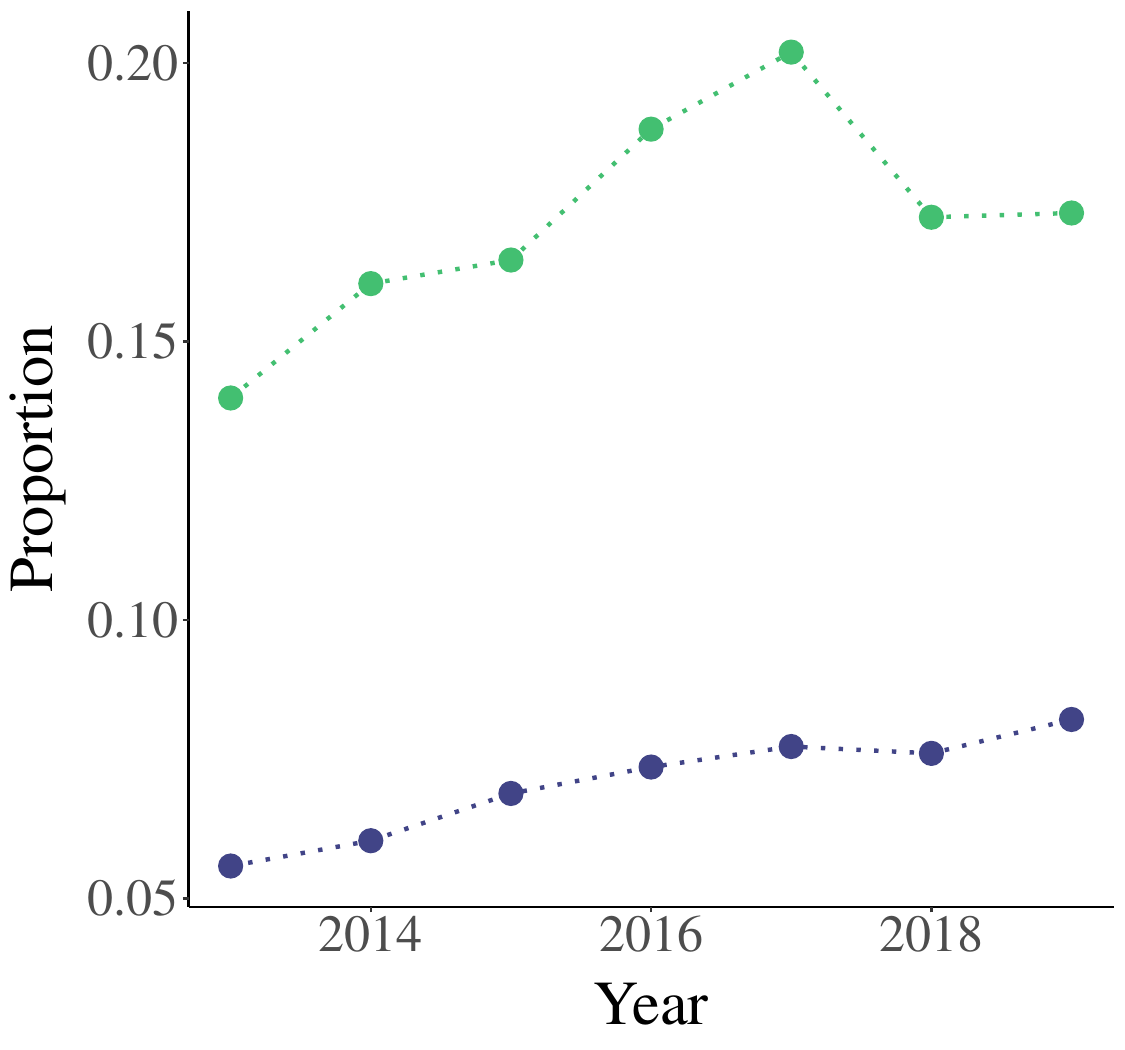}
\tabularnewline
\end{tabular}
\par\end{centering}
\medskip{}
\justifying
{\footnotesize{}Notes: \cref{fig: meta} displays the results of two exercises aimed at characterizing the change in the quantity of meta-analyses in the set of publications that cite clinical trials (three-year sample) over time. Estimates associated with the keyword search discussed in \cref{sec: meta keyword} are displayed in purple. Estimates associated with the `NLM' tag query discussed in \cref{sec: meta nlm} are displayed in purple. Panel A displays counts of the number of publications that satisfy both sets of criteria. Panel B displays proportions of publications that satisfy both sets of criteria.}{\footnotesize\par}
\end{figure}

The objective of this exercise is to yield a conservative estimate of the change in the quantity of meta-analyses over our period of interest. In \cref{fig: meta}, we plot this trend. Panel A plots the raw count of meta-analyses flagged by this procedure. Panel B plots the proportion of records in our citing sample that are designated as meta-analyses via this search. As expected, this approach captures a small number of records: in 2013, there are 1,606 records in our citing sample flagged as meta-analyses. This rises to 3,372 records in 2019. Although the baseline level this small, this corresponds to a 109 percent increase over this period. 

\subsubsection{NLM Tag Queries for Meta-Analyses\label{sec: meta nlm}} 

Next, we observe that NLM tags, added when publication records are indexed in PubMed, also record information that may signal whether a record is a meta-analysis. On inspection, three of the 77 NLM tags are relevant: ``literature review,'' ``meta-analysis,'' and ``systematic review.'' Of course, as \cref{sec: objective} makes clear, NLM tags may overcount records. As \cref{sec: meta keyword} generates what is likely a lower bound, we proceed with this strategy, noting that it is likely to yield an upper bound. 

We flag all records in our citing sample indexed with one of these three NLM tags. We plot the resulting trend---as a count in Panel A and as a proportion of records in our citing sample in Panel B---of \cref{fig: meta}. In 2013, there are 9,407 records satisfying these criteria. In 2019, there are 14,991 such records. Over our period of interest, there is, thus, a 59 percent increase. 

\section{Prompt Repository\label{sec: prompt repository}} 
\begin{figure}
\begin{centering}
\caption{Prompt 1.0: True/False, Base}
\label{prompt: 1.0}
\medskip{}
\begin{tabular}{c}
\includegraphics[scale=0.25]{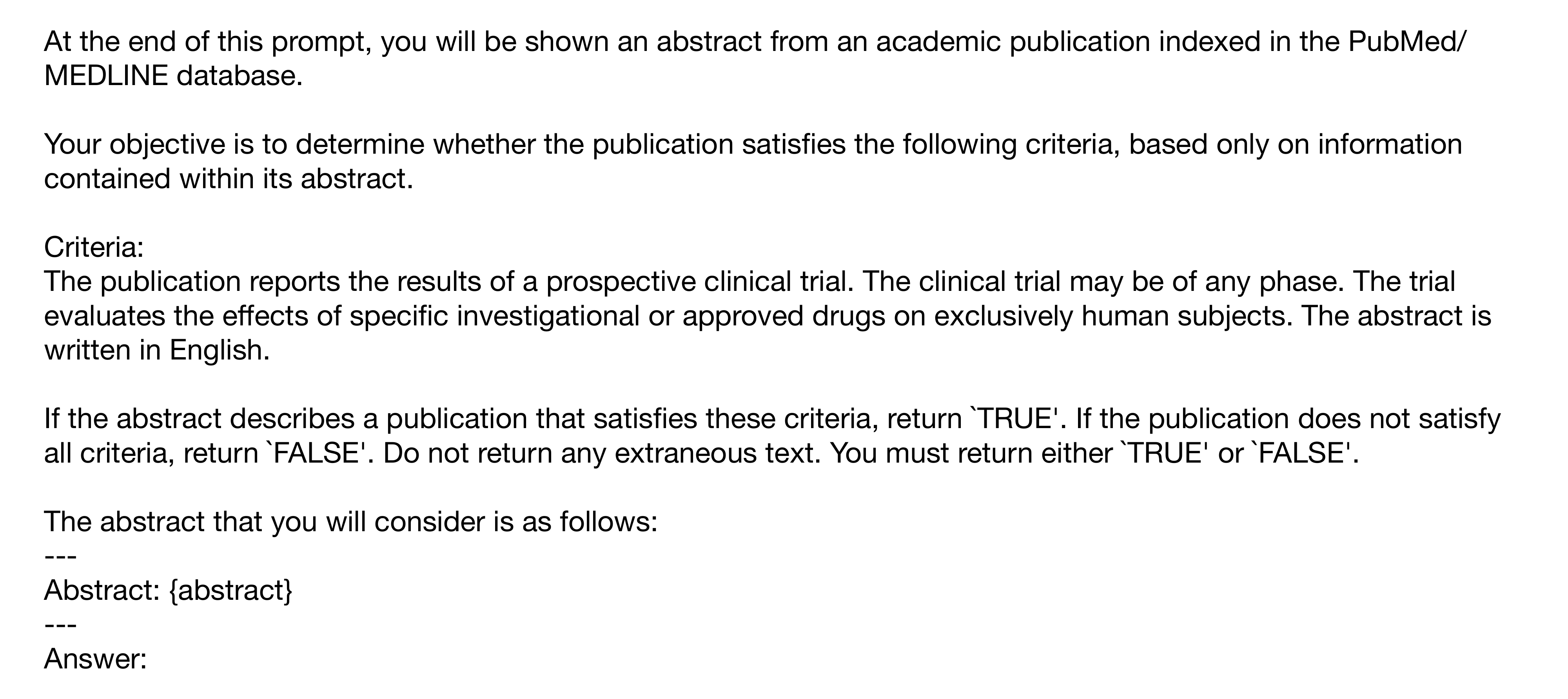}\tabularnewline
\end{tabular}
\par\end{centering}
\medskip{}
\justifying
{\footnotesize\par}
\end{figure}

\begin{figure}
\begin{centering}
\caption{Prompt 1.1: True/False, Initial Examples}
\label{prompt: 1.1}
\medskip{}
\begin{tabular}{c}
\includegraphics[scale=0.25]{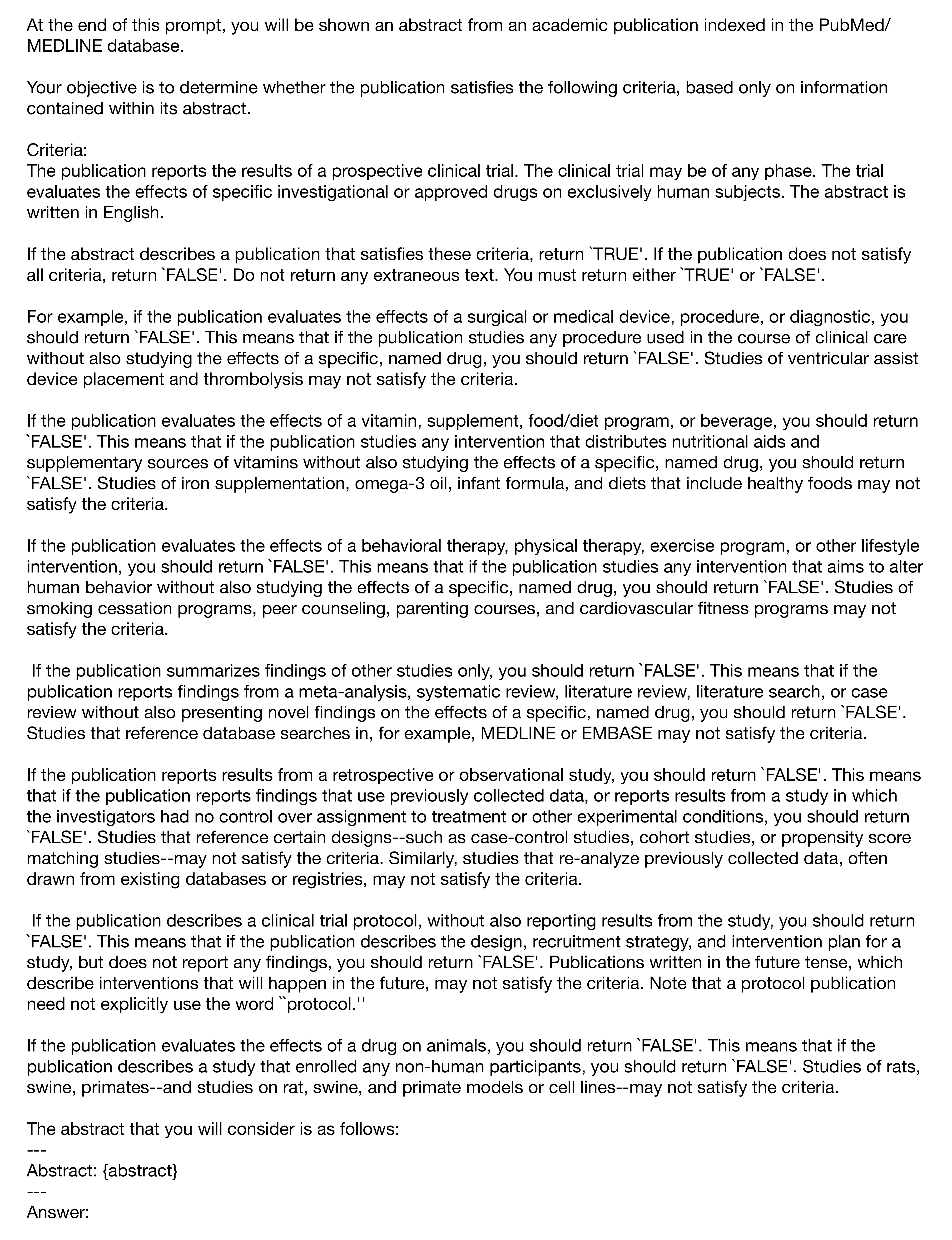}\tabularnewline
\end{tabular}
\par\end{centering}
\medskip{}
\justifying
{\footnotesize\par}
\end{figure}

This appendix serves as a repository for figures displaying each of the prompts considered in \cref{sec: prompt design}. Throughout, the dummy text ``\{abstract\}'' is placed in the location where the abstract of a publication would be input. The baseline ``True/False'' Prompt 1.0 is displayed in \cref{prompt: 1.0}. Refinements to this prompt are displayed in \cref{prompt: 1.1,prompt: 1.2a,prompt: 1.3a}. The prompt that asks the model to categorize excluded publications, i.e., Prompt 2.0, is displayed in \cref{prompt: 2.0}. A refinement to this prompt is given in \cref{prompt: 2.1a}. The prompt that asks the model to explain why it has excluded a publication, i.e., Prompt 3.0, is displayed in \cref{prompt: 3.0}. In turn, a refinement to this prompt is given in \cref{prompt: 3.1a}.

\clearpage

\begin{figure}
\begin{centering}
\caption{Prompt 1.2: True/False, Extended Examples, Short}
\label{prompt: 1.2a}
\medskip{}
\begin{tabular}{c}
\includegraphics[scale=0.25]{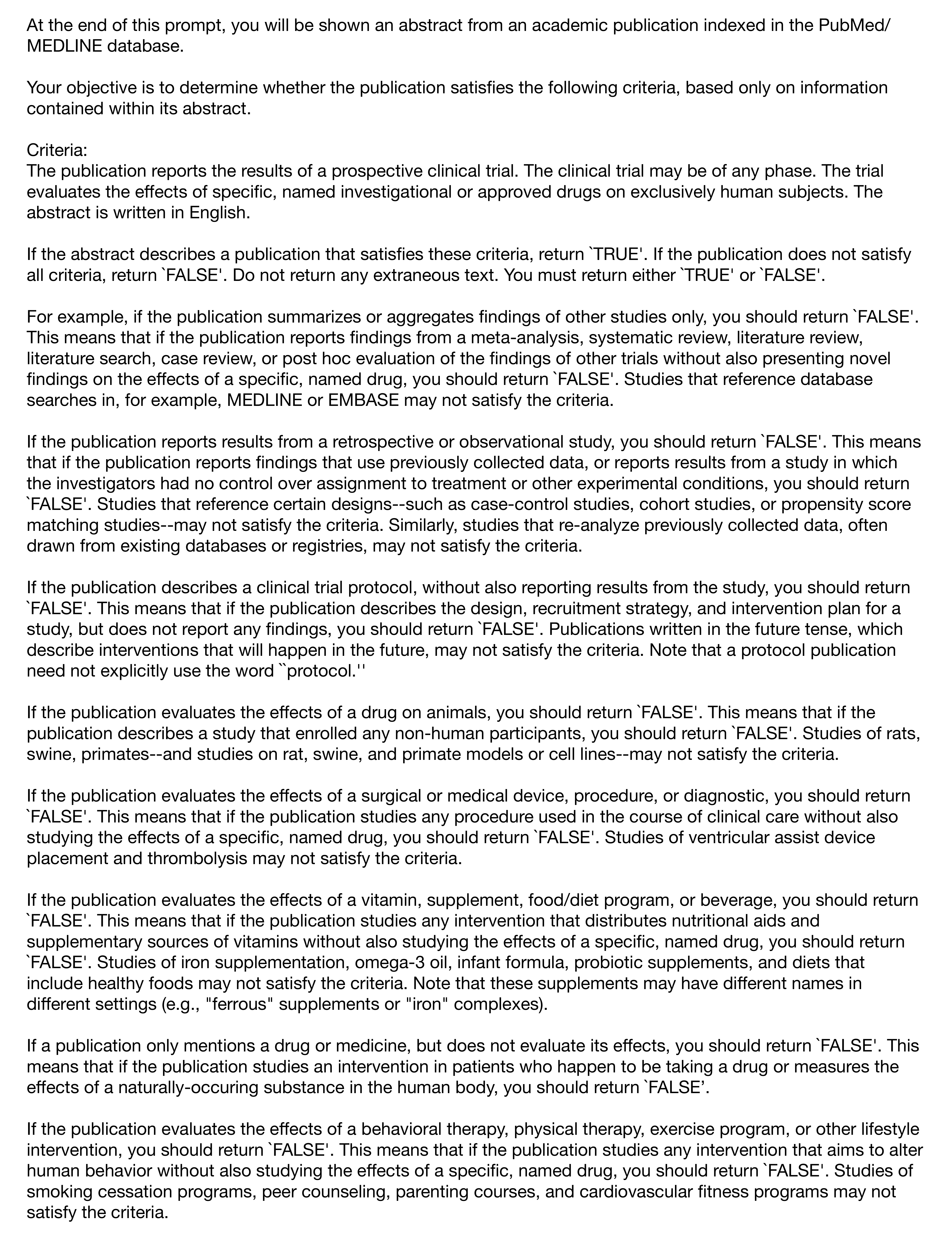}\tabularnewline
\end{tabular}
\par\end{centering}
\medskip{}
\justifying
{\footnotesize\par}
\end{figure}

\begin{figure}
\ContinuedFloat
\begin{centering}
\caption{Prompt 1.2: True/False, Extended Examples, Short (Cont.)}
\medskip{}
\begin{tabular}{c}
\includegraphics[scale=0.25]{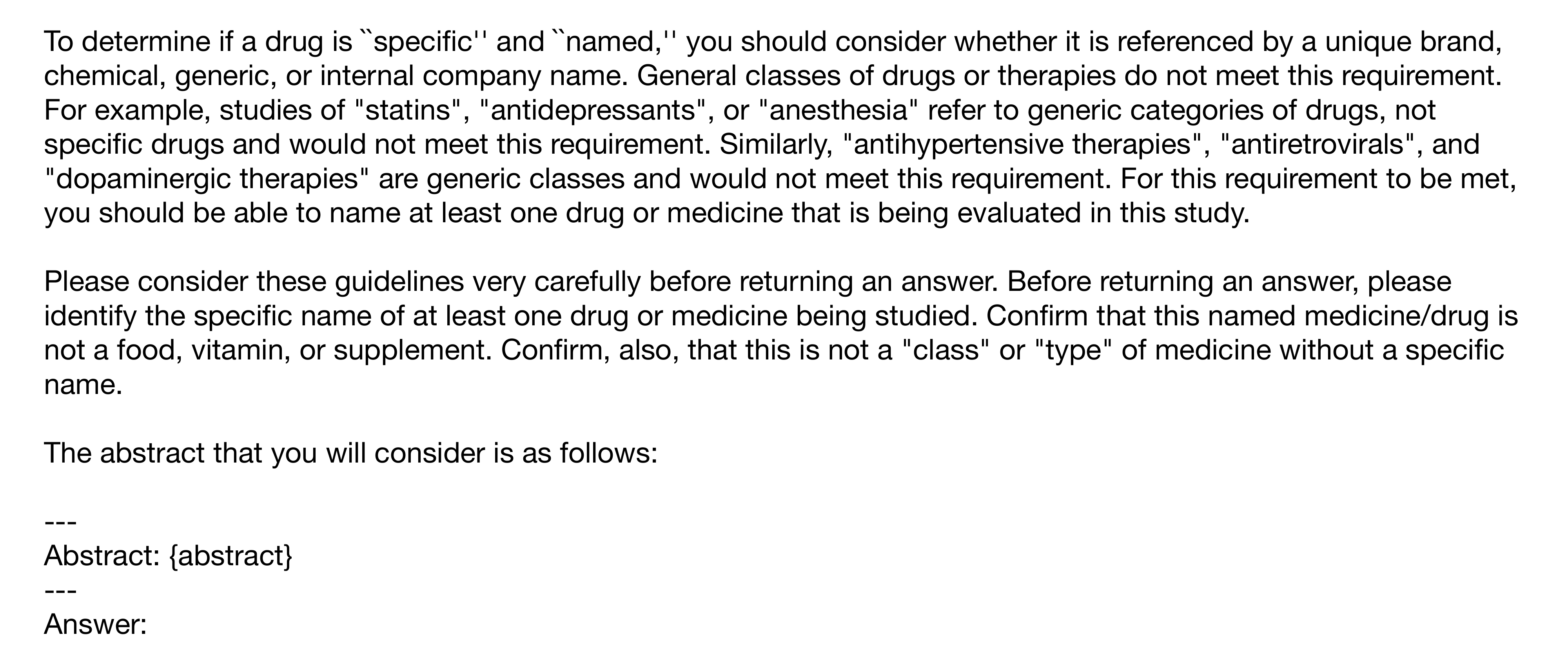}\tabularnewline
\end{tabular}
\par\end{centering}
\medskip{}
\justifying
{\footnotesize\par}
\end{figure}

\begin{figure}
\begin{centering}
\caption{Prompt 1.3: True/False, Extended Examples, Long}
\label{prompt: 1.3a}
\medskip{}
\begin{tabular}{c}
\includegraphics[scale=0.25]{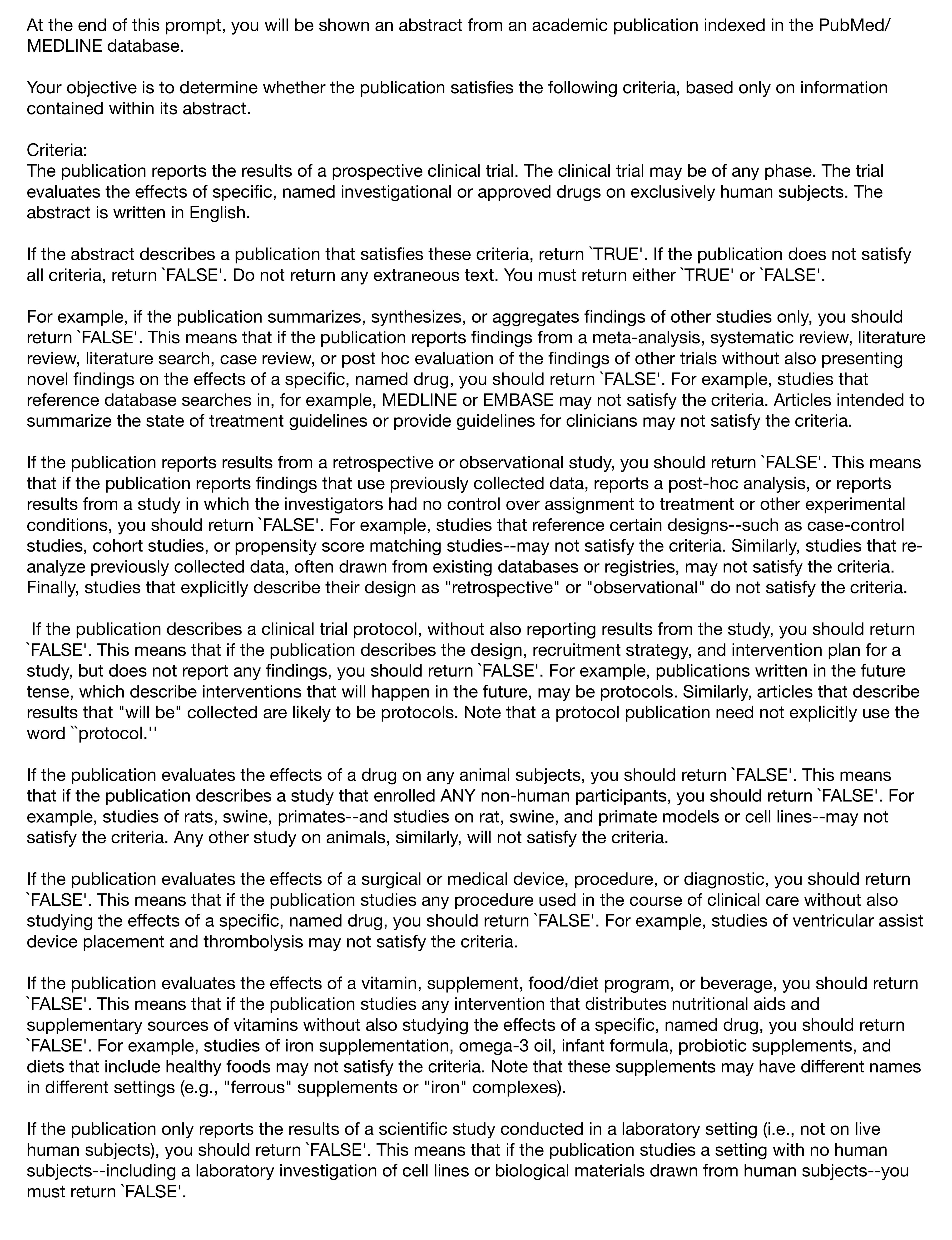}\tabularnewline
\end{tabular}
\par\end{centering}
\medskip{}
\justifying
{\footnotesize\par}
\end{figure}

\begin{figure}
\ContinuedFloat
\begin{centering}
\caption{Prompt 1.3: True/False, Extended Examples, Long (Cont.)}
\medskip{}
\begin{tabular}{c}
\includegraphics[scale=0.25]{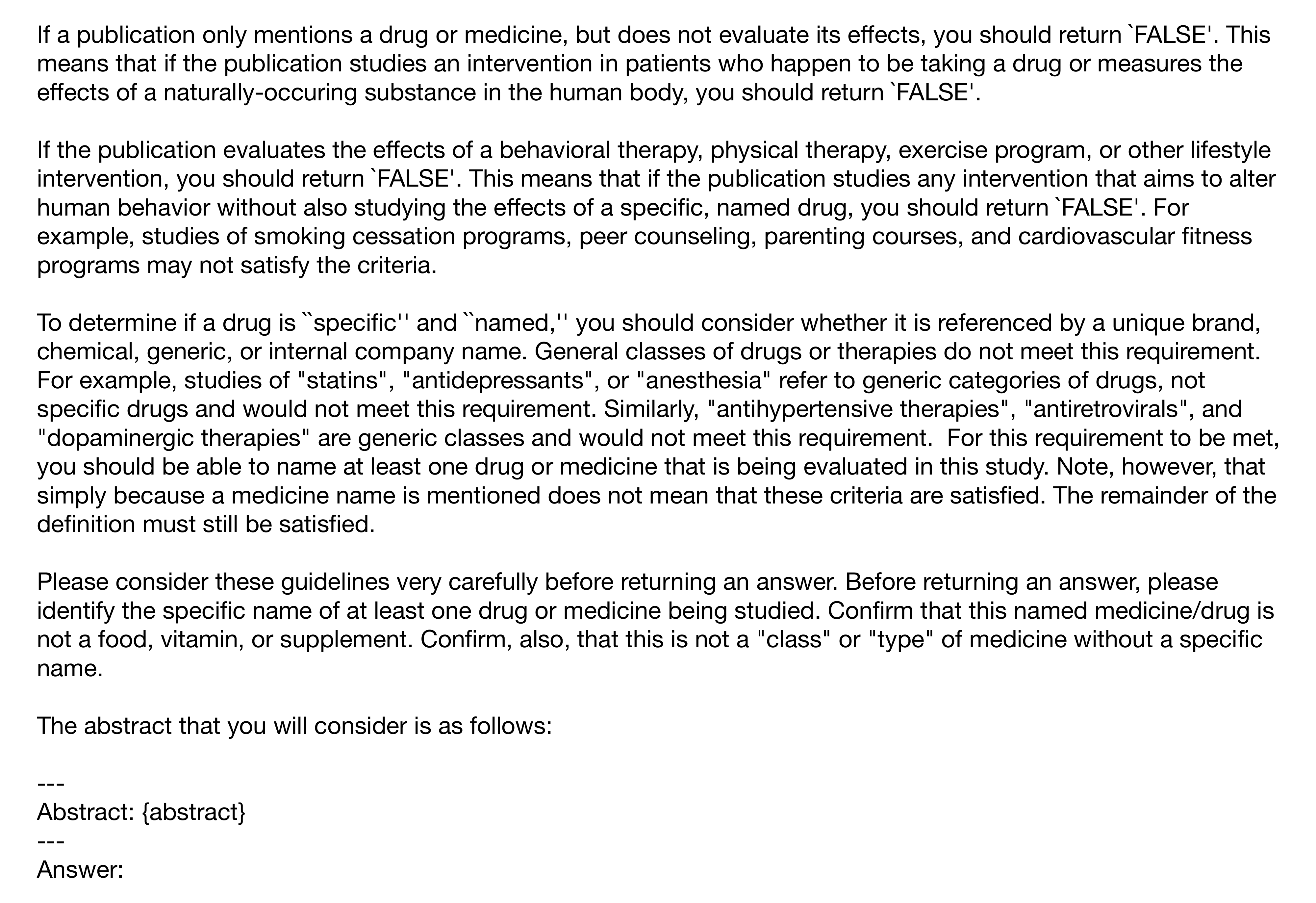}\tabularnewline
\end{tabular}
\par\end{centering}
\medskip{}
\justifying
{\footnotesize\par}
\end{figure}

\begin{figure}
\begin{centering}
\caption{Prompt 2.0: Categorize Exclusion Restriction}
\label{prompt: 2.0}
\medskip{}
\begin{tabular}{c}
\includegraphics[scale=0.25]{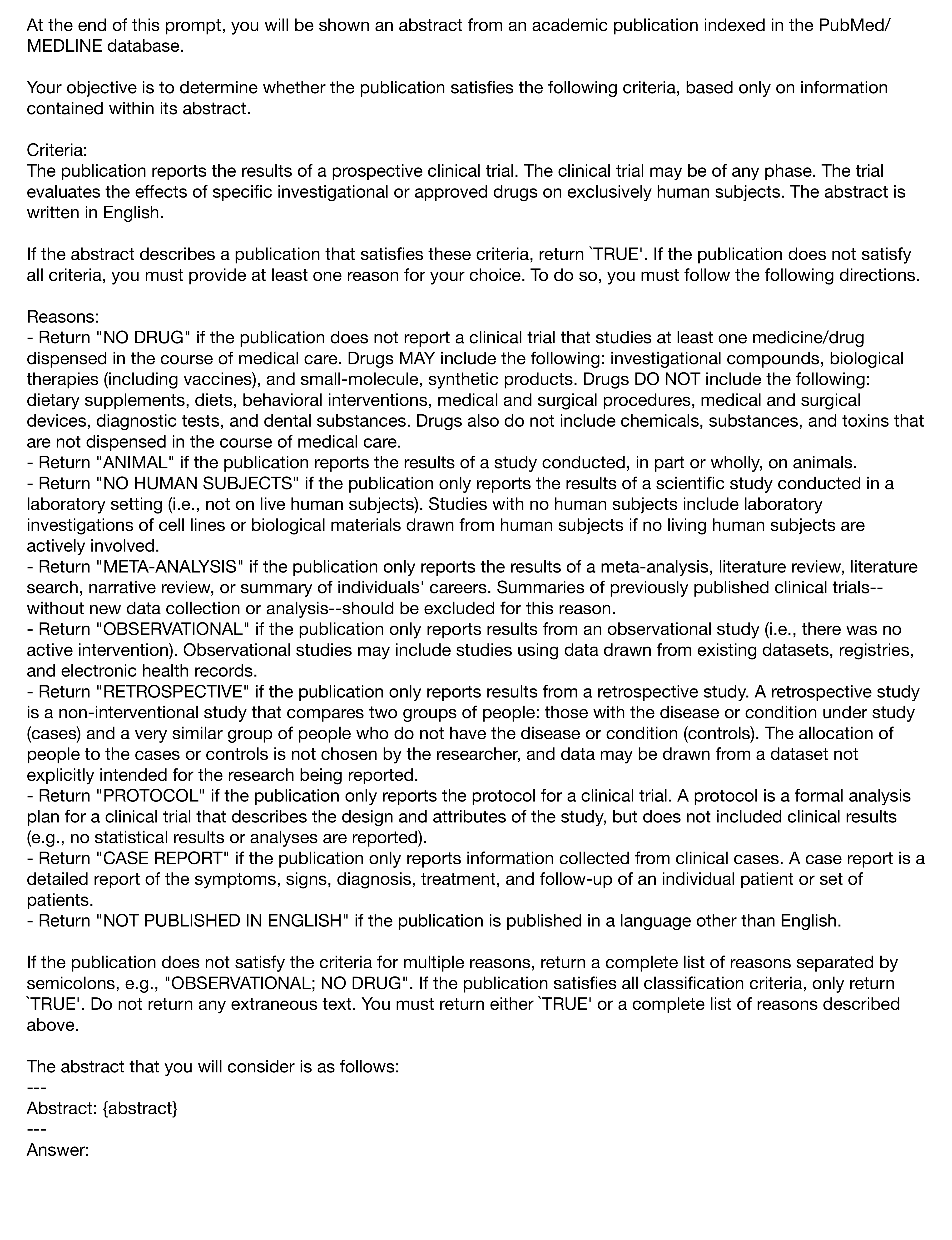}\tabularnewline
\end{tabular}
\par\end{centering}
\medskip{}
\justifying
{\footnotesize\par}
\end{figure}

\begin{figure}
\begin{centering}
\caption{Prompt 2.1: Categorize Exclusion Restriction, Examples}
\label{prompt: 2.1a}
\medskip{}
\begin{tabular}{c}
\includegraphics[scale=0.25]{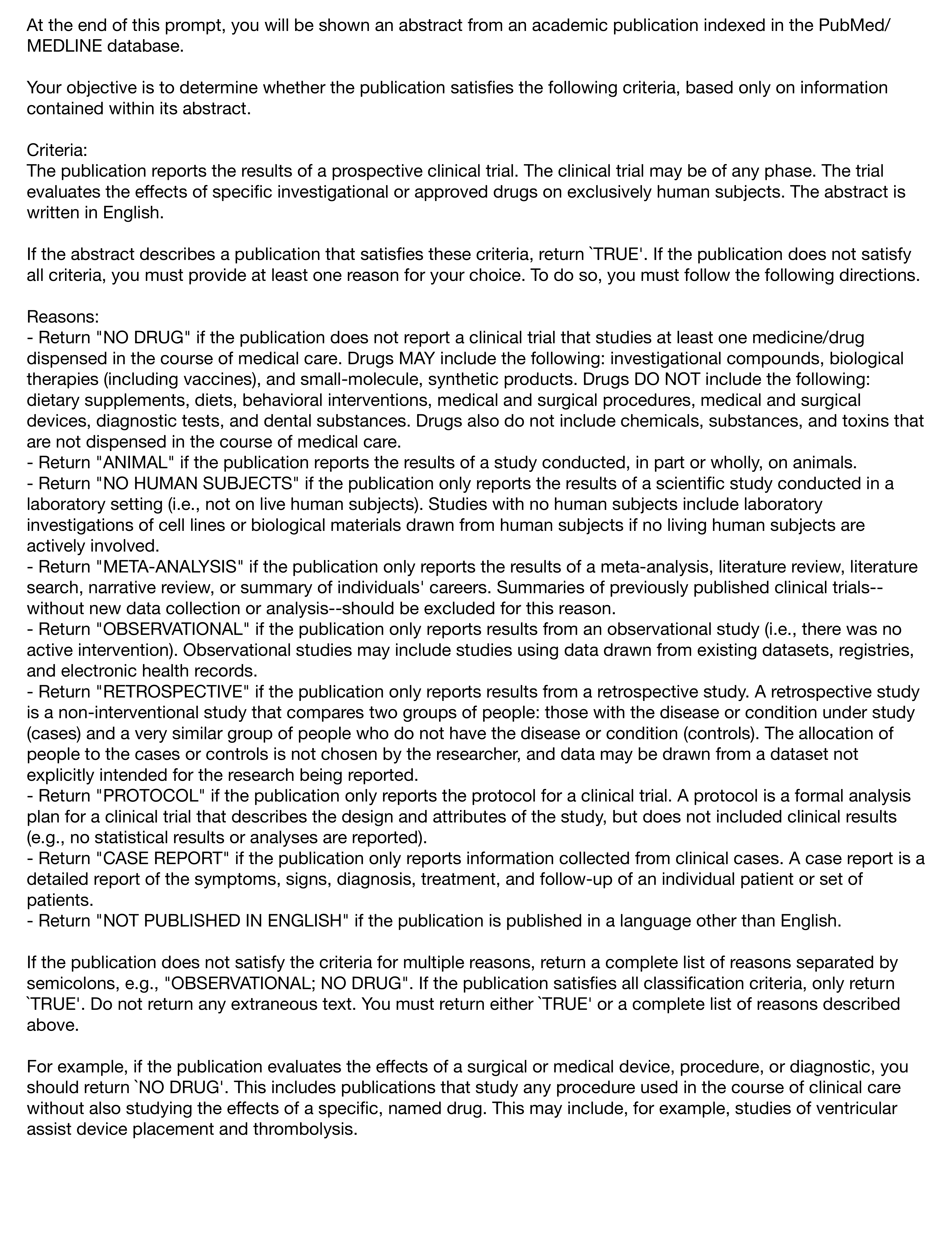}\tabularnewline
\end{tabular}
\par\end{centering}
\medskip{}
\justifying
{\footnotesize\par}
\end{figure}

\begin{figure}
\begin{centering}
\ContinuedFloat
\caption{Prompt 2.1: Categorize Exclusion Restriction, Examples (Cont.)}
\medskip{}
\begin{tabular}{c}
\includegraphics[scale=0.25]{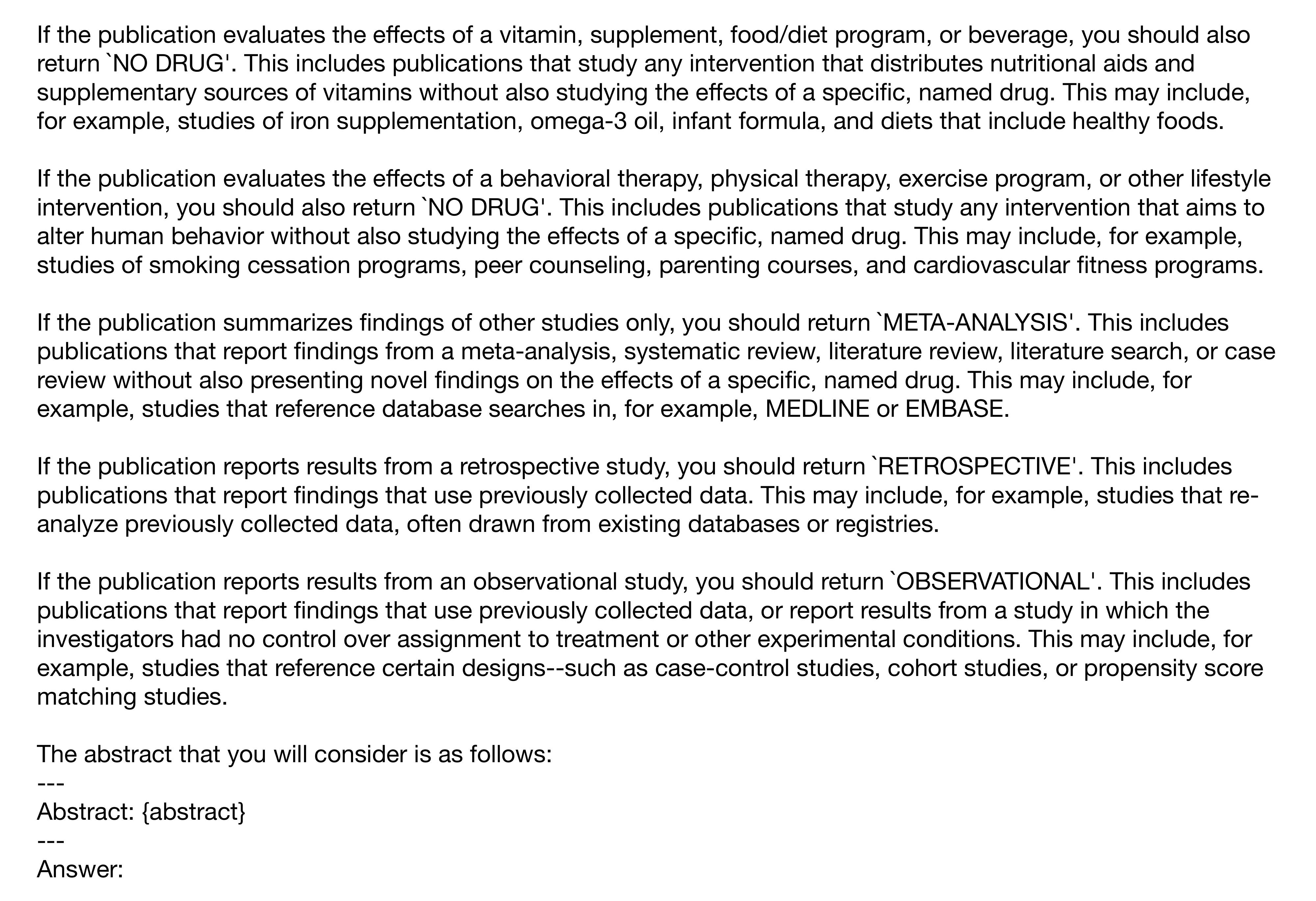}\tabularnewline
\end{tabular}
\par\end{centering}
\medskip{}
\justifying
{\footnotesize\par}
\end{figure}

\begin{figure}
\begin{centering}
\caption{Prompt 3.0: Provide Reason for Exclusion}
\label{prompt: 3.0}
\medskip{}
\begin{tabular}{c}
\includegraphics[scale=0.25]{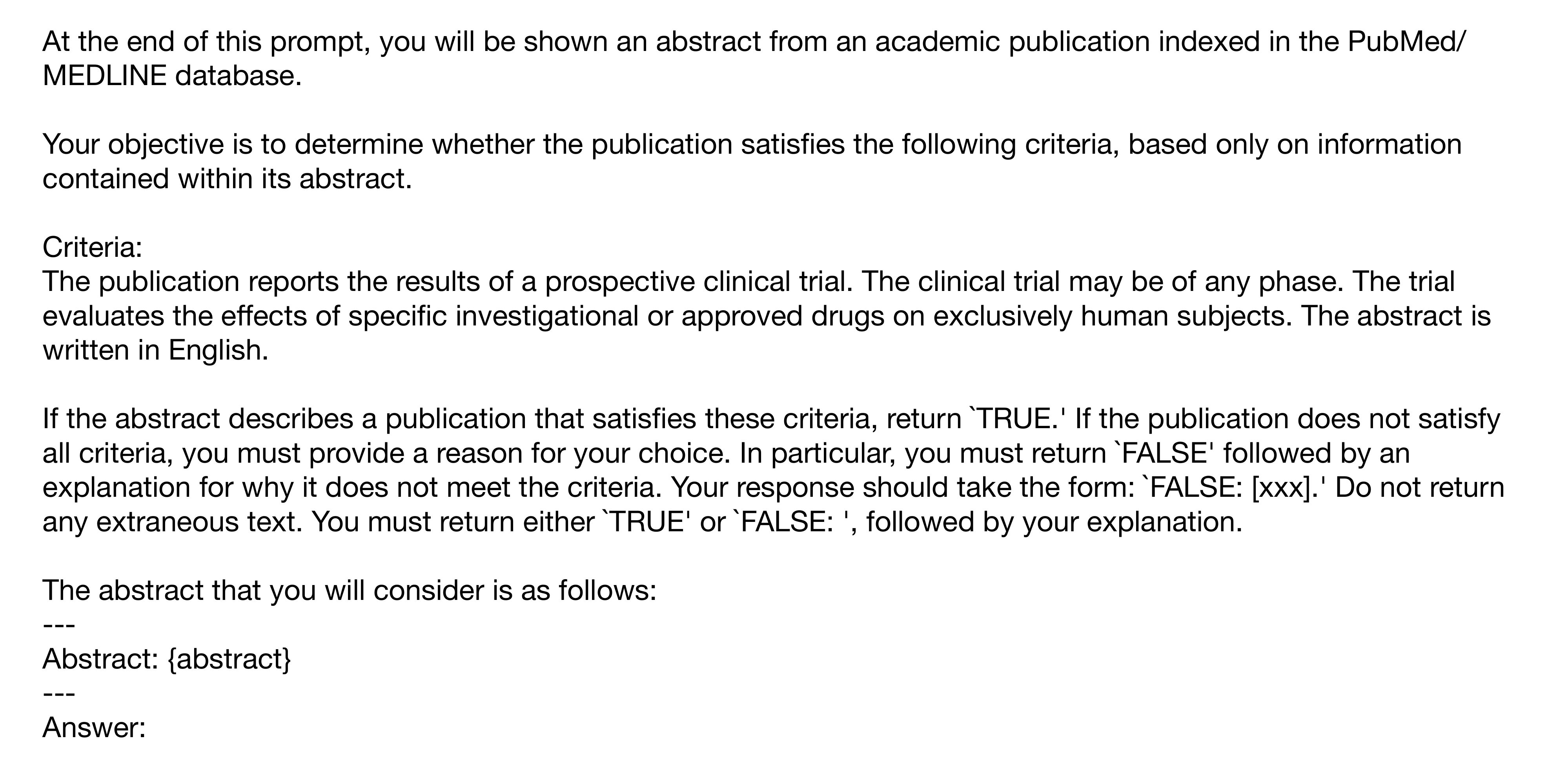}\tabularnewline
\end{tabular}
\par\end{centering}
\medskip{}
\justifying
{\footnotesize\par}
\end{figure}

\begin{figure}
\begin{centering}
\caption{Prompt 3.1: Provide Reason for Exclusion, Examples}
\label{prompt: 3.1a}
\medskip{}
\begin{tabular}{c}
\includegraphics[scale=0.25]{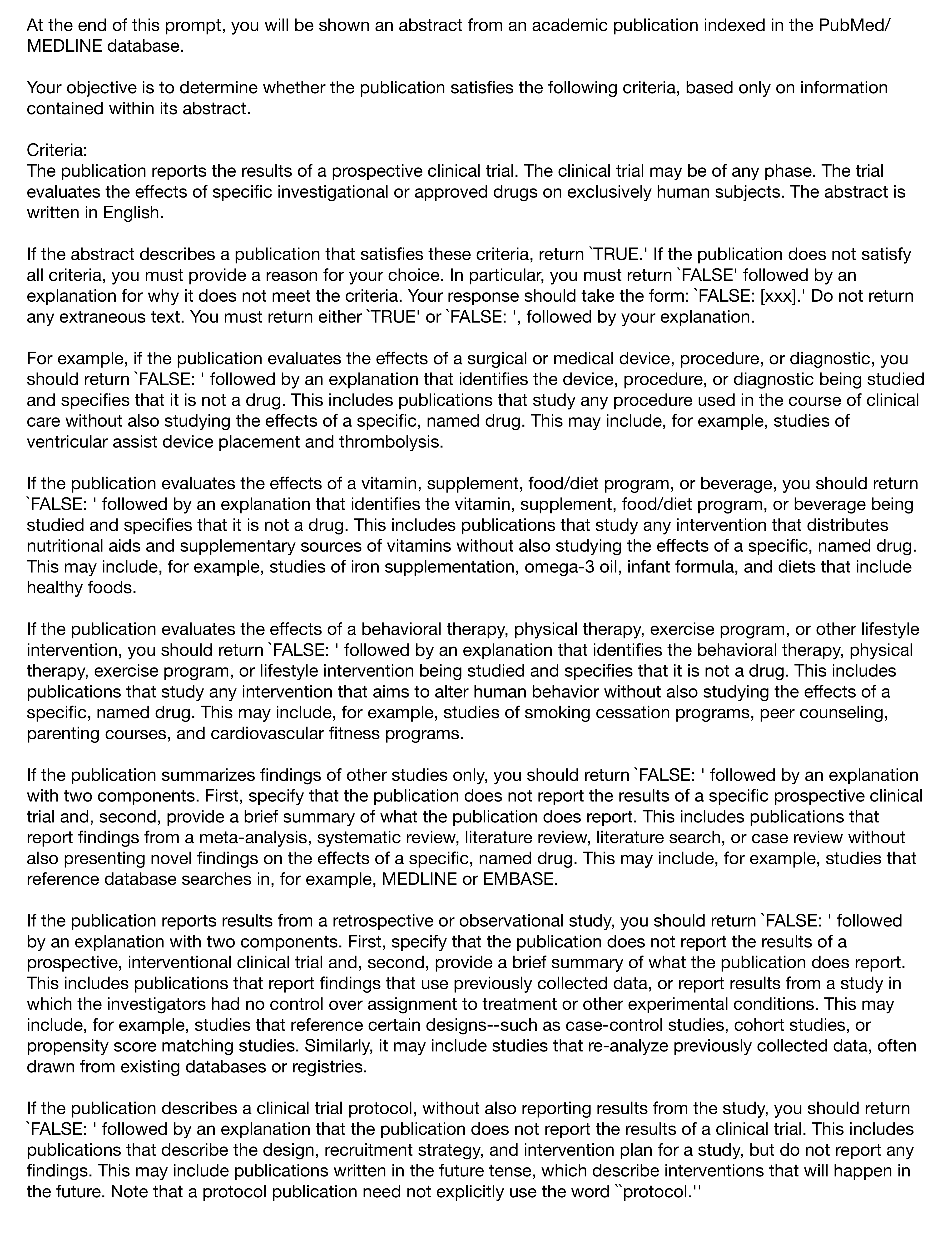}\tabularnewline
\end{tabular}
\par\end{centering}
\medskip{}
\justifying
{\footnotesize\par}
\end{figure}

\begin{figure}
\ContinuedFloat
\begin{centering}
\caption{Prompt 3.1: Provide Reason for Exclusion, Examples (Cont.)}
\medskip{}
\begin{tabular}{c}
\includegraphics[scale=0.25]{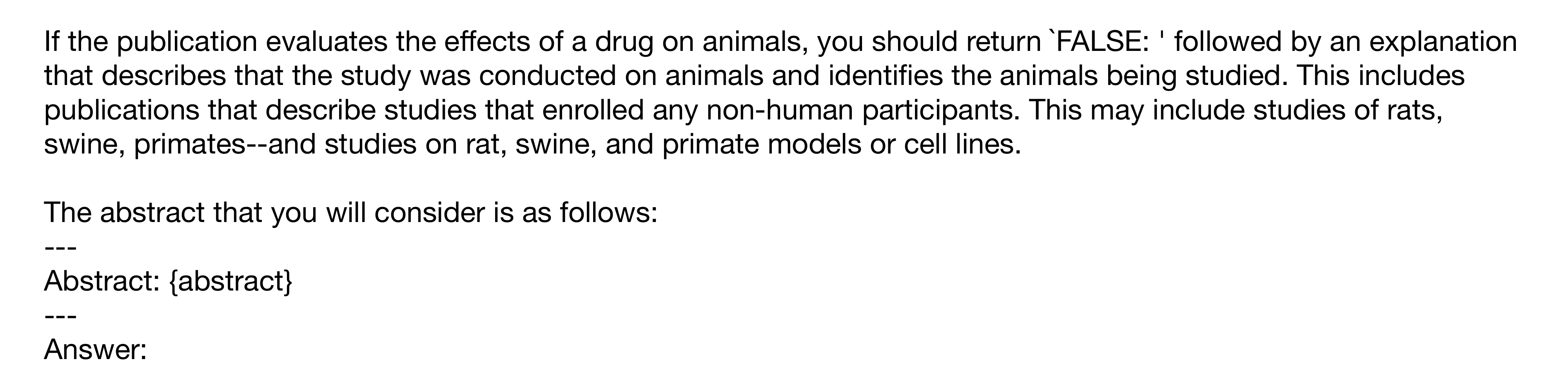}\tabularnewline
\end{tabular}
\par\end{centering}
\medskip{}
\justifying
{\footnotesize\par}
\end{figure}

\end{spacing}
\end{appendix}
\end{document}